\documentclass[useAMS,usenatbib]{mn2e}

\usepackage{amsmath}
\usepackage{natbib, txfonts, latexsym}
\usepackage{graphicx}

\newcommand{\owls}{\textit{OWLS}}
\newcommand{\msun}{M$_{\odot}$}

\title[Properties of simulated galaxy populations at $z=2$ -- II]{Physical
  properties of simulated galaxy populations at $z=2$ -- \\II. Effects of cosmology, reionization and ISM physics}
\author[Marcel R. Haas et al.]{Marcel~R.~Haas$^{1, 2, 3}$\thanks{E-mail: mhaas@physics.rutgers.edu (MRH)},
Joop~Schaye$^2$, C.~M.~Booth$^{4, 5, 2}$, Claudio~Dalla~Vecchia$^6$,
\newauthor Volker~Springel$^{7, 8}$,
Tom~Theuns$^{9, 10}$ and 
Robert~P.~C.~Wiersma$^2$\thanks{Current address: Atomic Energy of Canada Limited, Chalk River Laboratories, Chalk River, Ontario, K0J1J0, Canada} \\
\\
$^{1}$Department of Physics and Astronomy, Rutgers University, 136 Frelinghuysen Rd., Piscataway, NJ 08854, USA \\
$^{2}$Leiden Observatory, Leiden University, P.O. Box 9513, NL-2300 RA, Leiden, The Netherlands \\
$^3$Space Telescope Science Institute, 3700 San Martin Drive, Baltimore, MD 21218, USA \\
$^4$Department of Astronomy and Astrophysics, The University of Chicago, Chicago, IL 60637, USA \\
$^5$Kavli Institute for Cosmological Physics and Enrico Fermi Institute, The University of Chicago, Chicago, IL 60637, USA \\
$^{6}$Max Planck Institute for Extraterrestrial Physics, Gissenbachstra\ss{}e, 85748 Garching, Germany \\
$^7$Heidelberger Institut f\"{u}r Theoretische Studien, Schloss-Wolfsbrunnenweg 35, 69118 Heidelberg, Germany\\
$^8$Zentrum f\"ur Astronomie der Universit\"at Heidelberg, Astronomisches Recheninstitut, M\"{o}nchhofstr. 12-14, 69120 Heidelberg, Germany \\
$^9$Institute for Computational Cosmology, Department of Physics, University of Durham, Science Laboratories, South Road, Durham DH1 3LE, UK \\
$^{10}$Department of Physics, University of Antwerp, Campus Groenenborger, Groenenborgerlaan 171, B-2020 Antwerp, Belgium \\
}
\begin{document}

\date{Submitted to MNRAS}

\pagerange{\pageref{firstpage}--\pageref{lastpage}} \pubyear{2011}

\maketitle

\label{firstpage}

%%%%%%%%%%%%%%%%%%%%%%%%%%%%%%%%%%%%%%%%%%%%%%%%%%%%%%%%%%%%%%%%%%%%%%%%%%%%%%%%%%%%%%%%%%%%%%%%

\begin{abstract}
We use hydrodynamical simulations from the \owls\ project to investigate the dependence of the physical properties of galaxy populations at redshift 2 on the assumed star formation law, the equation of state imposed on the unresolved interstellar medium, the stellar initial mass function, the reionization history, and the assumed cosmology. This work complements that of Paper I, where we studied the effects of varying models for galactic winds driven by star formation and AGN. The normalisation of the matter power spectrum
strongly affects the galaxy mass function, but has a relatively small effect on the physical properties of galaxies residing in haloes of a fixed mass. Reionization suppresses the stellar masses and gas fractions of low-mass galaxies, but by $z=2$ the results are insensitive to the timing of reionization. The stellar initial mass function mainly determines the physical properties of galaxies through its effect on the efficiency of the feedback, while changes in the recycled mass and metal fractions play a smaller role. If we use a recipe for star formation that reproduces the observed star formation law independently of the assumed equation of state of the unresolved ISM, then the latter is unimportant. The star formation law, i.e.\ the gas consumption time scale as a function of surface density, determines the mass of dense, star-forming gas in galaxies, but affects neither the star formation rate nor the stellar mass. This can be understood in terms of self-regulation: the gas fraction adjusts until the outflow rate balances the inflow rate. 
\end{abstract}

%%%%%%%%%%%%%%%%%%%%%%%%%%%%%%%%%%%%%%%%%%%%%%%%%%%%%%%%%%%%%%%%%%%%%%%%%%%%%%%%%%%%%%%%%%%%%%%%


%%%%%%%%%%%%%%%%%%%%%%%%%%%%%%%%%%%%%%%%%%%%%%%%%%%%%%%%%%%%%%%%%%%%%%%%%%%%%%%%%%%%%%%%%%%%%%%%

\section{Introduction}
Understanding the star formation (SF) history of the Universe
represents one of the most fundamental pieces of the galaxy formation
puzzle.  There exists significant uncertainty as to which physical
processes control the rate of star-formation in galaxies of a given
mass.  At the coarsest level, the rate at which gas enters galaxies is
controlled, at high redshift, by the rate of growth of the dark matter
halo, while at low redshift it also depends sensitively on the rate at
which gas can cool into galaxies \citep[e.g.][]{whiterees78, hernquistspringel03,
choinagamine10, owls, vandevoort11}.  However, matching observational constraints on galaxy masses, ages and star formation rates (SFRs) is much more complex than this simple picture would suggest, and although many attempts have been made to model the formation of a galaxy population using complex hydrodynamical simulations and semi-analytic models, there exists no clear consensus on what combination of physical processes are required to explain the distribution of galaxies seen in the local Universe.

Difficulties in the simulations begin where gas becomes dense enough to form stars. This interstellar medium (ISM) gas has an extremely complex structure. Amongst other processes, magnetic fields, turbulence, cosmic rays and radiative transfer may all play some role in determining the rate at which stars form \citep[e.g.][]{mckeeostriker07}. For this reason, most cosmological simulations treat the gas in the ISM with very simple ``sub-grid'' prescriptions.  Additionally, the limited numerical resolution of the simulations prevents a detailed modelling of SF, and they need to rely on empirical laws \citep[e.g.][]{kennicutt98}. However, recent observations \citep{kennicutt07,bigiel08} show that the star formation rate surface density is a function of the molecular hydrogen density, or the surface density of cold gas \citep{schaye04, krumholz11, gloverclark12}. Simulations with sufficient resolution to resolve gas with temperatures $\ll 10^4$ K can include a more detailed treatment of the ISM \citep[e.g.][]{gnedin09, ceverinoklypin09, agertz09, christensen12}, but are currently limited to a very small number of galaxies.

Furthermore, the stellar initial mass function (IMF) assumed in the simulations is not predicted from first principles, but is imposed.  Observational determinations of the stellar IMF are very challenging outside the solar neighbourhood so the local IMF is usually assumed to be universal and is applied to all galaxies at all redshifts. In addition to uncertainties related to processes that take place inside galaxies, the ultra-violet (UV) background produced by the first generation of galaxies, ionizes the Universe and bathes gas in ionizing radiation. This ionizing UV background can then strongly suppress the infall of gas into low-mass haloes \citep[e.g.][]{quinn96, okamoto08, pawlikschaye09, hambrick11}.

It is therefore important to assess, independently, how each of these uncertain physical processes affects the properties of galaxies formed in cosmological simulations.  In \citet[][hereafter Paper I]{PaperI} of this series we use cosmological hydrodynamical simulations from the OverWhelmingly Large Simulations \citep[OWLS;][]{owls} to investigate the effects of cooling and feedback on the galaxy population at $z=2$. In this companion paper we turn our attention to other physical processes, namely cosmology, reionization, the treatment of the high density gas, the star formation law and the stellar initial mass function.  This work (together with Paper I) complements that of \citet{owls}, where the cosmic SF histories predicted by the \owls\ simulations were analysed.  

In Paper I we describe the behaviour of the simulations in detail, and in particular we focus there on the effects of radiative cooling and energetic feedback from star formation and AGN. One of the main conclusion from Paper I is that the SF is self-regulated by the interplay between gas cooling onto galaxies and the feedback that offsets this gas accretion. The cooling rate onto galaxies is, for a fixed halo mass and redshift, mainly determined by the radiative cooling rate, which itself depends on chemical composition.  We found that for many integrated galaxy properties the uncertainties in models for cooling and feedback give rise to a very large spread in galaxy physical properties (see Fig.~2). In this paper we discuss the importance of cosmology, reionization, the prescription for the ISM, the SF law and the stellar initial mass function. We will show that uncertainties in these processes are of secondary importance for the total amount of stars formed and for the SFR of simulated galaxies. However, we will also show that some of the processes discussed in this paper are important for the amount of mass residing in the star forming ISM of galaxies, which we will argue can be explained in terms of self-regulated star formation. 

The structure of this paper is follows:  in Sec.~\ref{sec:owls_sims} we briefly describe the simulations used in this study and the numerical techniques we employ.  In Sec.~\ref{sec:physicsvars} we describe how galaxy properties depend upon the physics included in the simulation.  In this paper we consider the effects of: cosmology (Sec~\ref{sec:cosmology}), reionization (Sec.~\ref{sec:reion}), the effective equation of state of the ISM (Sec.~\ref{sec:eos}), the star formation law (Sec.~\ref{sec:sf}), and the stellar IMF (Sec.~\ref{sec:imf}).  Finally, in Sec.~\ref{sec:conclusions} we summarize our findings and conclude.

%%%%%%%%%%%%%%%%%%%%%%%%%%%%%%%%%%%%%%%%%%%%%%%%%%%%%%%%%%%%%%%%%%%%%%%%%%%%%%%%%%%%%%%%%%%%%%%%

\section{Numerical techniques} \label{sec:owls_sims}
The simulations comprising the \owls\ project are described fully in \citet[][]{owls}. Here we briefly summarize the reference simulation, with a focus on the physical prescriptions relevant for this paper.  This simulation will be referenced throughout this paper as `\textit{REF}'.

\begin{table}
\caption{Overview of the values of the cosmological parameters according to WMAP3 (\textit{OWLS} reference) and WMAP1 (as used in the Millennium Simulation). Symbols have their usual meaning.}             % title of Table
\label{tab:cosmology}      % is used to refer this table in the text
\centering                          % used for centering table
\begin{tabular}{l l l}        % left aligned columns (2 columns)
\hline                % inserts one horizontal line
 & WMAP3 & WMAP1 \\
\hline
$\Omega_{\rm m}$ & 0.238 & 0.25 \\      
$\Omega_{\rm b}$ & 0.0418 & 0.045 \\
$\Omega_\Lambda$ & 0.762 & 0.75 \\
$\sigma_8$ & 0.74 & 0.9 \\
$n$  & 0.951 & 1.0 \\
$h = H_0$ / (100 km s$^{-1}$ Mpc$^{-1}$) & 0.73 & 0.73 \\
\hline                                   %inserts single line
 \end{tabular}
\end{table}

\begin{figure*}
  \centering
   \resizebox{0.9\hsize}{!}{\includegraphics{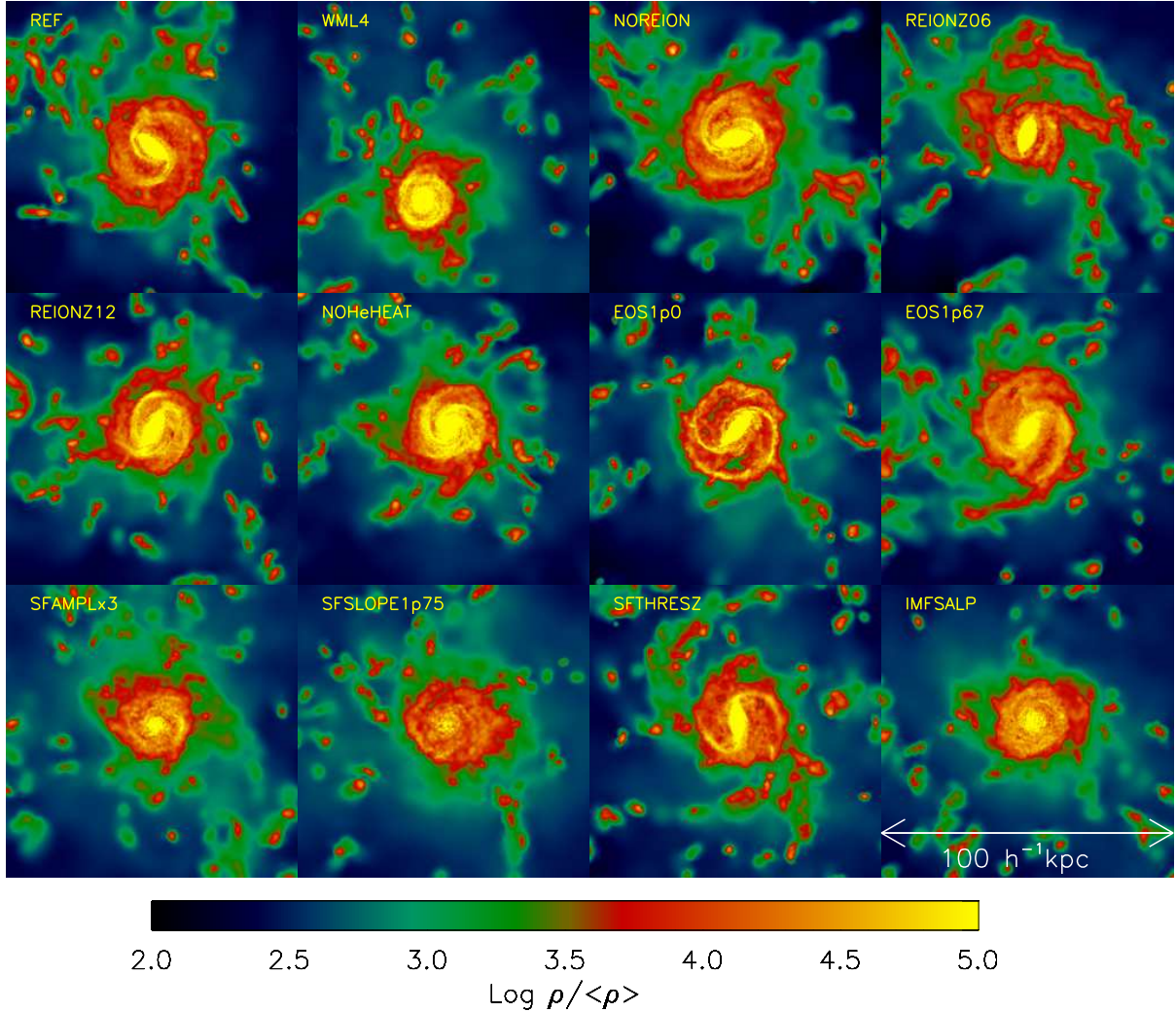}}
    \caption{A graphical representation of a galaxy in $10^{12.5}$ \msun\ halo in 13 of our simulations at redshift 2. The colour coding shows the gas density in a slice of 100 $h^{-1}$ kpc thickness, divided by the mean density of the universe. All frames are of size 100 co-moving kpc/h and are centered on the position of the galaxy in the `\textit{REF}' simulation. All frames have a thickness of 100 co-moving $h^{-1}$kpc. The images are oriented so that the galaxy is face on in the `\textit{REF}' simulation. Although all simulations form a disk, it is clear that some of the physics prescriptions considered in this paper can have a significant effect on galaxy morphology.}
   \label{fig:prettypics}
\end{figure*}

All simulations are performed with an extended version of the N-Body Tree/SPH code \textsc{Gadget3} \citep[last described in][]{gadget2} in periodic boxes of 25 co-moving $h^{-1}$Mpc and contain $512^3$ dark matter and the same number of baryonic particles (which can be either collisionless `stars' or collisional `gas' particles). The particle mass of the simulations we use here is $8.68 \times 10^{6}$ \msun\ for dark matter and $1.85 \times 10^{6}$ \msun\ for baryons (initially, the baryonic particle masses  change in the course of the simulation due to mass transfer from star particles to gas particles). The gravitational softening length is initially fixed in co-moving coordinates at 1/25 the mean inter-particle spacing (1.95 co-moving $h^{-1}$kpc). Below $z=2.91$ the softening is fixed in proper units, at 0.5 $h^{-1}$kpc. We provide tests showing convergence of our results with respect to simulation box size and the particle number in the appendix of Paper I.

The cosmology assumed in the reference simulation is summarized in Table~\ref{tab:cosmology} and is deduced from the WMAP 3 year results \citep{WMAP3}. The results are largely consistent with the more recent WMAP7 results \citep{komatsu11}, the most notable differences are in $\sigma_8$, which is 2.3 $\sigma$ lower in WMAP3 than in WMAP7 and in the Hubble parameter which is 1$\sigma$ lower in the WMAP3 than in the WMAP7 data. The primordial helium mass fraction is set to 0.248.

As the subgrid model variation is the main power of the \owls\ suite, we will now describe the parameters and subgrid models used in the reference simulation, before varying them in later sections. Radiative cooling and heating are calculated element-by-element by explicitly following the 11 elements H, He, C, N, O, Ne, Mg, Si, S, Ca and Fe in the presence of the Cosmic Microwave Background and the \citet{haardtmadau} model for the UV/X-ray background radiation from quasars and galaxies, as described in \citet{wiersma09cooling}. The timed release of these elements by stars is followed as described by \citet{wiersma09chemo}. The gas is assumed to be optically thin and in photo-ionization equilibrium.  The simulations model hydrogen reionization by switching on the \citet{haardtmadau} background at $z=9$. Helium reionization is modelled by heating the gas by an extra amount of 2 eV per atom. This heating takes place at z = 3.5, with the heating spread in redshift with a Gaussian filter with $\sigma(z) = 0.5$. As shown by \citet{wiersma09chemo},  the reionization prescription used in these simulations roughly matches the temperature history of the intergalactic medium (IGM) inferred from observations by \citet{schaye00}.

In the centres of haloes the density and pressure are so high, that the gas is expected to be in multiple phases, with cold and dense molecular clouds embedded in a warmer, more tenuous gas. This multi-phase structure is not resolved by our simulations (and the simulations lack the physics to describe these phases), so we impose a polytropic effective equation of state for particles with densities $n_{\textrm{\scriptsize H}} > 10^{-1}$ cm$^{-3}$. These particles are also assumed to be star forming, as this is the density required to form a molecular phase \citep{schaye04}. We set the pressure of these particles to $P \propto \rho^{\gamma_{\textrm{\scriptsize eff}}}$, where $\gamma_{\textrm{\scriptsize eff}}$ is the polytropic index and $\rho$ is the physical proper mass density of the gas. In order to prevent spurious fragmentation due to a lack of numerical resolution we set $\gamma_{\textrm{\scriptsize eff}} = 4/3$, as then the ratio of the Jeans length to the SPH kernel and the Jeans mass are independent of density \citep{schayedallavecchia08}. The normalization of the polytropic equation of state is such that for atomic gas with primordial composition, the energy per unit mass corresponds to $10^4$ K, namely ($P/k = 1.08 \times 10^{3}$ K cm$^{-3}$ for $n_{\textrm{\scriptsize H}} = 10^{-1}$ cm$^{-3}$). The implementation of SF is stochastic, as described in \citet{schayedallavecchia08}, with a pressure-dependent SFR, obtained from the assumption of local hydrostatic equilibrium and the observed Kennicutt-Schmidt law \citep{kennicutt98}.

Energetic feedback from star formation is implemented kinetically. On average, we give 2 of the SPH neighbours of each
newly formed star particle a `kick' such that the total energy in the outflow corresponds to roughly 40\% of the energy available from
supernovae (SNe) of type II (including Ib,c). $\eta = 2$  is the mass loading factor. For our assumed Chabrier (2003) IMF this corresponds total wind velocity of 600 km s$^{−1}$ . See \citet{dallavecchiaschaye08} for more details on the kinetic implementation of SN feedback.

Haloes are identified using a Friends-of-Friends algorithm, as described in Sec.~2.2 of Paper I and we only show results over mass ranges where the simulation achieves numerical convergence (Appendix A of Paper I).  This corresponds to haloes that contain at least 100 star particles when considering properties as a function of stellar mass and a minimum of 2000 dark matter particles when we plot properties against total halo mass.

%%%%%%%%%%%%%%%%%%%%%%%%%%%%%%%%%%%%%%%%%%%%%%%%%%%%%%%%%%%%%%%%%%%%%%%%%%%%%%%%%%%%%%%%%%%%%%%%

\section{Isolating the effects of the input physics}\label{sec:physicsvars}

\begin{table*}
\caption{Overview of the simulations and the input physics that is varied in this paper. Bold face values represent departures from the reference model (`\textit{REF}'). The first column gives the name of the simulation; the second column indicates the assumed cosmological parameters; the third lists the redshift of reionization; the fourth column specifies whether or not extra heat was injected during He reionization happens; the fifth column specifies the adiabatic index of the equation of state imposed on star forming gas; the sixth, seventh and eighth columns indicate the amplitude, slope and threshold of the SF law \citep[K98 indicates the value from][]{kennicutt98}; the ninth column specifies the IMF; the tenth column indicates whether the simulation includes SN feedback and the final column specifies the section in which the simulation is discussed.}.
\label{tab:simulations}      % is used to refer this table in the text
\centering                          % used for centering table
\begin{tabular}{llllllllllll}        % left aligned columns
\hline
\hline
Name &				Cosmology &	Reion. &	He &		EoS index &	KS-law &	KS-law &	KS-law &	IMF & 		SN  & 	Sect. \\
     &                                    &	$z$ &		reion. & 	$\gamma_\textrm{eff}$ &		ampl. &		slope &		threshold &	&		feedback & \\
(1) & (2) & (3) & (4) & (5) & (6) & (7) & (8) &(9) & (10) & (11) \\
\hline
\emph{REF}	&		WMAP3 &		9 &		yes &		4/3 &		K98 &		1.4 (K98) &	$n_\textrm{H} = 0.1$ cm$^{-3}$ &		Chabrier &	yes &	All \\
\emph{MILL} &			{\bf WMAP1} &	9 &		yes &		4/3 &		K98 &		1.4 (K98) &	$n_\textrm{H} = 0.1$ cm$^{-3}$ &		Chabrier &{\bf energy x2}&	3.1 \\
\emph{WML4} &			WMAP3 &	9 &		yes &		4/3 &		K98 &		1.4 (K98) &	$n_\textrm{H} = 0.1$ cm$^{-3}$ &		Chabrier &	{\bf energy x2} &	3.1 \\
\emph{NOREION} &		WMAP3 &		{\bf n.a.} &	yes &		4/3 &		K98 &		1.4 (K98) &	$n_\textrm{H} = 0.1$ cm$^{-3}$ &		Chabrier &	yes &	3.2 \\
\emph{REIONZ06} &		WMAP3 &		{\bf 6} &	yes &		4/3 &		K98 &		1.4 (K98) & 	$n_\textrm{H} = 0.1$ cm$^{-3}$ &		Chabrier &	yes &	3.2 \\
\emph{REIONZ12} &		WMAP3 &		{\bf 12} &	yes &		4/3 &		K98 &		1.4 (K98) &	$n_\textrm{H} = 0.1$ cm$^{-3}$ &		Chabrier &	yes &	3.2 \\
\emph{NOHeHEAT} &		WMAP3 &		9 &		{\bf no} &	4/3 &		K98 &		1.4 (K98) &	$n_\textrm{H} = 0.1$ cm$^{-3}$ &		Chabrier &	yes &	3.2 \\
\emph{EOS1p0} &		        WMAP3 &			9 &		yes &		{\bf 1} &	K98 &		1.4 (K98) &	$n_\textrm{H} = 0.1$ cm$^{-3}$ &		Chabrier &	yes &	3.3 \\
\emph{EOS503} &		        WMAP3 &			9 &		yes &		{\bf 5/3} &	K98 &		1.4 (K98) &	$n_\textrm{H} = 0.1$ cm$^{-3}$ &		Chabrier &	yes &	3.3 \\
\emph{SFAMPLx3} &		WMAP3 &		9 &		yes &		4/3 &		{\bf K98 x 3} &	1.4 (K98) &	$n_\textrm{H} = 0.1$ cm$^{-3}$ &		Chabrier &	yes &	3.4 \\
\emph{SFSLOPE1p75} &		WMAP3 &		9 &		yes &		4/3 &		K98 &		{\bf 1.75} &	$n_\textrm{H} = 0.1$ cm$^{-3}$ &		Chabrier &	yes &	3.4 \\
\emph{SFTHRESZ} &		WMAP3 &		9 &		yes &		4/3 &		K98 &		1.4 (K98) &	{\bf Z-dep.} &	Chabrier &	yes &	3.4 \\
\emph{IMFSALP} &		WMAP3 &		9 &		yes &		4/3 &		K98 &		1.4 (K98) &	$n_\textrm{H} = 0.1$ cm$^{-3}$ &		{\bf Salpeter} & yes &	3.5 \\
\hline
\hline
\end{tabular}
\end{table*}

In this section we discuss each of the variations to the input physics in turn. Table~\ref{tab:simulations} summarizes the simulations used in this paper and indicates in which subsection they are discussed. Bold-face values indicate departures from the reference model. In the subsequent sections we will discuss the sensitivity of galaxy properties to cosmology (Section~\ref{sec:cosmology}), reionization (Section~\ref{sec:reion}), the polytropic equation of state of the ISM (Section~\ref{sec:eos}), the assumed SF law (Section~\ref{sec:sf}) and the IMF (Section~\ref{sec:imf}).

A graphical representation of the gas density of a $z=2$ galaxy formed in the different simulations (except `\textit{MILL}', see below) is shown in Fig.~\ref{fig:prettypics}. The galaxy resides in a halo of total mass $\sim 10^{12.5}$ \msun. It was first identified in the `\textit{REF}' simulation, where its position (defined as the centre of mass of all particles within 10\% of the virial radius) was determined. The line of sight is along the z-axis, which is almost perfectly aligned with the angular momentum vector of the gas within 10\% of the virial radius ($\cos(\phi) = 0.994$). For the other simulations the image is centered on the same position, illustrating the remarkable similarity in the positions and orientations of the galaxies. The `\textit{MILL}' simulation was run with different cosmological parameters, resulting in a different distribution of galaxies over the volume. This model is therefore not plotted in Fig.~\ref{fig:prettypics}.  The physics variations discussed in this paper do not lead to large differences in the total stellar content of galaxies.  Nevertheless, we see in Fig.~\ref{fig:prettypics} that there are significant differences in galaxy morphologies, although a gaseous disk forms in all cases.  These differences will be discussed throughout the remainder of this section.

Fig. 2 of paper I, as well as Figs.~\ref{fig:sims_cosmology} through \ref{fig:sims_imf} show, as a function of total halo mass the nine different galaxy properties we consider in this paper: medians of stellar mass fraction ($f_\textrm{star} = M_\textrm{star} / M_\textrm{tot}$, panel A), SFR (panel B), baryon fraction ($f_\textrm{baryon} = M_\textrm{baryon} / M_\textrm{tot}$, panel C), fraction of star-forming gas ($f_\textrm{ISM} = M_\textrm{ISM} / M_\textrm{tot}$, panel D) and gas mass fraction ($f_\textrm{gas, halo} = (M_\textrm{gas, total} - M_\textrm{ISM}) / M_\textrm{tot}$, panel E). Then, as a function of stellar mass:  medians of the molecular gas mass in the ISM (panel F), specific SFR (sSFR = SFR/$M_{*}$, panel G), the inverse of the gas consumption timescale (SFR/$M_\textrm{ISM}$, panel H) and galaxy number density (the galaxy stellar mass function, panel I). In panels F, G and I we compare to observations of, respectively, \citet{genzel10}, \citet{daddi07} and \citet{marchesini09}, as described in detail in Paper I. The black line in all panels is the `\textit{REF}' simulation. The physics variations presented in Paper I strongly influence the amount of stars formed, the star formation rate and the gas and baryon fractions in the halo. As discussed below, the physics variations described in this paper mainly influence gas consumption time scales and ISM mass fractions.

\subsection{Cosmology} \label{sec:cosmology}

\begin{figure*}
\centering
\includegraphics[width=0.33\linewidth]{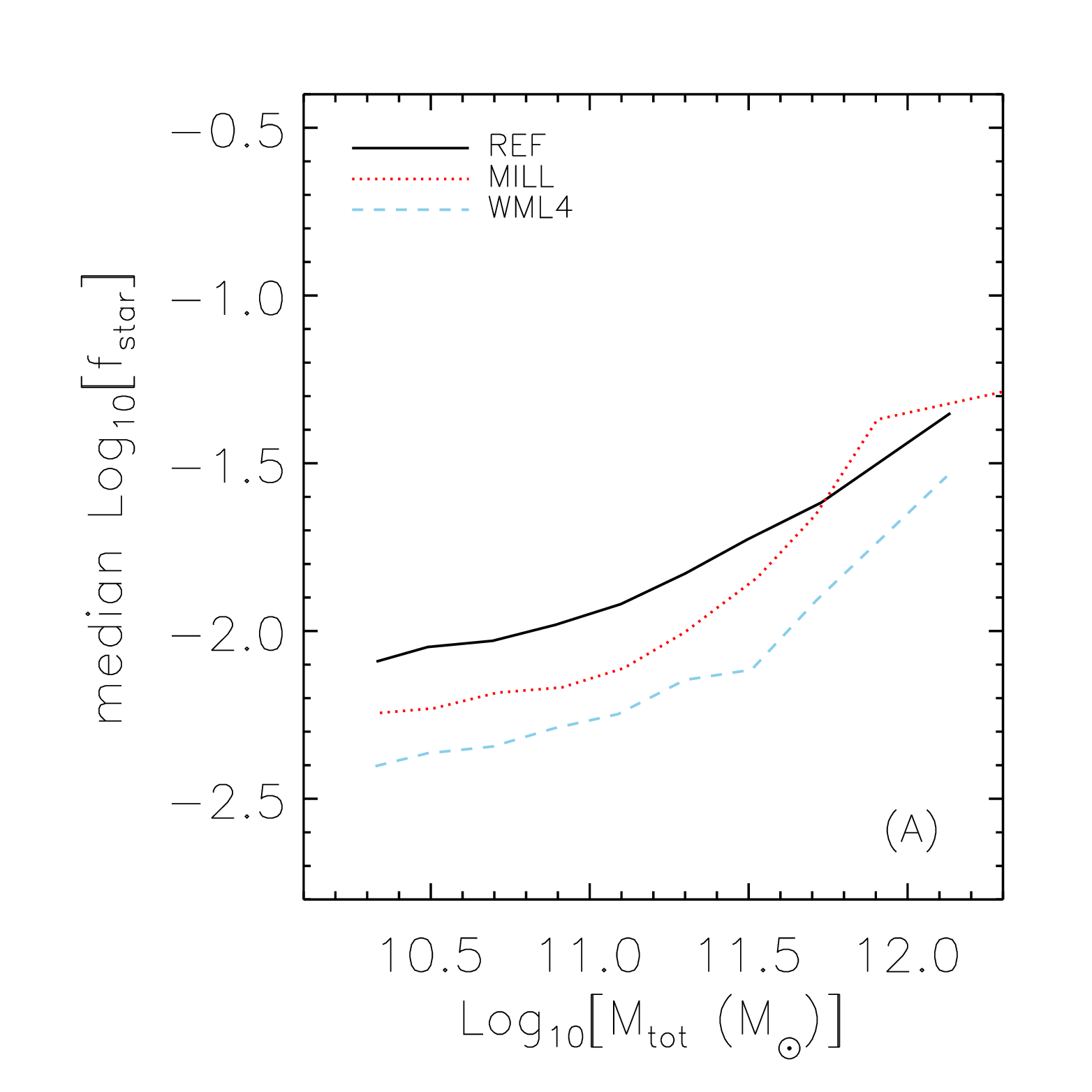}
\includegraphics[width=0.33\linewidth]{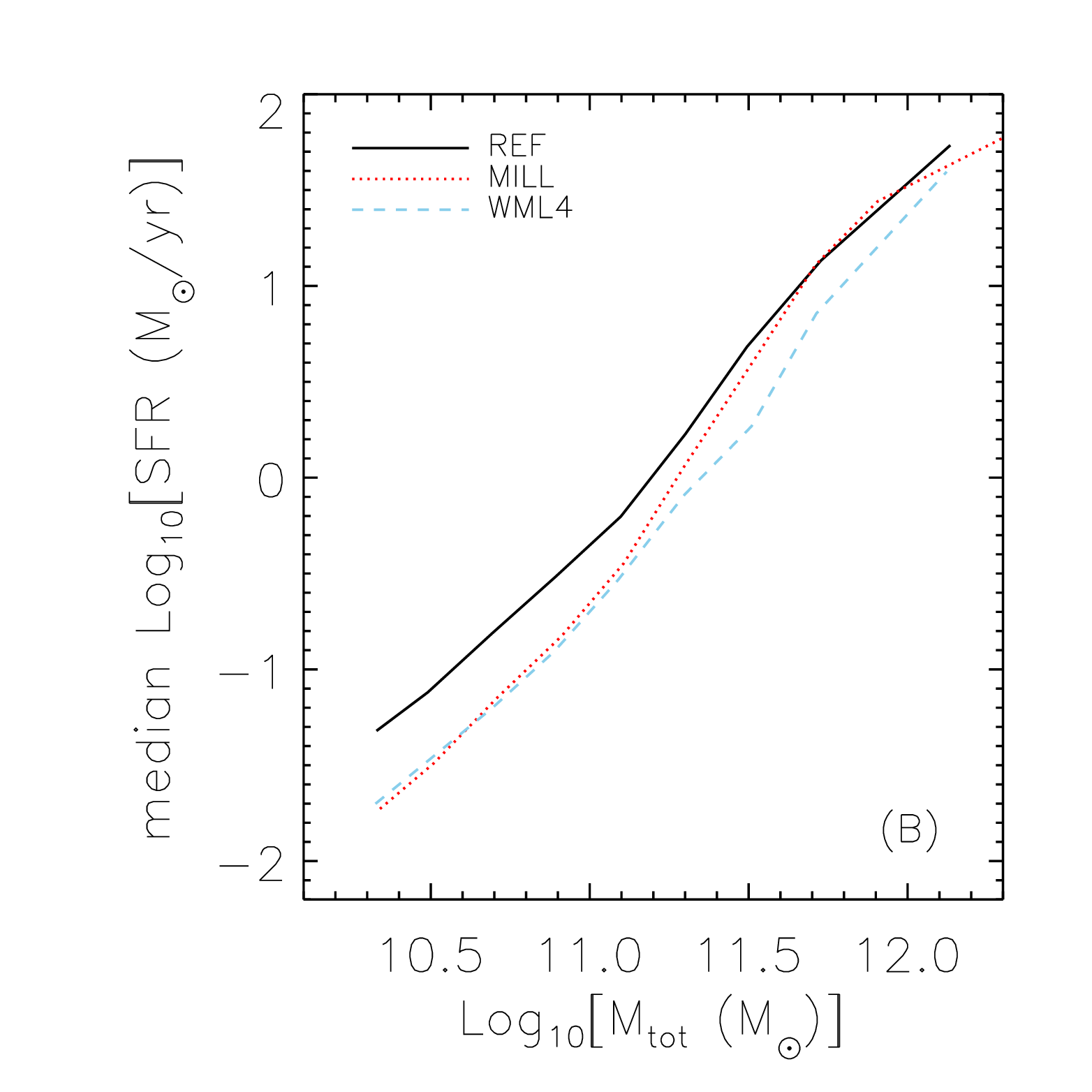}
\includegraphics[width=0.33\linewidth]{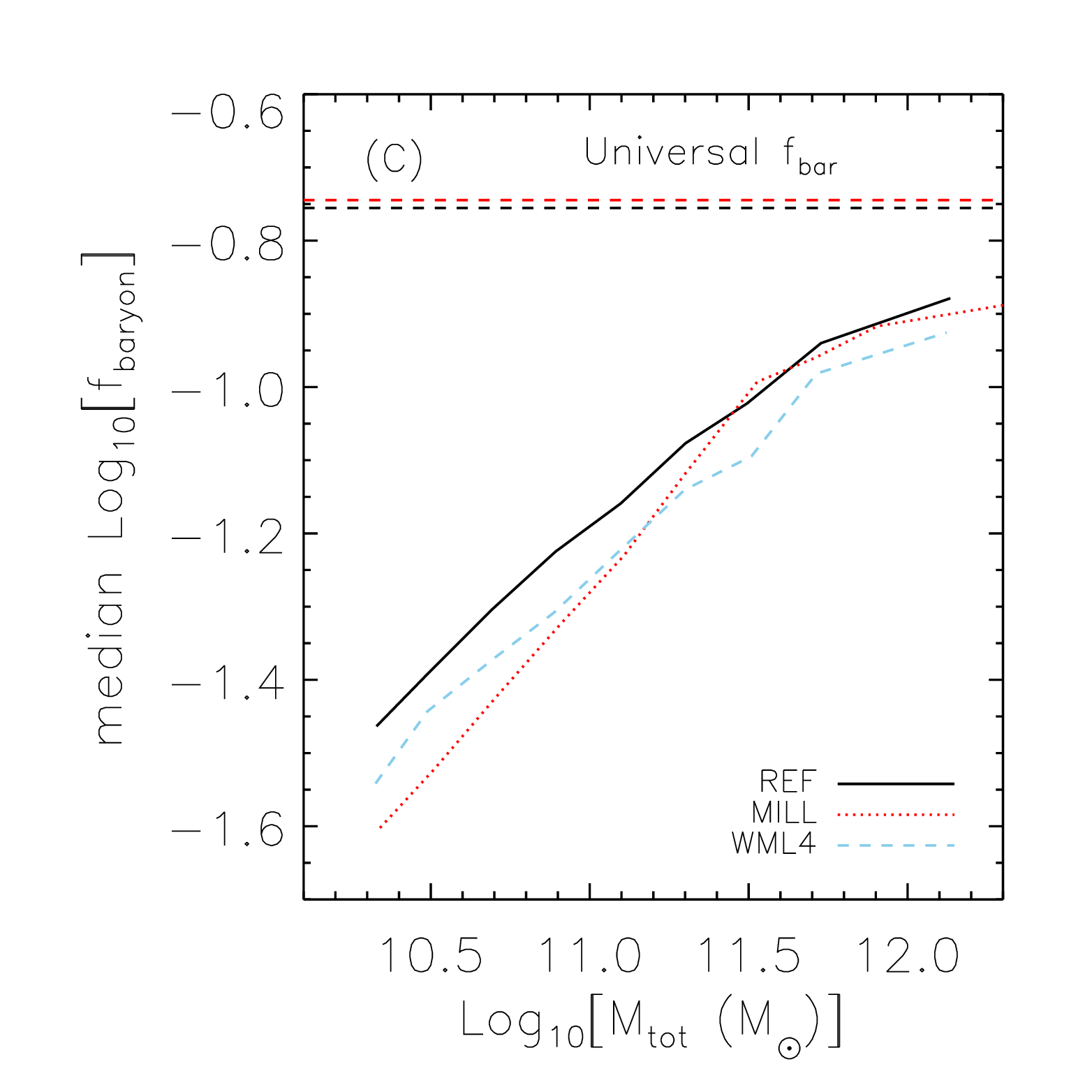} \\
\includegraphics[width=0.33\linewidth]{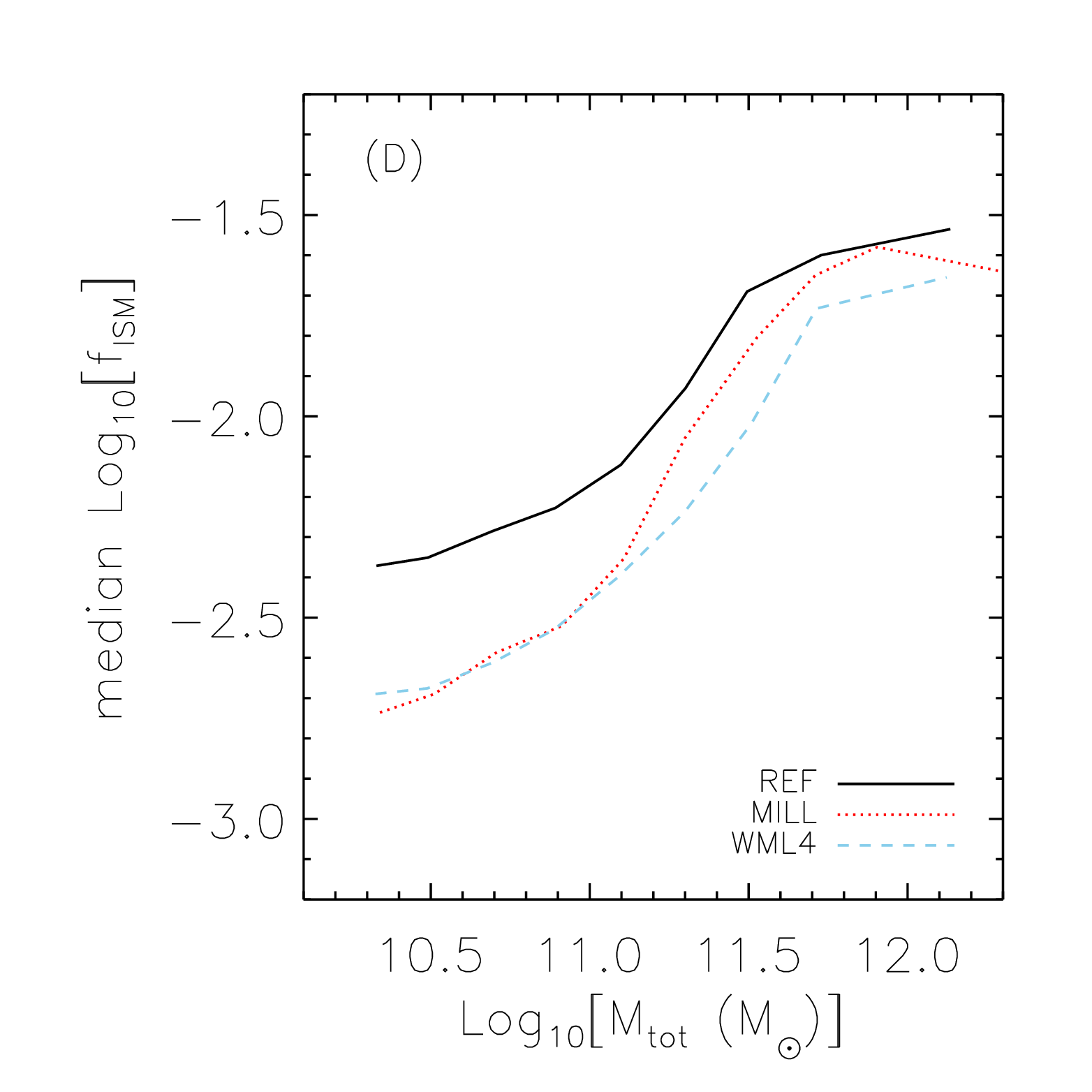}
\includegraphics[width=0.33\linewidth]{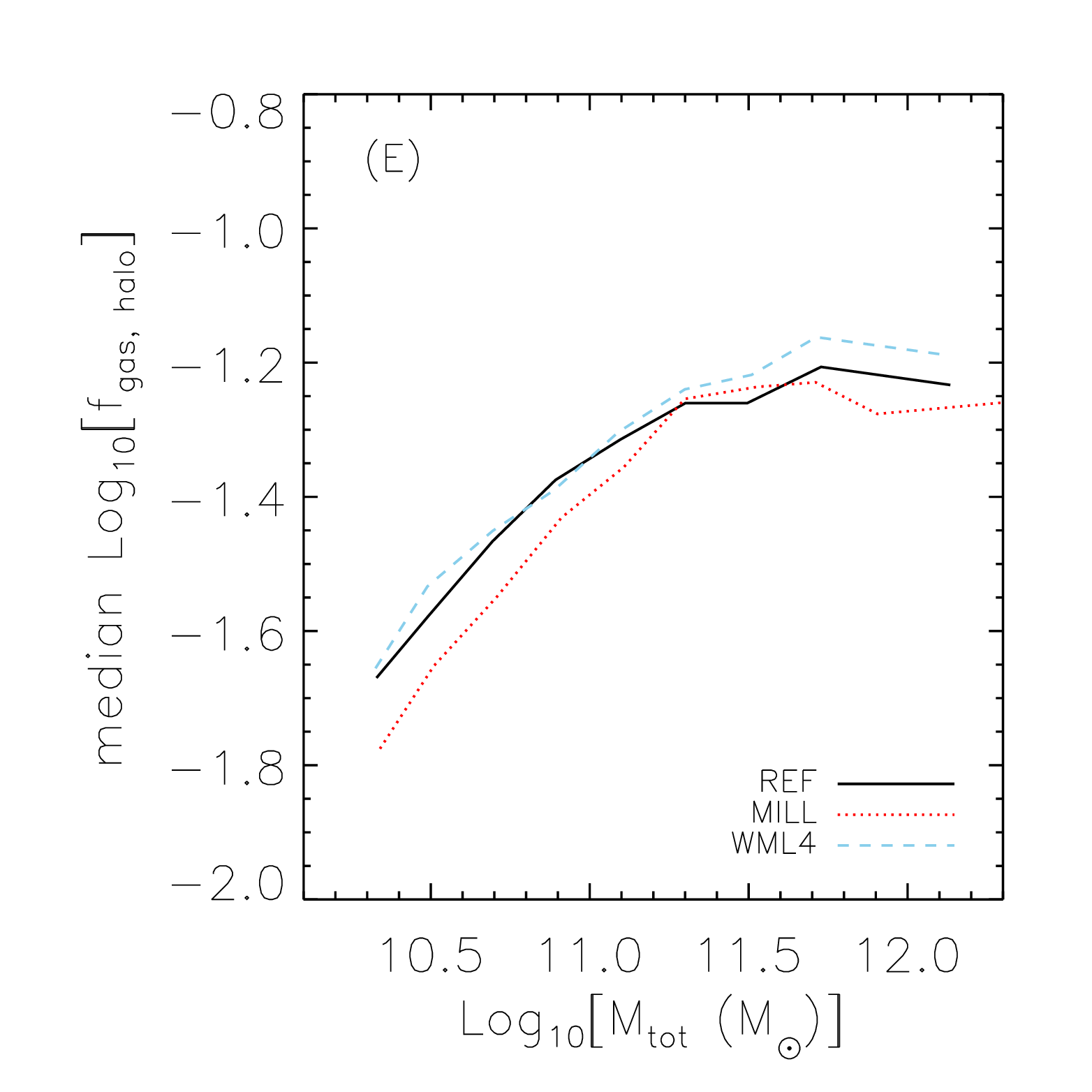}
\includegraphics[width=0.33\linewidth]{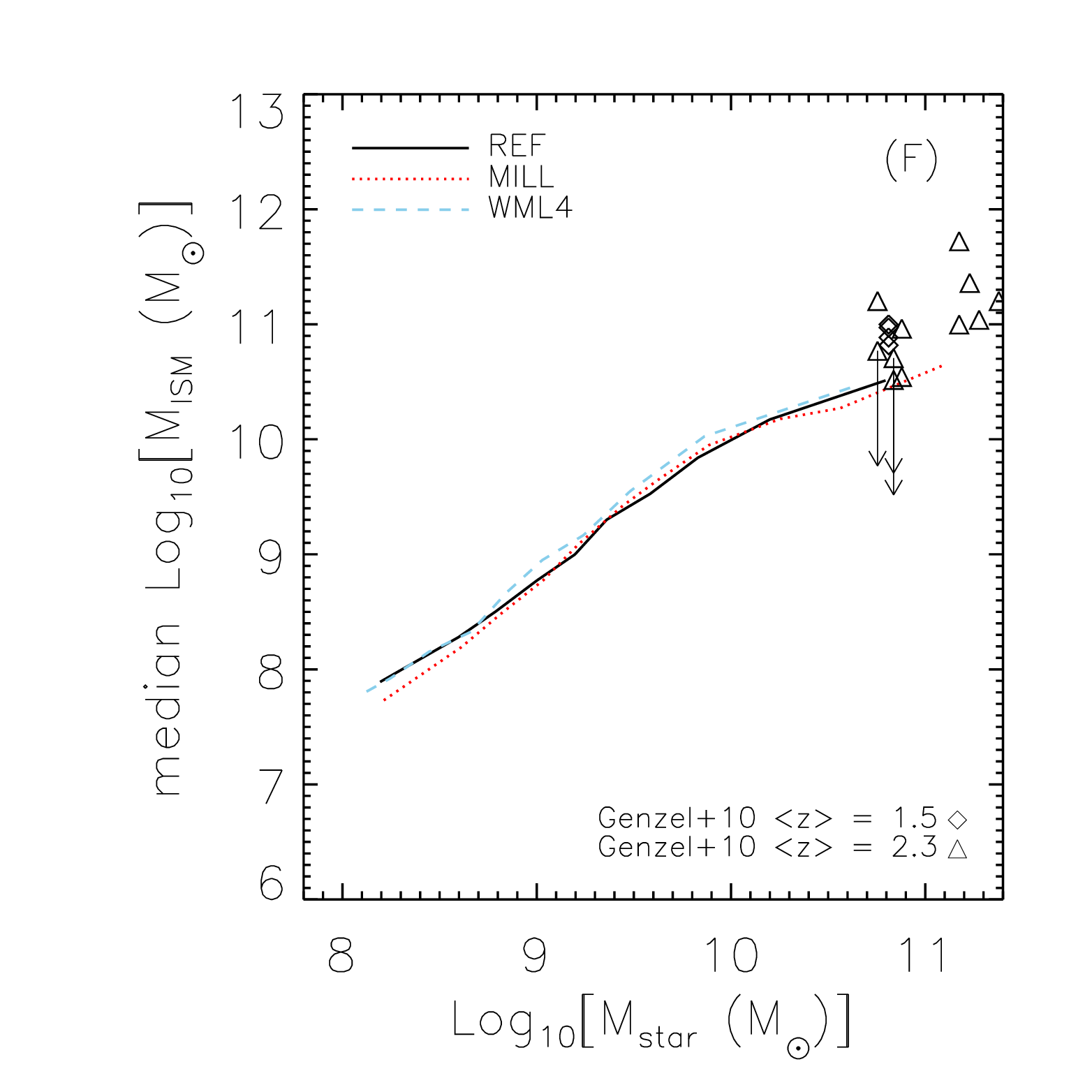} \\
\includegraphics[width=0.33\linewidth]{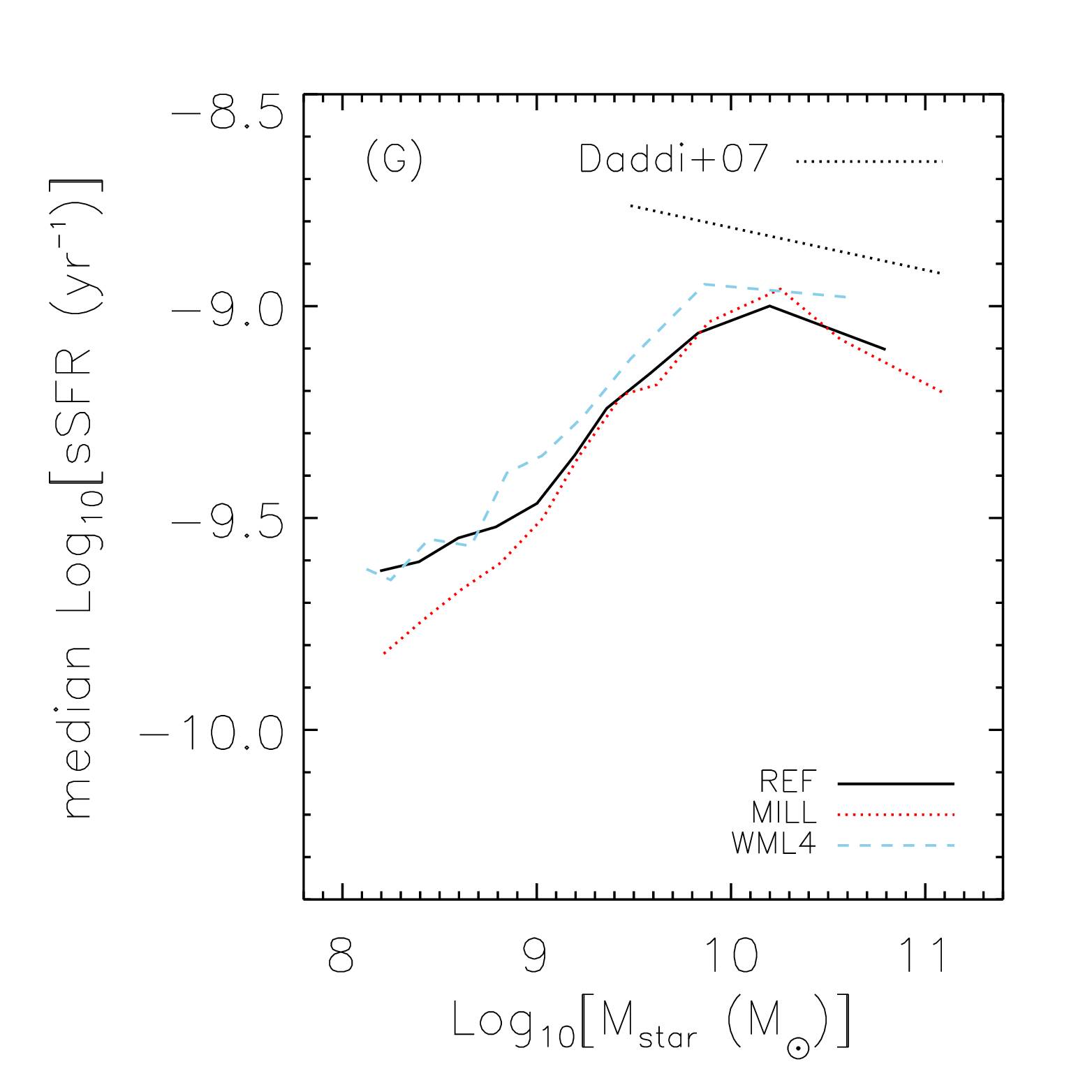}
\includegraphics[width=0.33\linewidth]{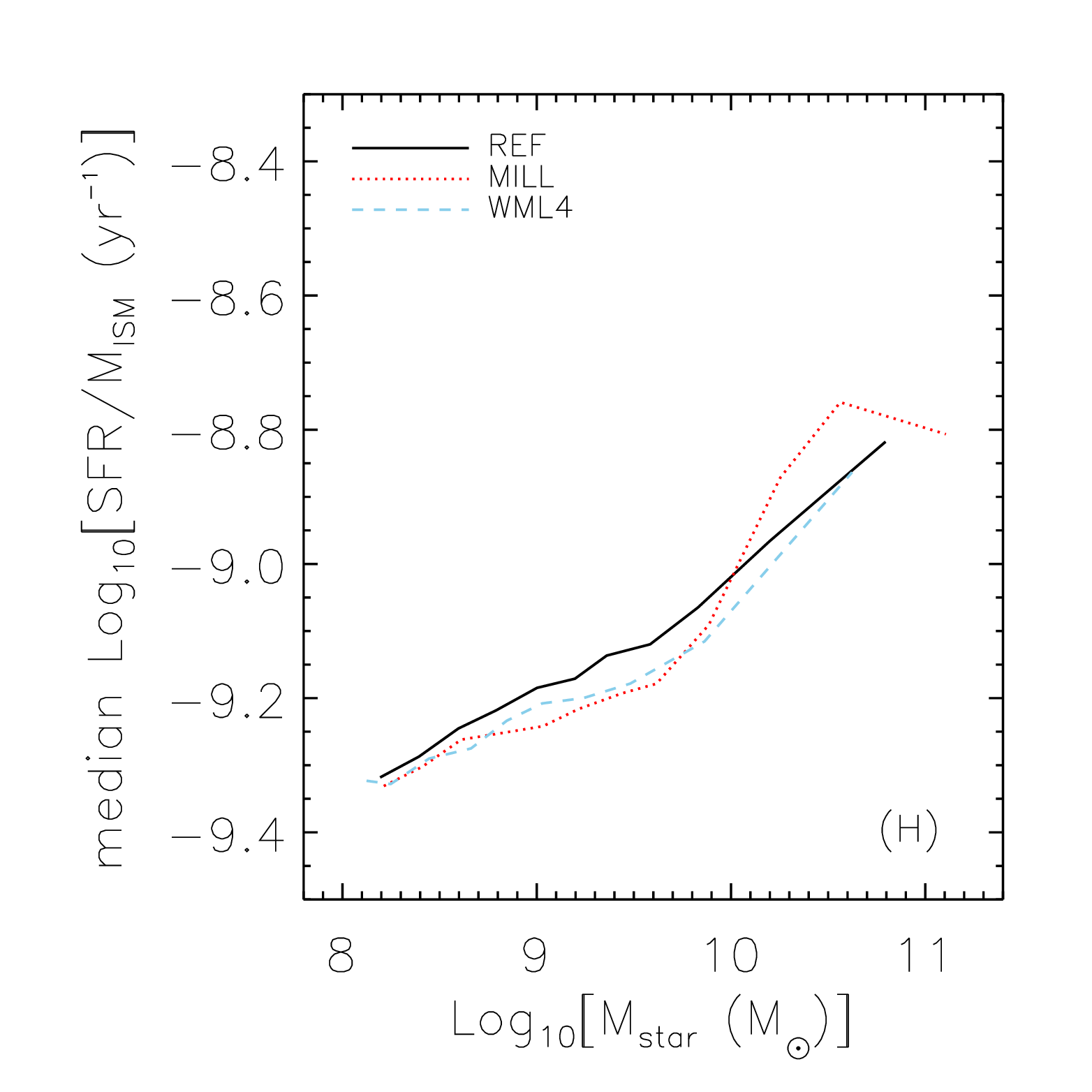}
\includegraphics[width=0.33\linewidth]{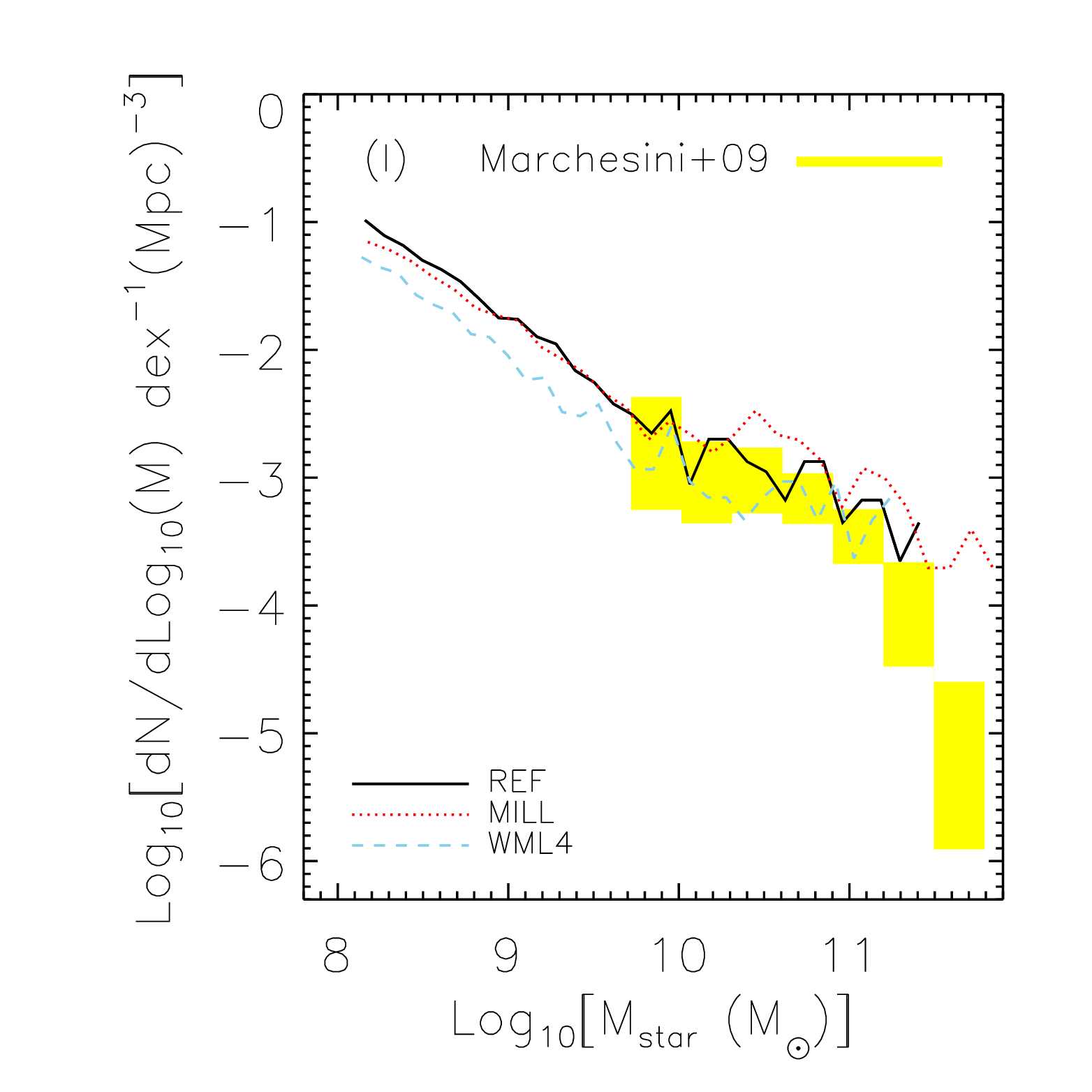} \\
\caption{Median relations between halo properties in the simulations described in Section~\ref{sec:cosmology}. The reference model is shown as a black curve in each panel. In the first five panels we show, as a function of total halo mass, medians of stellar mass fraction ($f_\textrm{star} = M_\textrm{star} / M_\textrm{tot}$, panel A), SFR (panel B), baryon fraction ($f_\textrm{baryon} = M_\textrm{baryon} / M_\textrm{tot}$, panel C), fraction of star-forming gas ($f_\textrm{ISM} = M_\textrm{ISM} / M_\textrm{tot}$, panel D) and gas mass fraction ($f_\textrm{gas, halo} = (M_\textrm{gas, total} - M_\textrm{ISM}) / M_\textrm{tot}$, panel E). The last four panels show, as a function of stellar mass, medians of the molecular gas mass in the ISM (panel F), specific SFR (sSFR = SFR/$M_{*}$, panel G), the inverse of the gas consumption timescale (SFR/$M_\textrm{ISM}$, panel H and galaxy number density (the galaxy stellar mass function, panel I). As described in the text, we show medians in bins along the horizontal axes for all haloes that satisfy the convergence criteria that apply to that specific panel.  The horizontal, dashed line in panel (C) shows the universal baryon fraction for the cosmology used in all simulations except `\textit{MILL}', the data points in panel (F) show a sub-set of the compilation studied in Genzel et al. (2010), the dotted black line in panel (G) shows the stellar mass - sSFR relation from the GOODS field (Daddi et al. 2007) and the shaded yellow region in panel (I) shows the galaxy stellar mass function of Marchesini et al. (2009). The \textit{`WML4'} (blue, dashed line) run uses the same cosmology as the `\textit{REF}' simulation, but the same feedback as the \textit{`MILL'} model. Therefore, to isolate the effect of the cosmology, the red, dotted curve should be compared to the blue, dashed curved.} 
\label{fig:sims_cosmology} 
\end{figure*}

\noindent To investigate the dependence of the galaxy properties on cosmology, and to facilitate comparisons to earlier work, we investigate the effects of changing the cosmological parameters from the WMAP 3-year results \citep{WMAP3} used in the `\textit{REF}' simulation to the so-called `concordance' or WMAP year 1 cosmology that was used in many previous studies including the Millennium Simulation \citep{millennium}. We will refer to this set of cosmological parameters as the `Millennium cosmology' and denote the model assuming this cosmology `\textit{MILL}'.  The main differences between the WMAP3 and Millennium cosmologies are firstly $\sigma_{8}$, which is 0.74 in WMAP3 and 0.9 in the Millennium cosmology and secondly, the universal baryon fraction $\Omega_{\rm b}$, which is 0.0418 in WMAP3 and 0.045 in the Millennium cosmology. Other parameter values are summarized in Table~\ref{tab:cosmology}. The WMAP3 cosmological parameters are largely consistent with the most recent 7-year results of WMAP \citep{komatsu11}, although in WMAP7 the value of $\sigma_8$ is 2.3$\sigma$ higher (0.8) and the Hubble parameter is 1$\sigma$ in WMAP7 than in WMAP3.

Increasing $\sigma_8$ and $\Omega_{\rm b}$ leads to the formation of many more stars than in the `\textit{REF}' simulation.  In order to roughly match the global SFR in the `\textit{MILL}' simulation, it is therefore necessary to double the amount of SN energy injected into the ISM per unit mass.  The `\textit{MILL}' simulation therefore uses a mass loading of $\eta = 4$ for SN-driven winds, rather than the $\eta = 2$ used in the reference model. To isolate the effect of cosmology, we therefore compare `\textit{MILL}' to a simulation that uses the WMAP 3 cosmology, but the same SN feedback model as `\textit{MILL}'.  We term this simulation `\textit{WML4}'. In Fig.~\ref{fig:sims_cosmology} the effect of the cosmology can thus be isolated by comparing the blue dashed and red dotted curves, which correspond to the WMAP3 (`\textit{WML4}') and `Millennium' (`\textit{MILL}') cosmological parameters (as indicated in Table~\ref{tab:cosmology}), respectively.

The main effect of $\sigma_8$ is to set the time-scale for structure formation. With a higher $\sigma_8$ structures of a given mass form earlier \citep[e.g.][]{peeblesbook}. The concentration of a DM halo at a given mass is set by formation time. Indeed, \citet{duffy08} showed that dark matter halo concentrations are significantly lower in the WMAP5 cosmology ($\sigma_8 = 0.77$) than in the WMAP1 cosmology ($\sigma_8 = 0.90$). Galaxy SFRs at a given halo mass could thus be influenced by the value of $\sigma_8$ through its effect on the halo gravitational potential. Comparing `\textit{MILL}' to `\textit{WML4}' in Fig.~\ref{fig:sims_cosmology} the total stellar mass (panel A), the SFR (panel B)  and the ISM mass (panel D) of high-mass haloes are all slightly higher for the WMAP1 cosmology than for the WMAP3 cosmology, due to the higher central densities in the WMAP1 cosmology at a given halo mass. Comparing these results to the global SFR \citep{owls}, we note that cosmology has a much larger effect on the global SFR than on individual objects.  This reflects the effect that the cosmological parameters have on the halo mass function. A higher $\sigma_8$ leads to more haloes of a given mass, and the larger number of haloes results in a much higher global SFR density.

The primary effect of cosmology on the galaxy stellar mass function (panel I) is through its effect on the number density of haloes at a given mass, i.e. the halo mass function, see also \citet{crain09}.  At low masses, the $M_{\rm tot}-f_{\rm star}$ (panel A) and the $M_*-M_{\rm ISM}$ (panel F) relations differ in `\textit{MILL}' and `\textit{WML4}' by $\sim0.1$\, dex). The stellar mass function at the low-mass end changes by up to 0.2 dex, because haloes of a given mass are more common in the simulation with the higher values of $\sigma_8$.  Panel I also demonstrates that a higher $\sigma_8$ also leads to the formation of more massive galaxies ($M_{\rm star}$ goes up to $\sim10^{12}$\,\msun\, for the `\textit{MILL}' simulation). We do not see an exponential cut-off in the predicted stellar mass function. This is because (as demonstrated by the gas consumption time-scale in panel H and the star forming gas mass in panel F), gas is being consumed rapidly in massive haloes where SN feedback is ineffective (see Paper I for more discussion on this point). Also, `radio-mode' AGN feedback, which is not considered here, is often assumed to cause this drop.

\subsection{Reionization} \label{sec:reion}

\begin{figure*}
\centering
\includegraphics[width=0.33\linewidth]{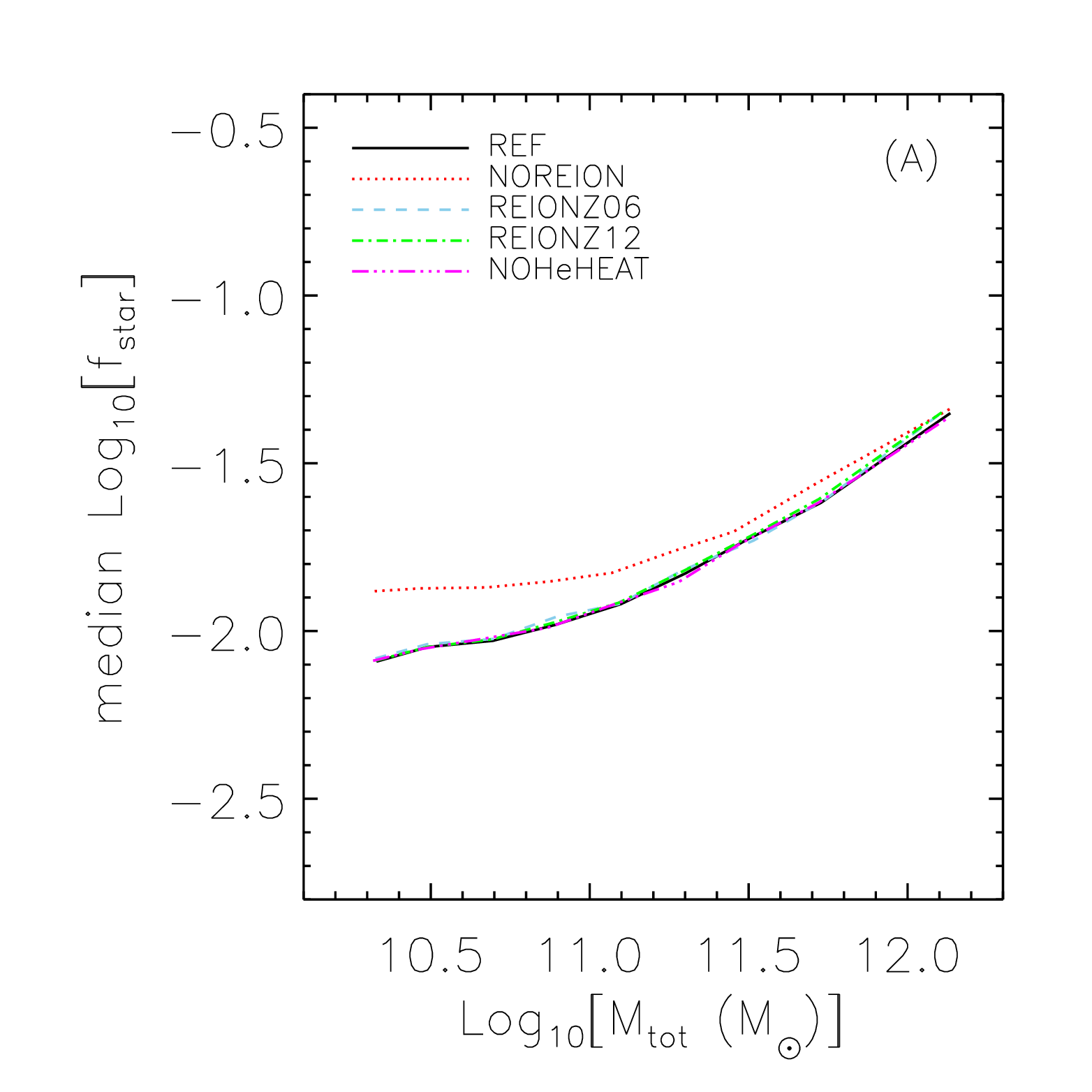}
\includegraphics[width=0.33\linewidth]{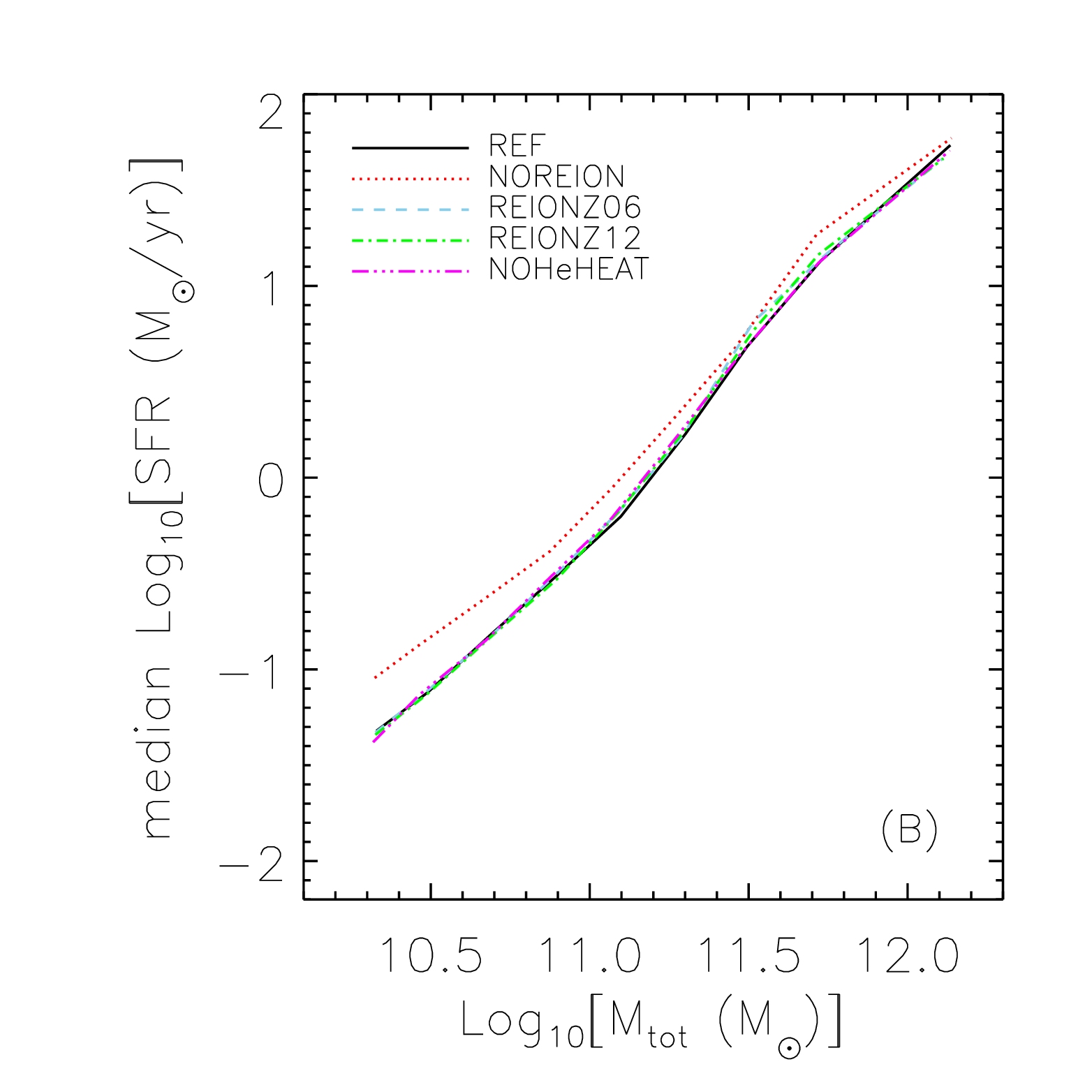}
\includegraphics[width=0.33\linewidth]{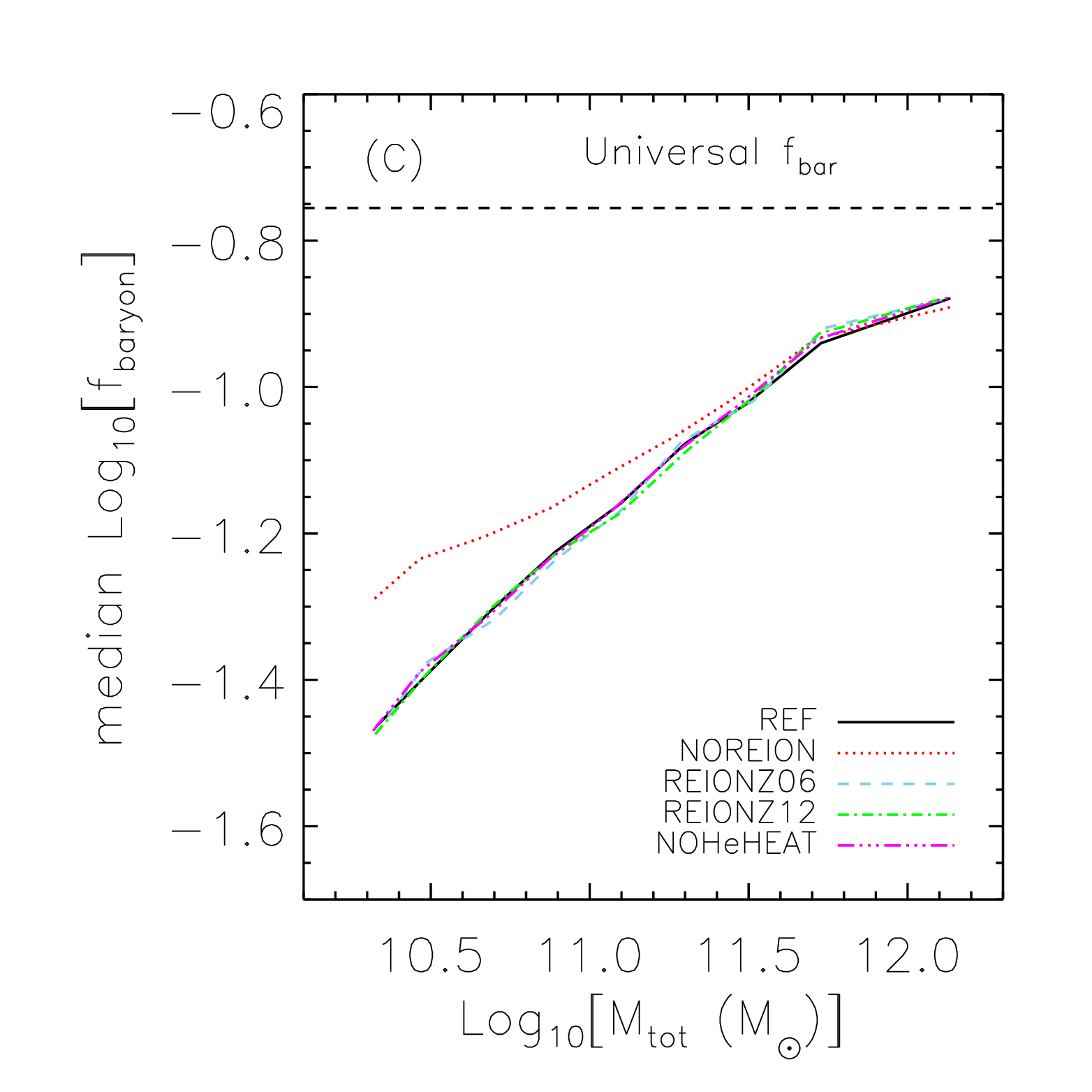} \\
\includegraphics[width=0.33\linewidth]{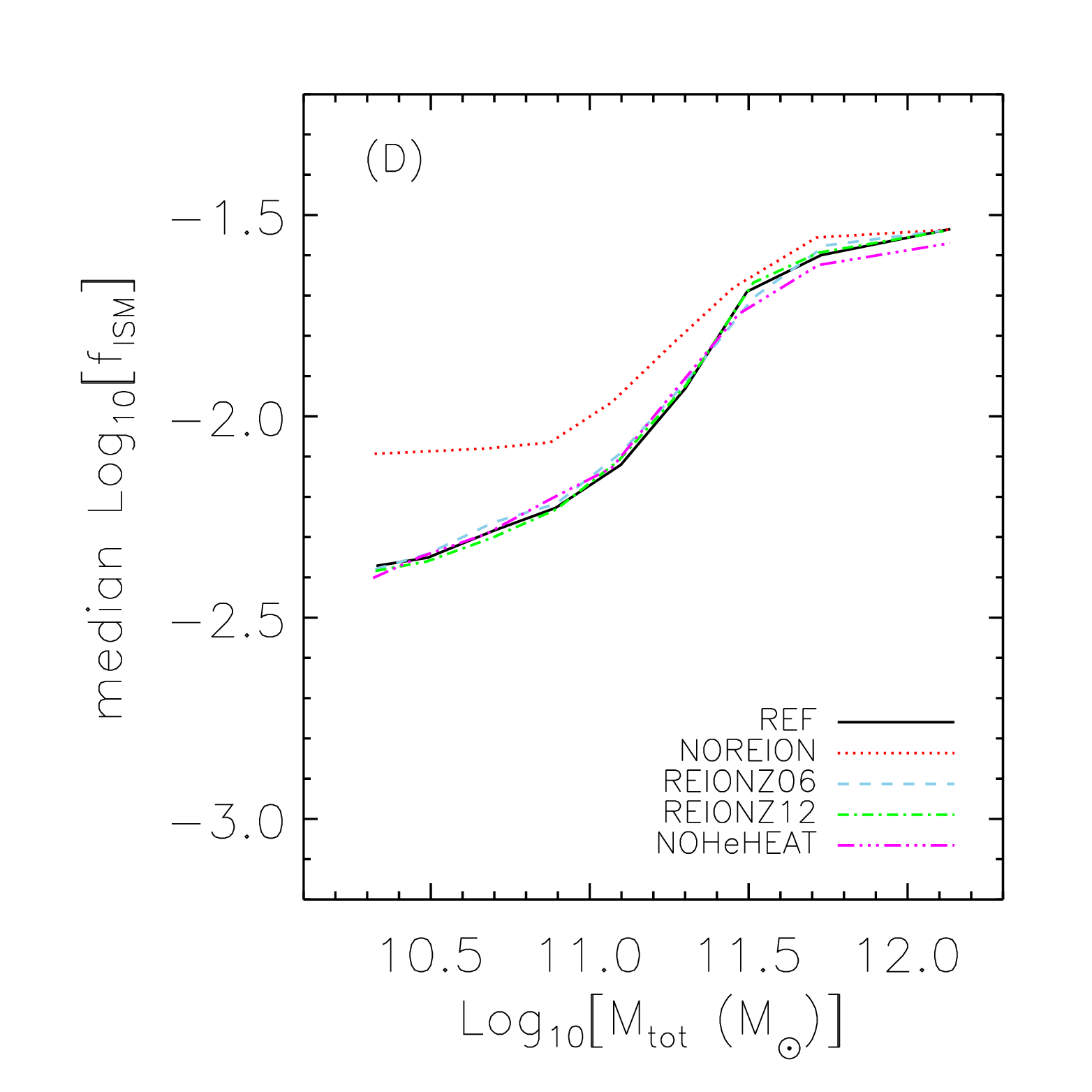}
\includegraphics[width=0.33\linewidth]{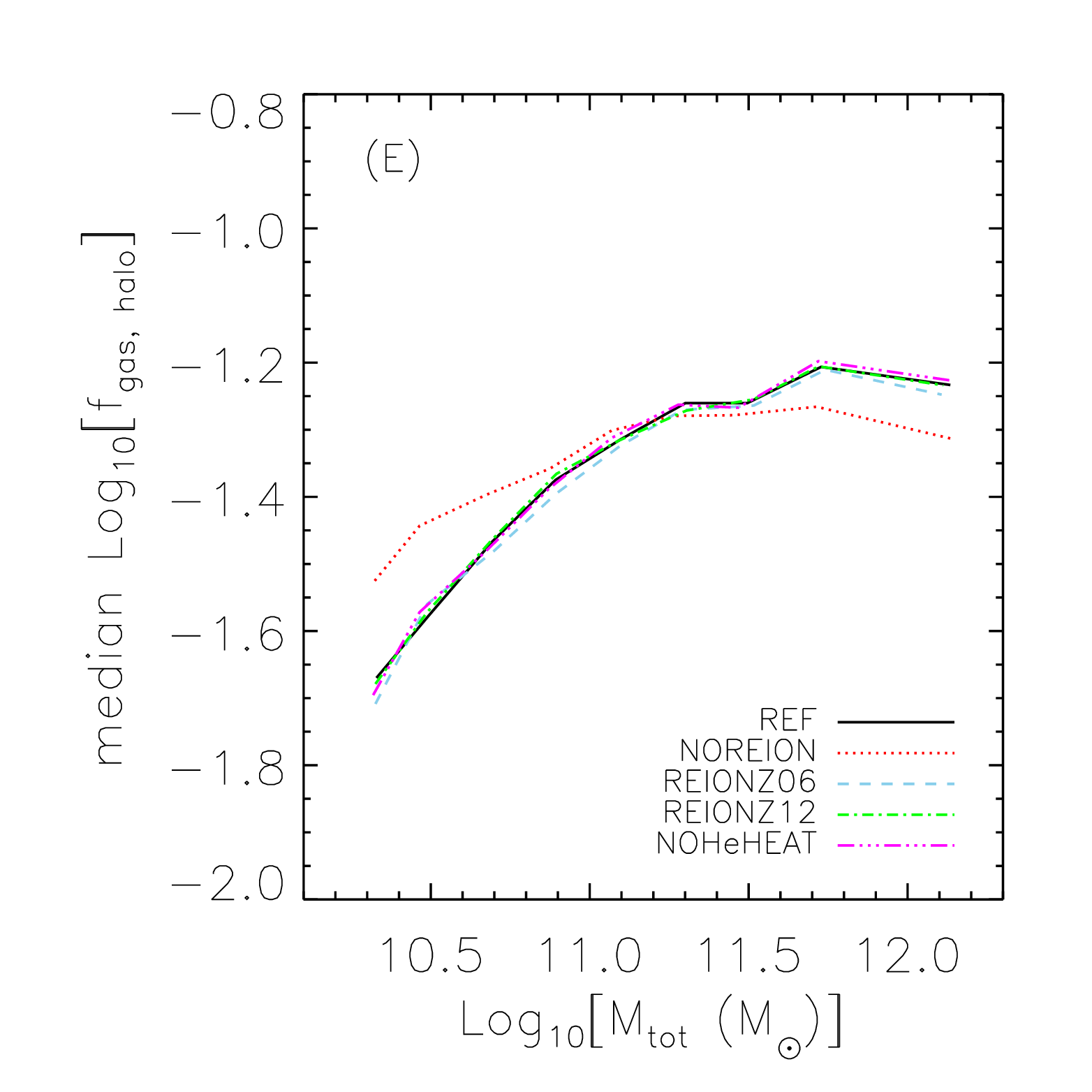}
\includegraphics[width=0.33\linewidth]{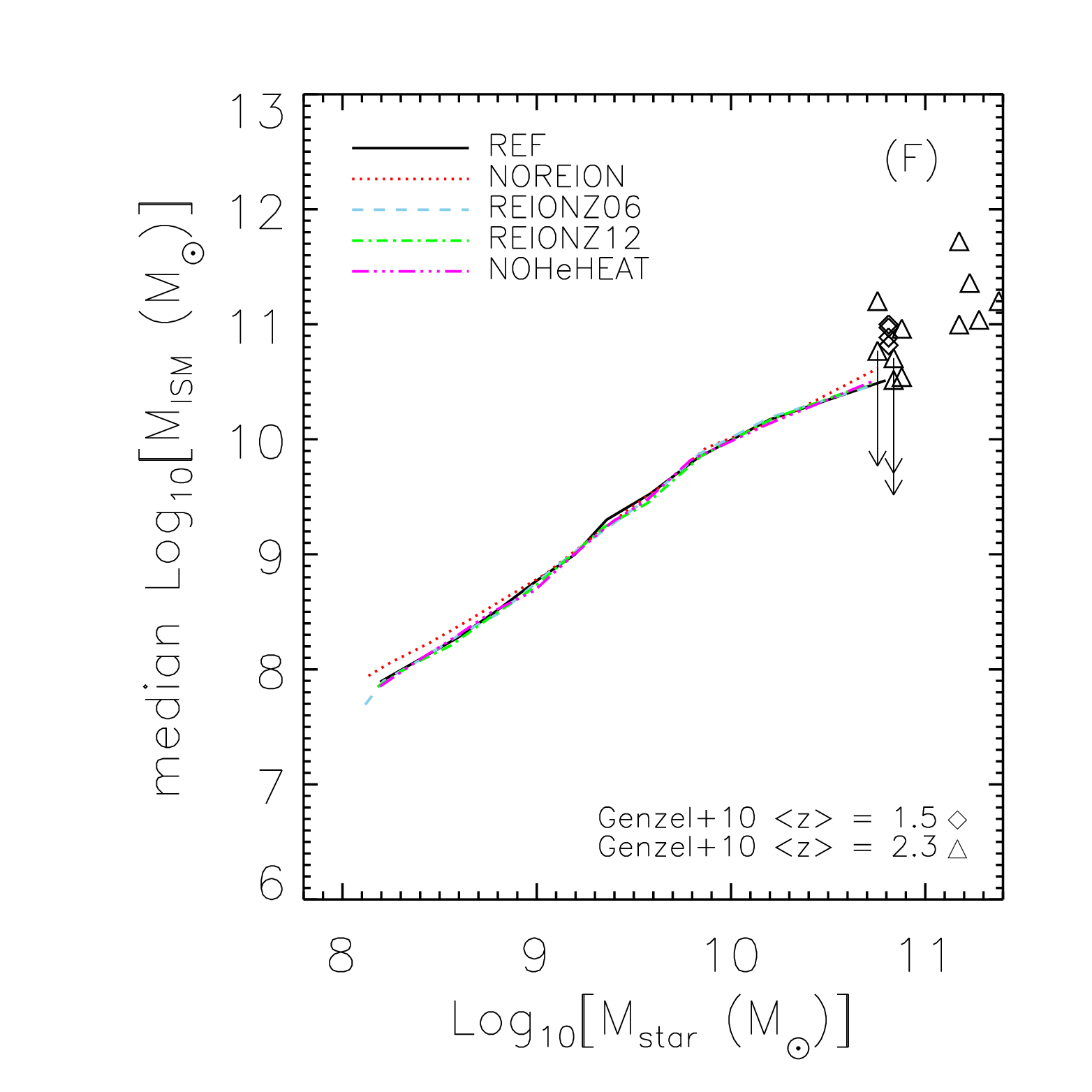} \\
\includegraphics[width=0.33\linewidth]{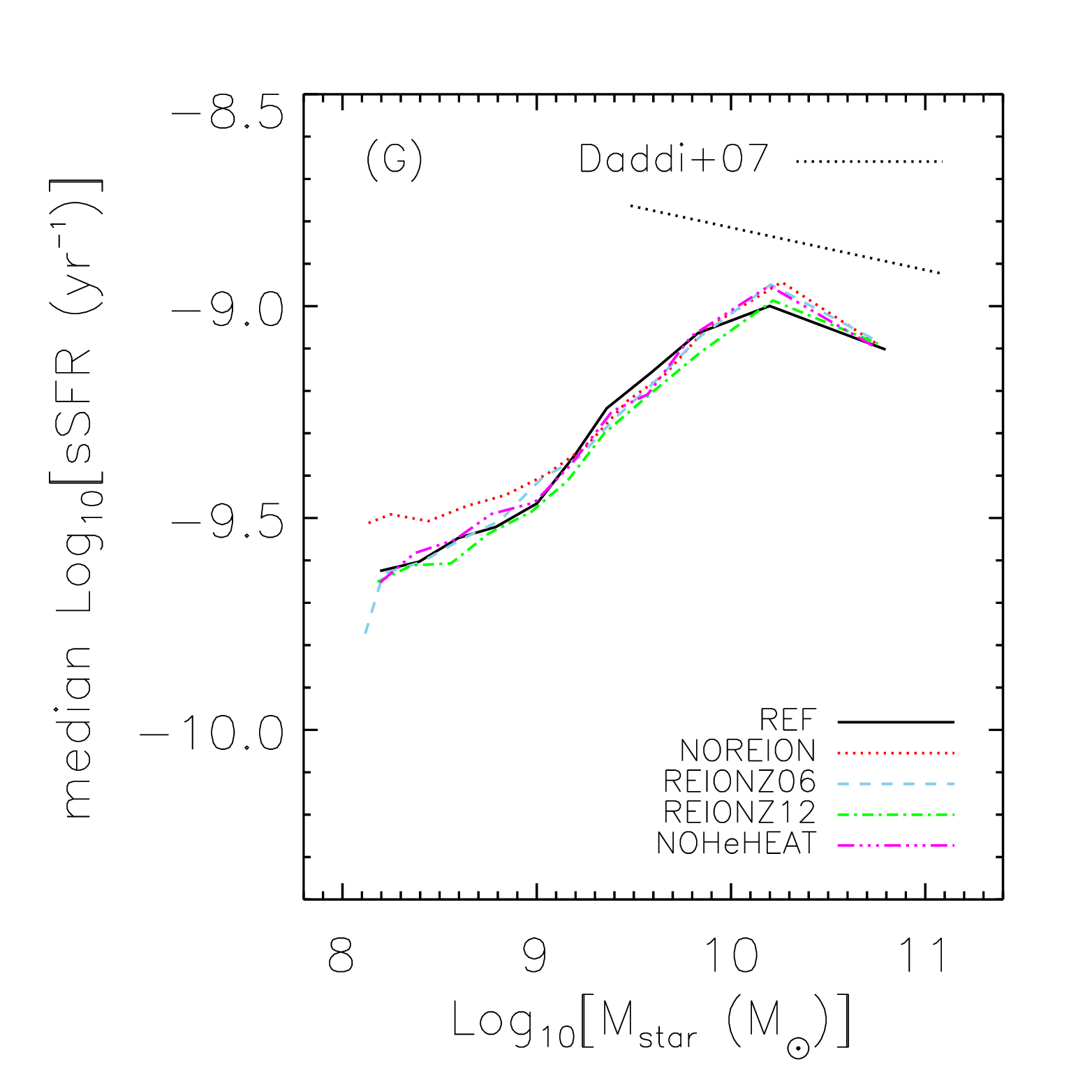}
\includegraphics[width=0.33\linewidth]{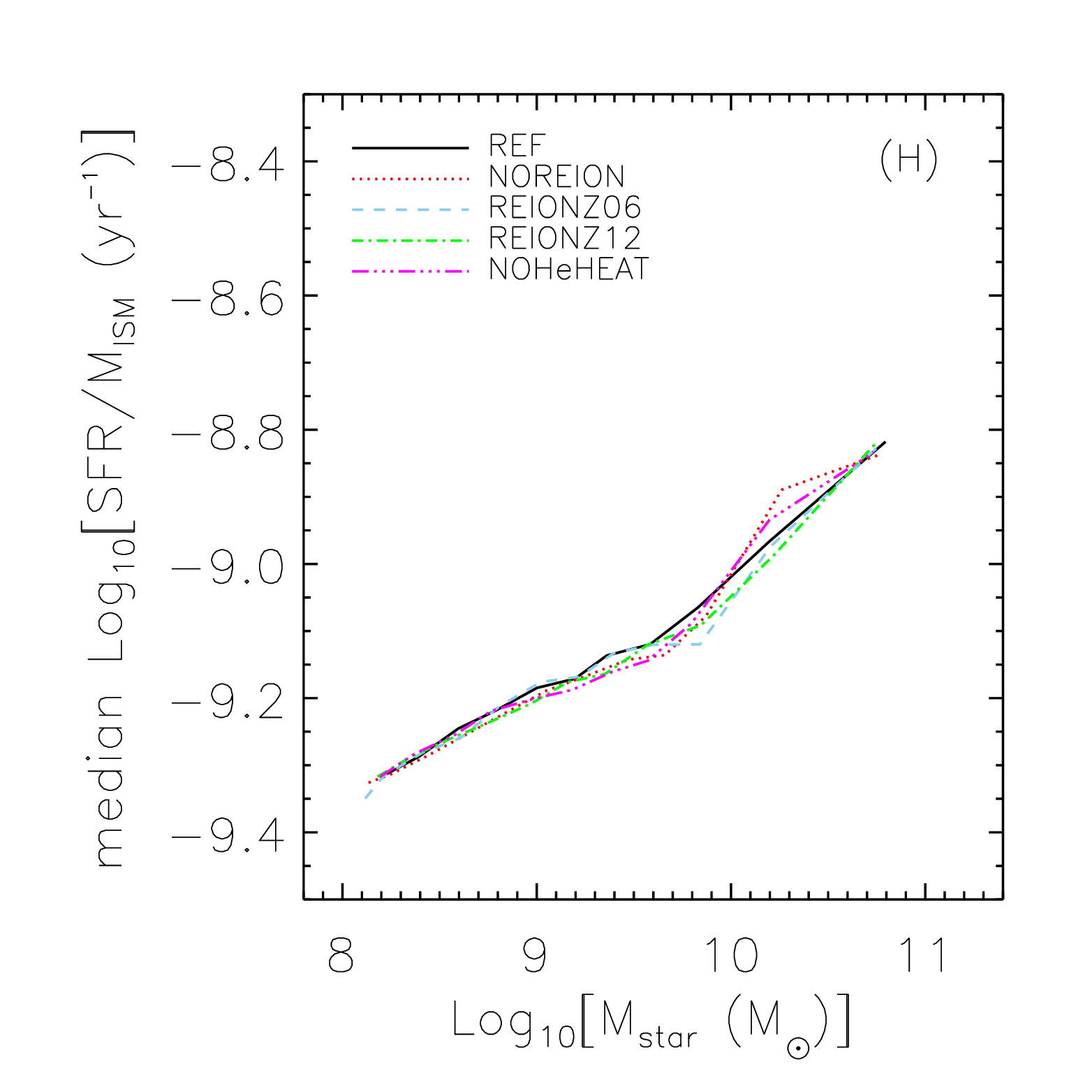}
\includegraphics[width=0.33\linewidth]{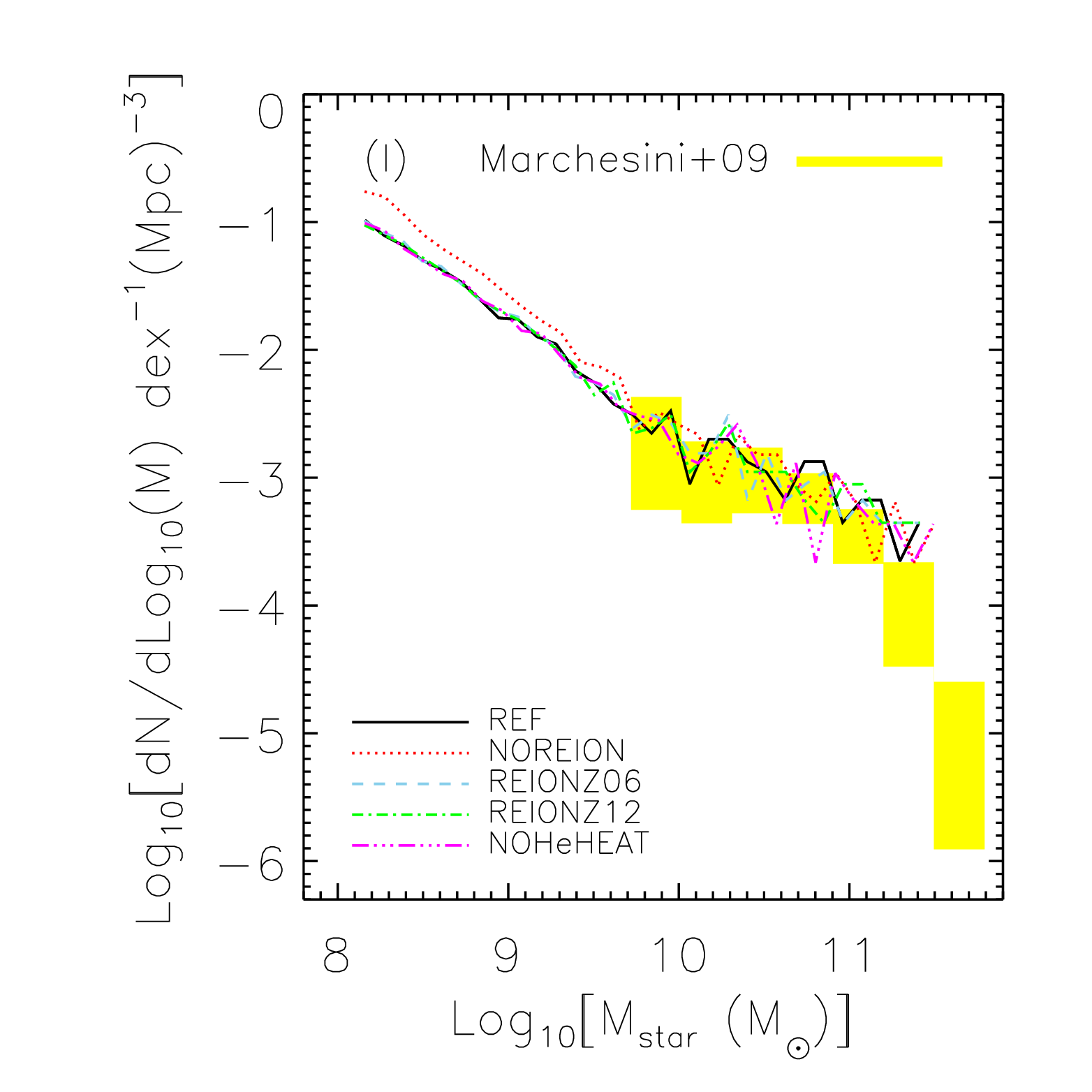} \\
\caption{As Fig.~\ref{fig:sims_cosmology}, but showing only the subset of simulations in which the reionization history is varied. In the `\textit{REF}' simulation (black, solid curve), reionization occurs at $z=9$. The red, dotted curve shows a simulation without reionization (`\textit{NOREION}'), whereas the blue, dashed and the green, dot-dashed curves have reionization redshifts of $z=6$ and $z=12$ respectively (`\textit{REIONZ06}' and `\textit{REIONZ12}'). The magenta dot-dot-dot-dashed curve shows a simulation in which the extra heat input due to helium reionization at $z=3.5$ is neglected (`\textit{NOHeHEAT}'). The predictions only differ for the simulation that neglects reionization altogether.} 
\label{fig:sims_reionization} 
\end{figure*}

\noindent The reionization of the Universe by quasars and galaxies has a profound effect on the temperature of the IGM and may impact the galaxy formation process \citep[e.g.][]{okamoto08}.  Reionization is modelled in the simulations by turning on a UV+X-ray background from galaxies and quasars \citep{haardtmadau}. All gas in the simulations is assumed to be optically thin to this radiation, which is assumed to be uniform and isotropic. In the `\textit{REF}' simulation, the redshift of reionization is set to $z_{\rm r}=9$.   As shown in \citet{wiersma09cooling}, the main effect of the UV background is to heat low-density, cold gas to $T \sim 10^4$ K.  This leads to gas in haloes with virial temperatures $\lesssim 10^4$\,K ($M_{\rm tot} \lesssim 10^8$\msun) being evaporated from their host haloes, and so effectively imposes a minimum halo mass in which star-formation can take place \citep[e.g.][]{quinn96, okamoto08, pawlikschaye09, hambrick11}. This mass scale corresponds to only $\sim 10$ particles in our simulations. Hence, our simulations do not probe the regime where photo-heating is expected to be most important and they may underestimate the effect for somewhat higher masses because the effect of photo-heating on their progenitors is not resolved.

To isolate the effects of reionization on the galaxy population, we compare the `\textit{REF}' simulation to a simulation without reionization (`\textit{NOREION}') and to two models in which the redshift of reionization is changed to $z_{\rm r}=6$ or $z_{\rm r}=12$ from its default value $z_{\rm r} = 9$, which are denoted `\textit{REIONZ06}' and`\textit{REIONZ12}' respectively. As described in Sec.~\ref{sec:owls_sims}, we match the observed temperature evolution of the IGM by modeling helium reionization as an additional heat input of 2 eV per atom around $z=3.5$.  The simulation `\textit{NOHeHEAT}' neglects this extra heating.

Inspection of Fig.~\ref{fig:sims_reionization} (all panels) reveals that by $z=2$
% , so long as a simulation includes reionization (all except for the red, dotted curve), the properties of galaxies at lower redshift are unaffected.  A
all knowledge of the redshift of reionization has been washed out of the resolved galaxy population, see also \citet{owls, wiersma11}.  Comparing `\textit{NOREION}' to the rest of the simulations, we see that the effects of neglecting the heating due to reionization are to increase the stellar masses (panel A), star forming gas masses (panel D) and gas masses (panels C and E), and hence the SFRs (panels B and G) of low-mass objects ($M_\textrm{tot} \lesssim 10^{11}$ \msun). Gas that should have been removed from these objects by the extra energy it receives from the photo-ionizing background is allowed to remain.  This effect is also visible in the stellar mass function (panel I), where at low masses ($M_\textrm{star} \lesssim 10^9$ \msun), a given stellar mass corresponds to a lower halo mass, making these objects more abundant.

At high halo masses ($M_\textrm{tot} \gtrsim 10^{11}$ \msun), neglecting reionization results in slightly lower halo gas fractions because more of the gas is in the ISM and in stars.

We can examine the effect of helium reionization by comparing `\textit{REF}' (solid, black curve) to `\textit{NOHeHEAT}' (dashed, purple curve).  From all panels we can immediately see that, consistent with \citet{owls}, the extra heat input due to helium reionization has no effect on the properties of resolved haloes and galaxies.   All massive objects have already formed a very significant proportion of their mass by the time that helium reionization takes place ($z=3.5$) and this heat input has a negligible effect at gas densities typical of haloes. The unimportance of helium reionization holds for all of the properties of galaxies we investigate in this work. The extra heat input to the IGM from helium reionization is therefore only important for controlling the temperature of the low-density gas of the IGM \citep{theuns02} and has no effect on the properties of galaxies formed in the simulations, at least for $M_* > 10^8$\msun.

\subsection{The polytropic equation of state for high-density gas} \label{sec:eos}

\begin{figure*}
\centering
\includegraphics[width=0.33\linewidth]{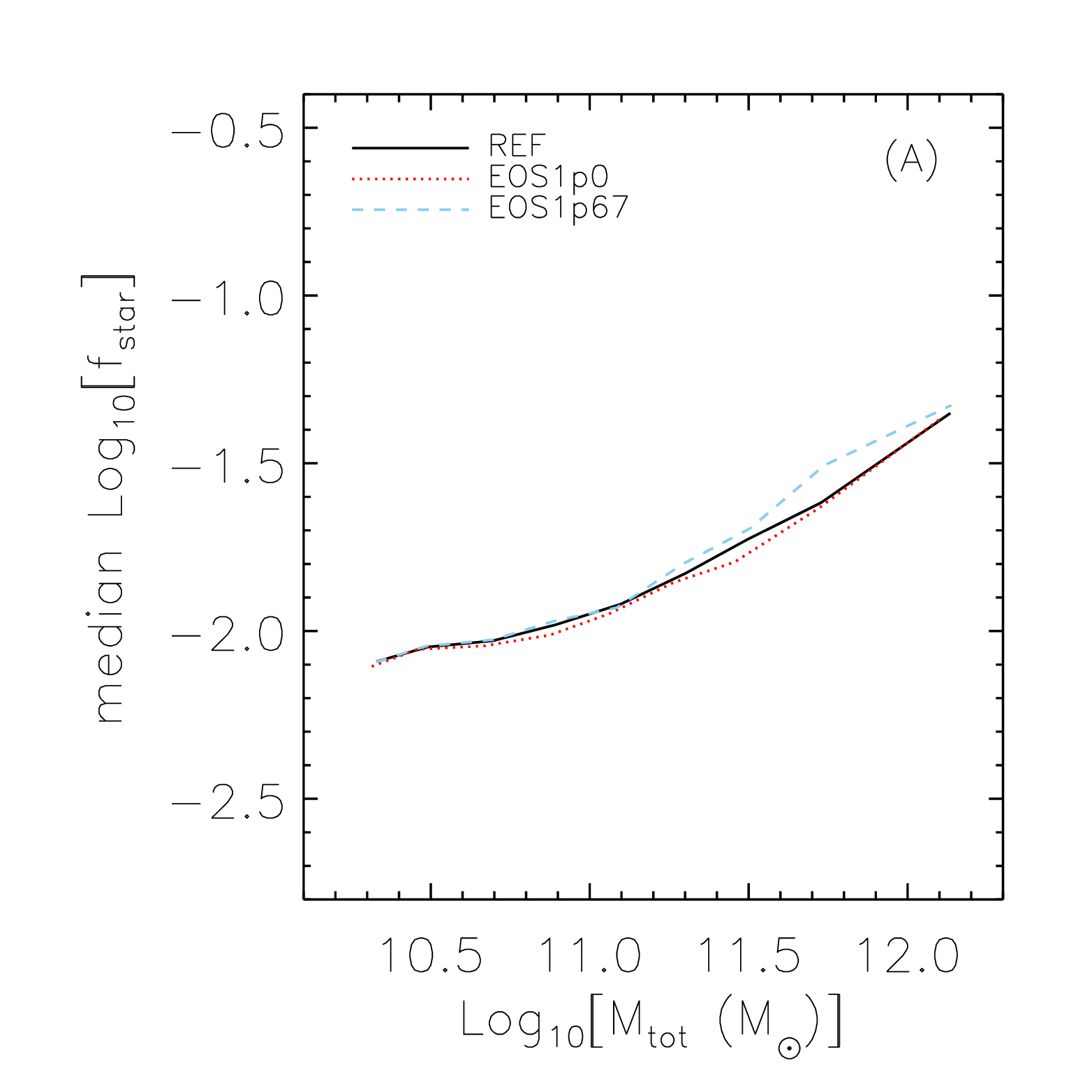}
\includegraphics[width=0.33\linewidth]{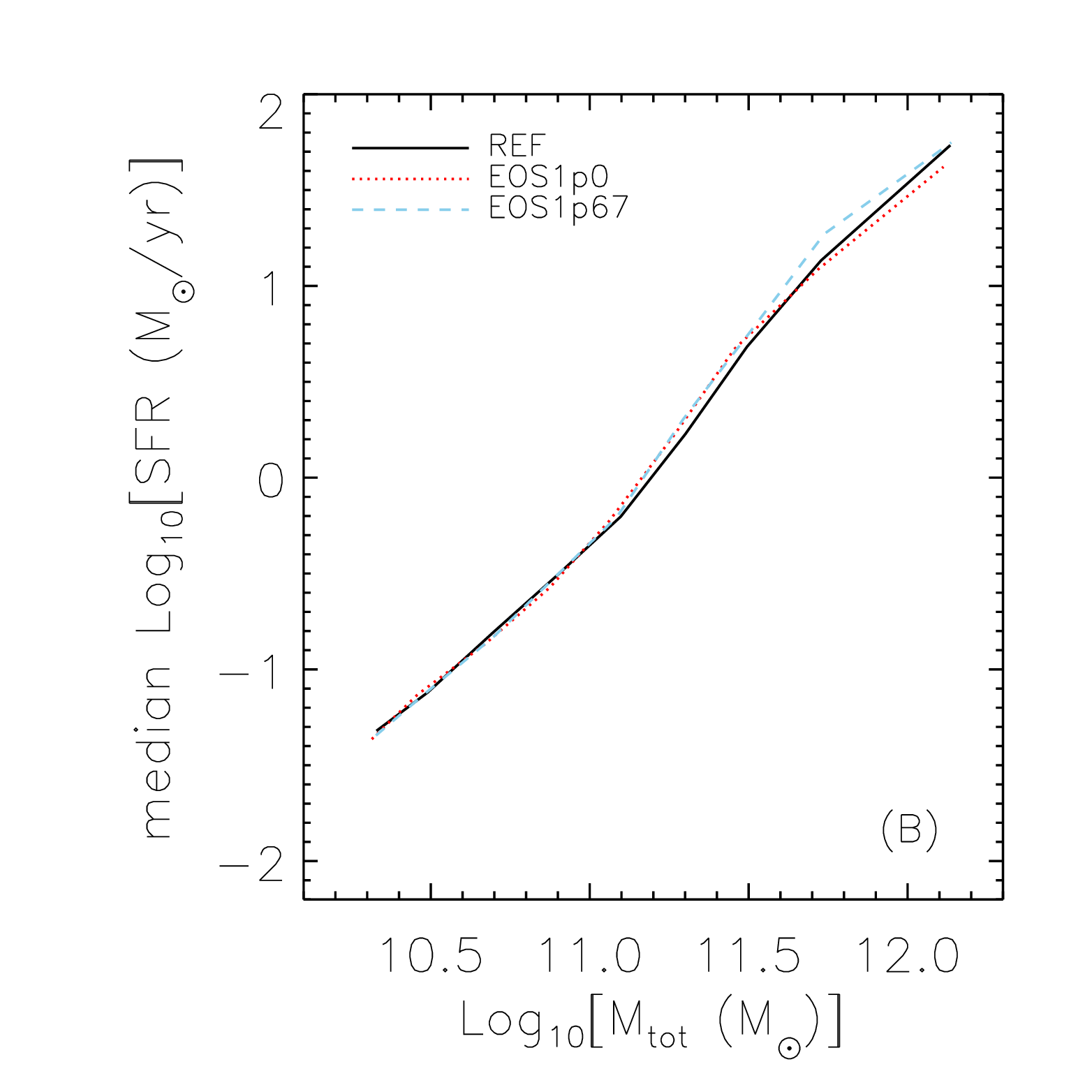}
\includegraphics[width=0.33\linewidth]{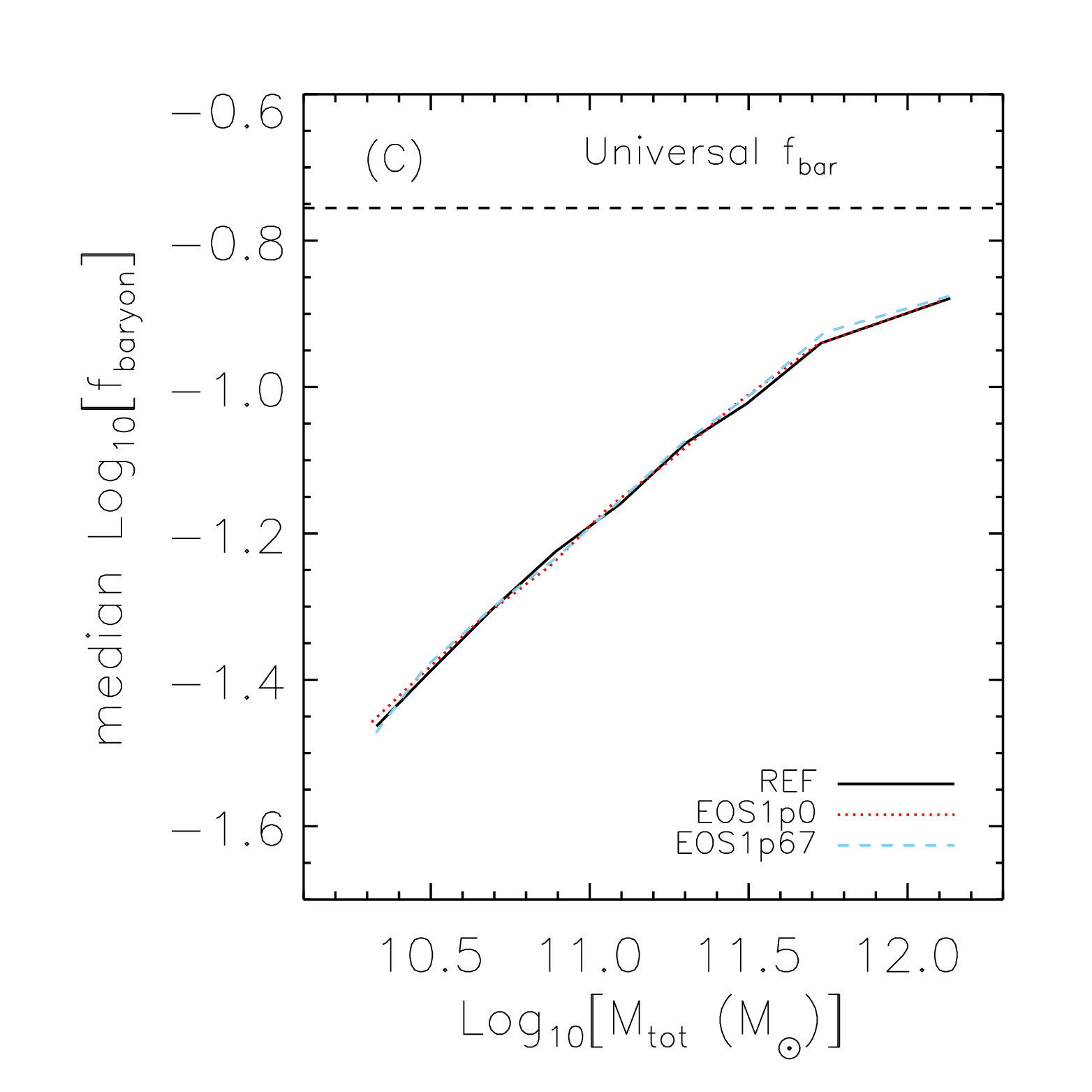} \\
\includegraphics[width=0.33\linewidth]{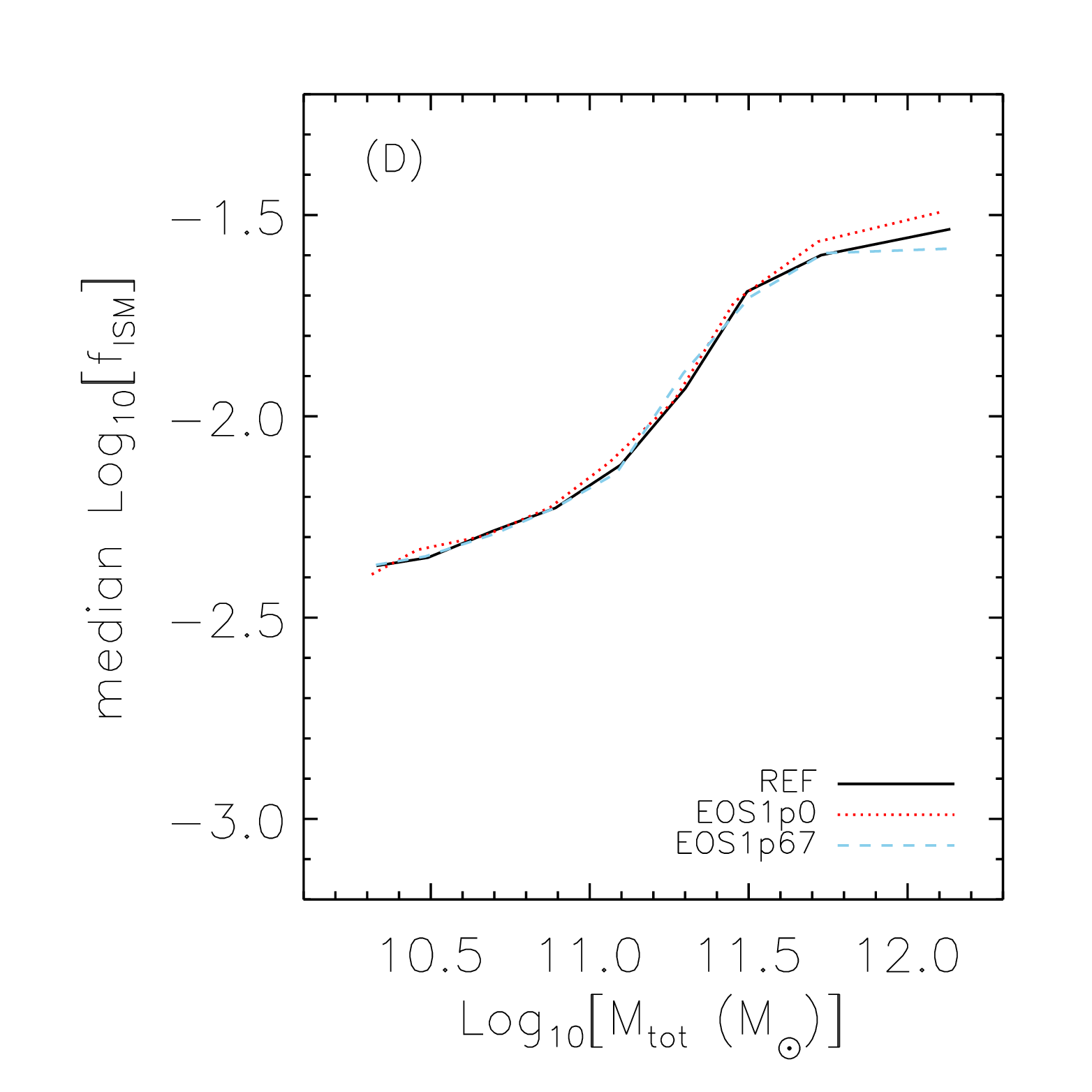}
\includegraphics[width=0.33\linewidth]{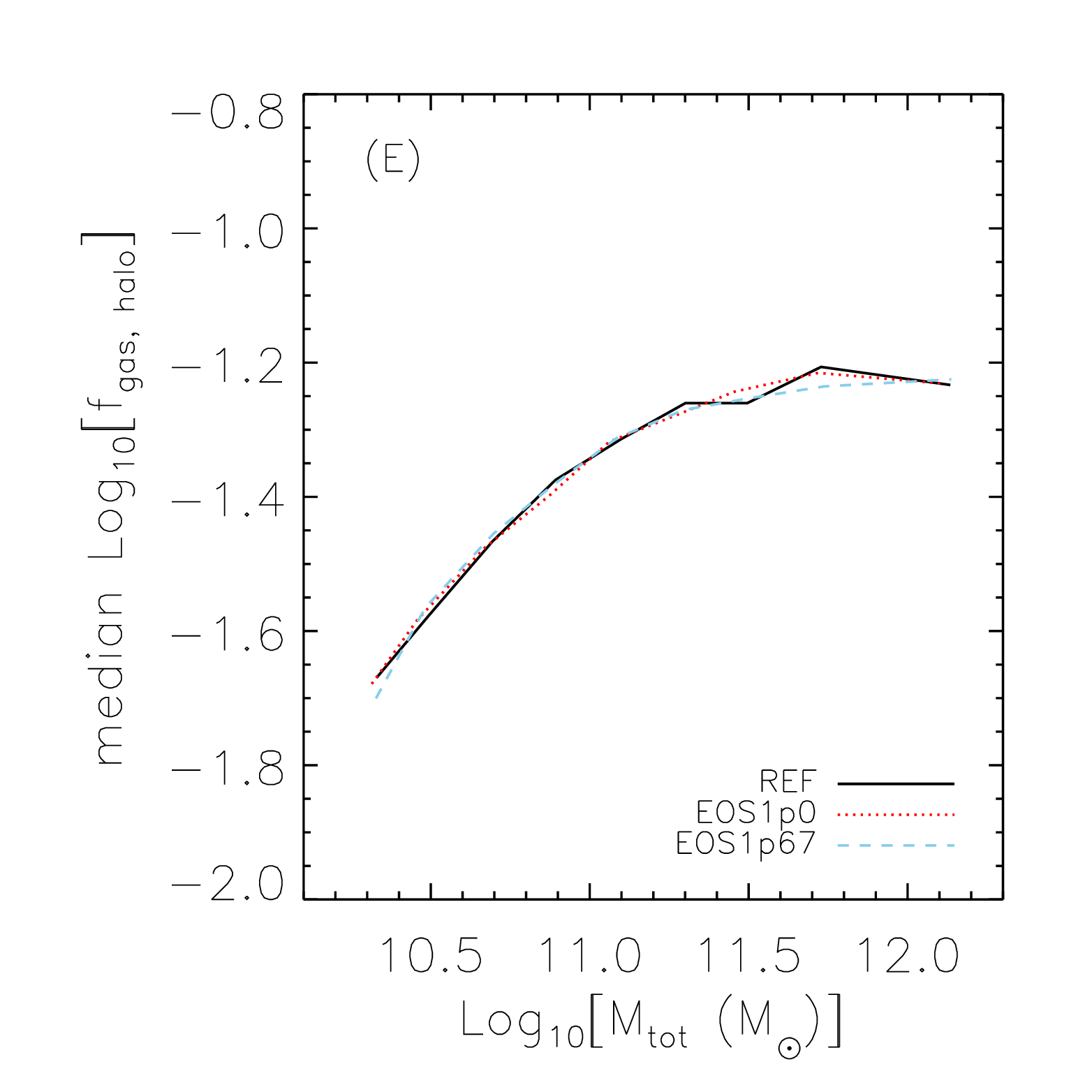}
\includegraphics[width=0.33\linewidth]{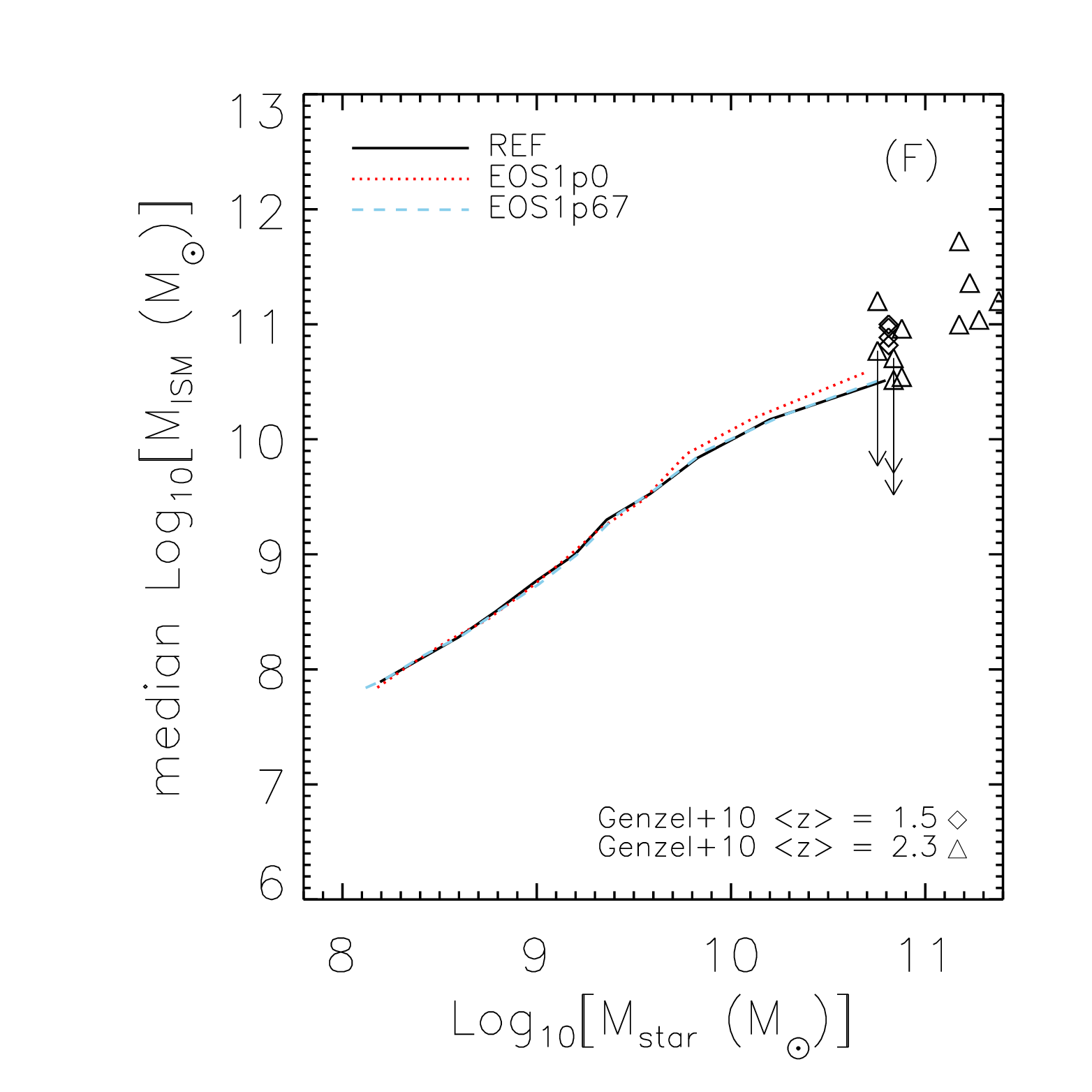} \\
\includegraphics[width=0.33\linewidth]{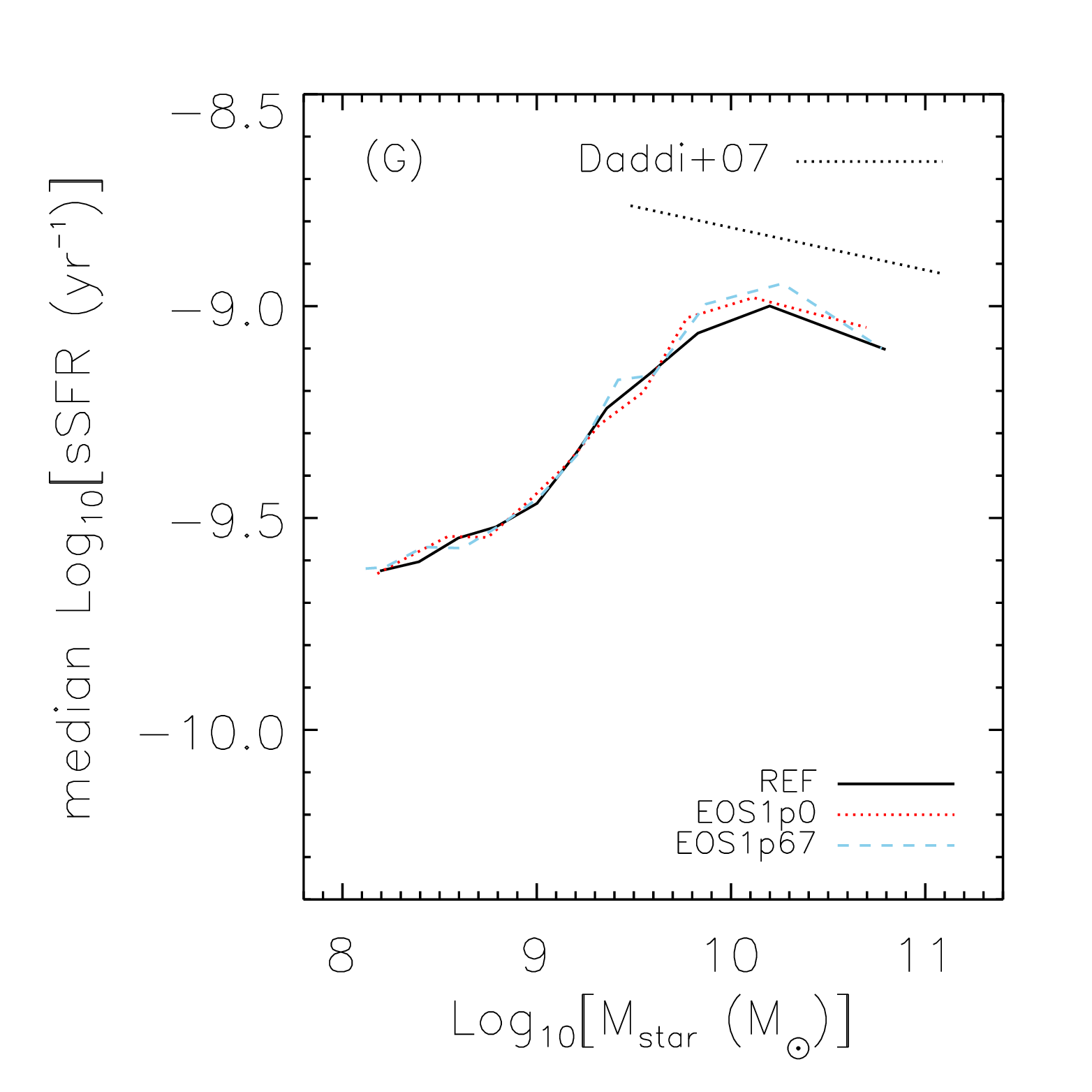}
\includegraphics[width=0.33\linewidth]{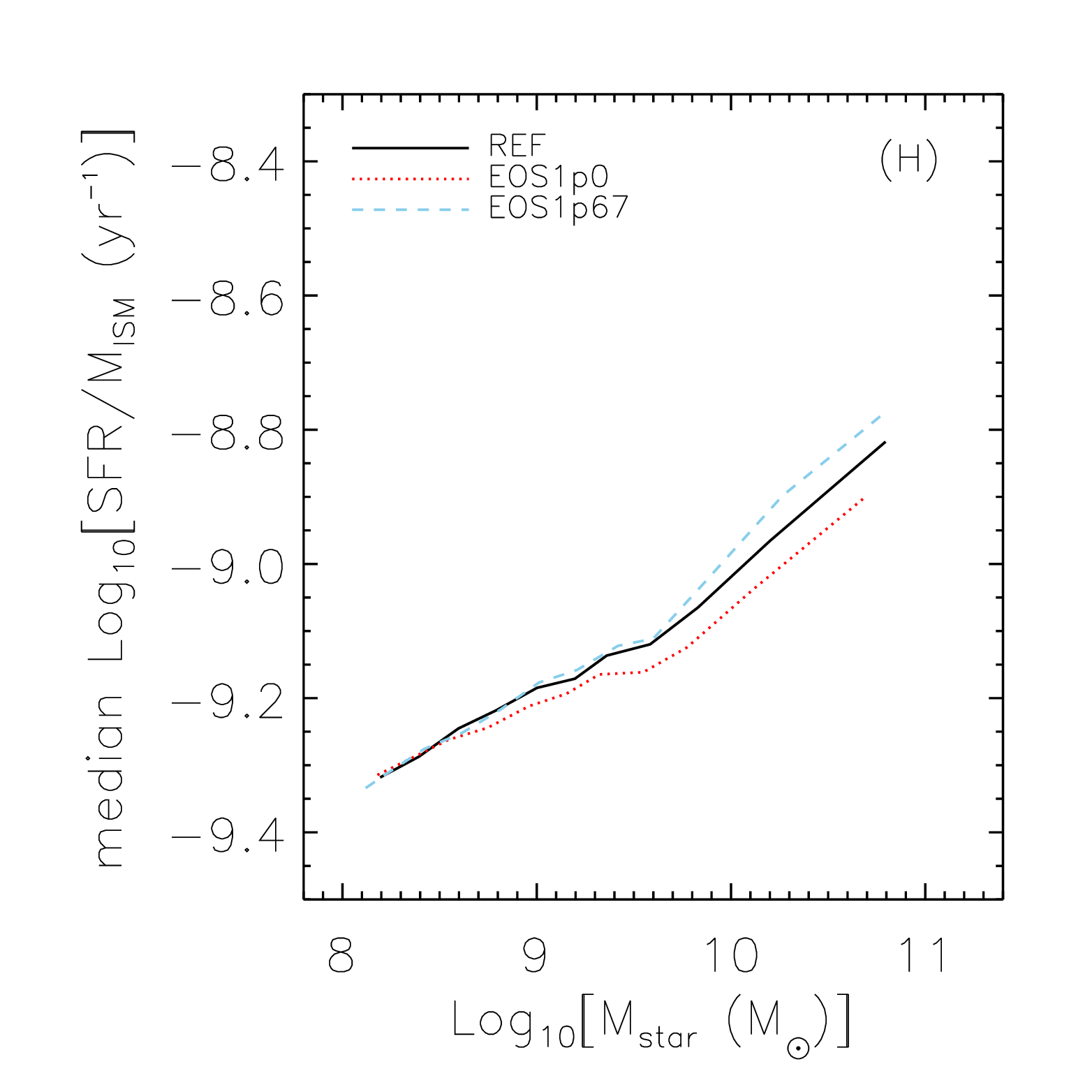}
\includegraphics[width=0.33\linewidth]{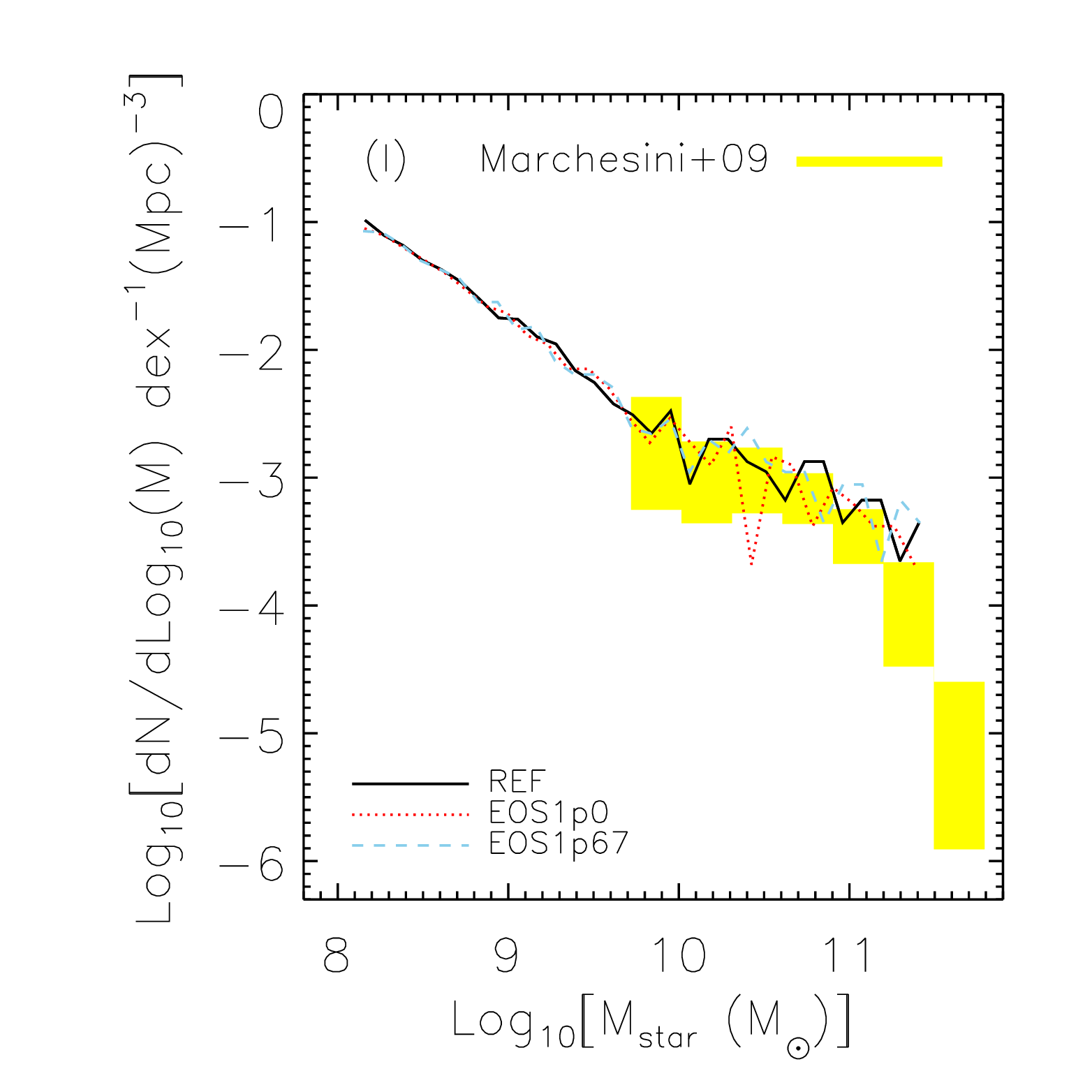} \\
\caption{As Fig.~\ref{fig:sims_cosmology}, but showing only the subset of simulations in which the polytropic equation of state (EoS) that is imposed on high-density ($n_\textrm{H} > 0.1$ cm$^{-3}$) gas is varied. In the `\textit{REF}' simulation (black, solid curve) we use a polytropic EoS with a power law index of $\gamma_{\rm eff} = 4/3$. The red, dotted curve shows a simulation with an isothermal EoS, $\gamma_{\rm eff}=1$ (`\textit{EOS1p0}'). The blue, dashed curve shows a simulation with an equation of state with an adiabatic EoS, $\gamma_{\rm eff} = 5/3$ (`\textit{EOS1p67}'). The predictions are insensitive to the assumed equation of state because our prescription for star formation depends on the pressure rather than the volume density.} 
\label{fig:sims_eos} 
\end{figure*}

\noindent Large volume cosmological simulations lack both the resolution and the physics to model the multiphase ISM. We therefore impose an effective equation of state (EoS) for all gas particles with densities higher than $n_{\textrm{\scriptsize H}} = 0.1$ cm$^{-3}$.  The EoS we use is polytropic, $P \propto \rho^{\gamma_{\textrm{\scriptsize eff}}}$ and in the `\textit{REF}' simulation $\gamma_{\textrm{\scriptsize eff}}= 4/3$.  This value of $\gamma_{\textrm{\scriptsize eff}}$ ensures that both the Jeans mass and the ratio of the Jeans length and the kernel of the SPH particles are independent of density \citep{schayedallavecchia08}, provided that they are resolved at  $n_{\textrm{\scriptsize H}} = 0.1$ cm$^{-3}$, preventing spurious numerical fragmentation due to a lack of resolution. In order to isolate the effect of the EoS, we compare `\textit{REF}' to two simulations in which the slope of the EoS changed to be either isothermal, with $\gamma_{\textrm{\scriptsize eff}} = 1$ (`\textit{EOS1p0}'), or adiabatic, with $\gamma_{\textrm{\scriptsize eff}}= 5/3$ (`\textit{EOS1p67}').

Changing the slope of the EoS has a significant effect on the visual appearance of the galaxy disk.  Comparing `\textit{EOS1p0}' and `\textit{EOS1p67}' in Fig.~\ref{fig:prettypics}, we can immediately see that using a steeper EoS pressurizes the gas more strongly, resulting in a smoother gas distribution.  Despite markedly changing the visual appearance of the disks, it is clear from Fig.~\ref{fig:sims_eos} that the galaxy stellar properties (panels A and I) as well as the total amount of baryons associated with the galaxies (panels C, D, E and F) and the SFRs (panels B and G) are virtually unaffected by the structure of the ISM on small scales.  The only noticeable (although very weak) difference between the simulations is that in the very most massive galaxies, steepening the EoS leads to slightly more efficient SF at a given halo mass. This is because, at a given density, pressures (and hence star-formation efficiencies) are higher for a steeper EoS.

It is striking that the properties of the galaxy are almost entirely insensitive to changes in the polytropic index (at least in the range $\gamma_{\textrm{\scriptsize eff}} = 1 - 5/3$). This is a consequence of our use of the prescription for SF of \citet{schayedallavecchia08}. In this prescription the observed Kennicutt-Schmidt surface density law is analytically converted into a pressure law under the assumption of vertical, local hydrostatic equilibrium. As demonstrated by \citet{schayedallavecchia08}, this enables the same recipe to reproduce the observed SF law regardless of the equation of state, and without tuning any parameters. Because the pressure profile is determined by the gravitational potential in hydrostatic equilibrium, which is generally not badly violated, the SF we predict is the same if the potential is the same, irrespective of the assumed equation of state (which does, however, change the thickness of the gas disk).

Note that, as far as we know, all other groups use Schmidt-type SF laws, i.e.\ volume density SF laws, rather than pressure laws. In that case the SFR is expected to depend on the assumed equation of state, even if the gravitational potential and the pressure profile remain unchanged. Because the observed SF law that the simulations are calibrated to match is a surface density rather than a volume density law, the calibration would have to be repeated if the equation of state is changed. This is, however, generally not done, which means that the simulations can no longer be expected to reproduce the observed SF law when $\gamma_\textrm{eff}$ is changed.

\subsection{The star formation law} \label{sec:sf}

\begin{figure*}
\centering
\includegraphics[width=0.33\linewidth]{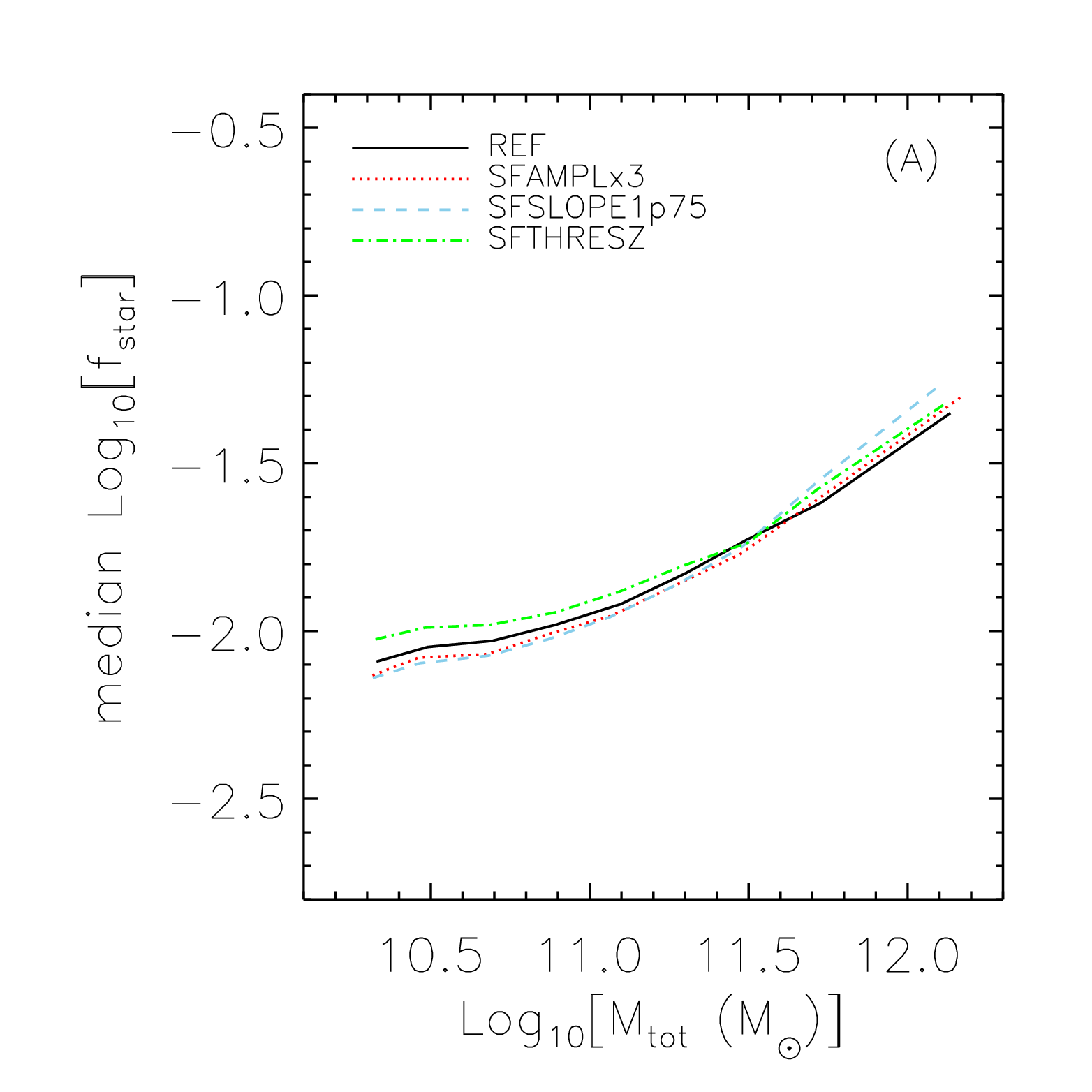}
\includegraphics[width=0.33\linewidth]{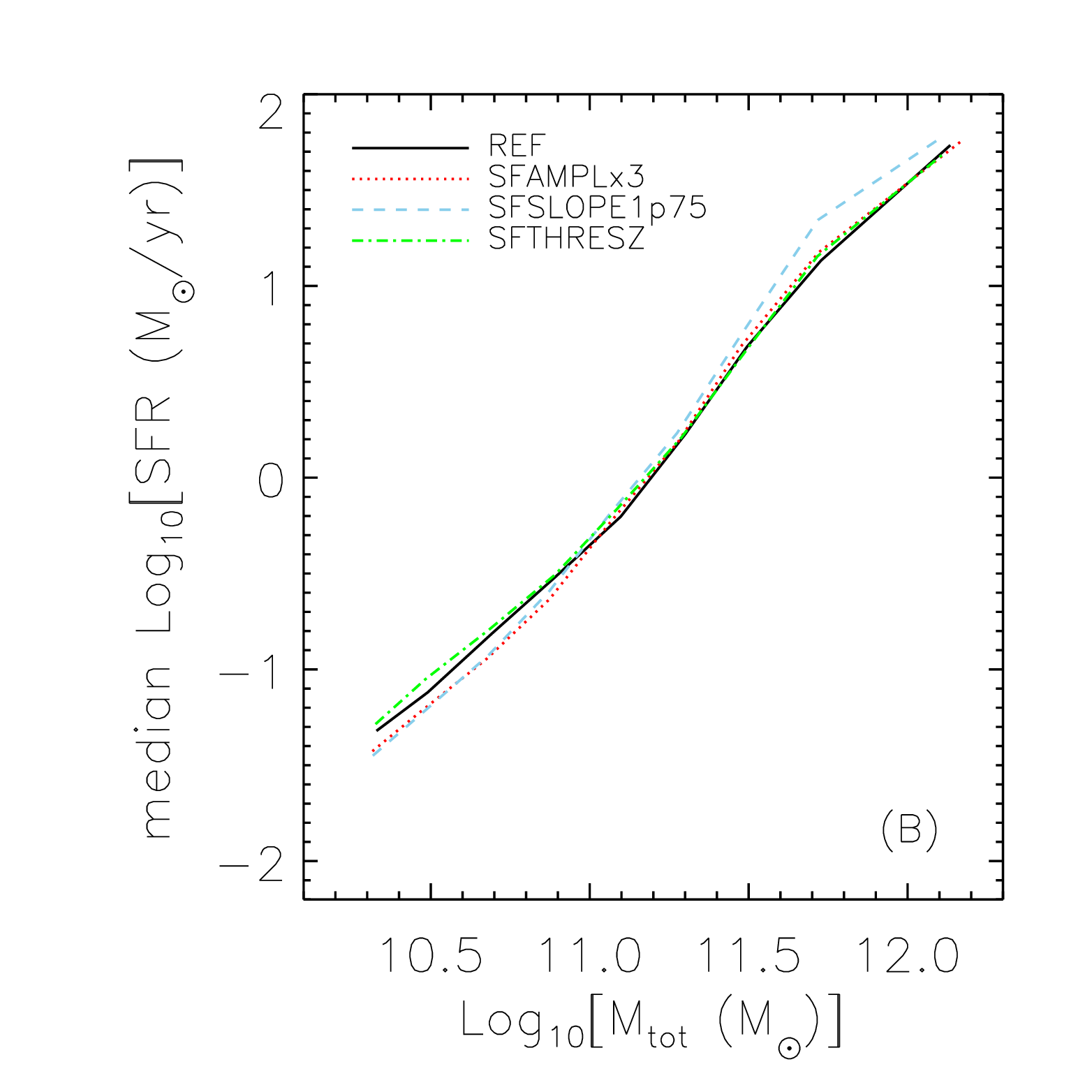}
\includegraphics[width=0.33\linewidth]{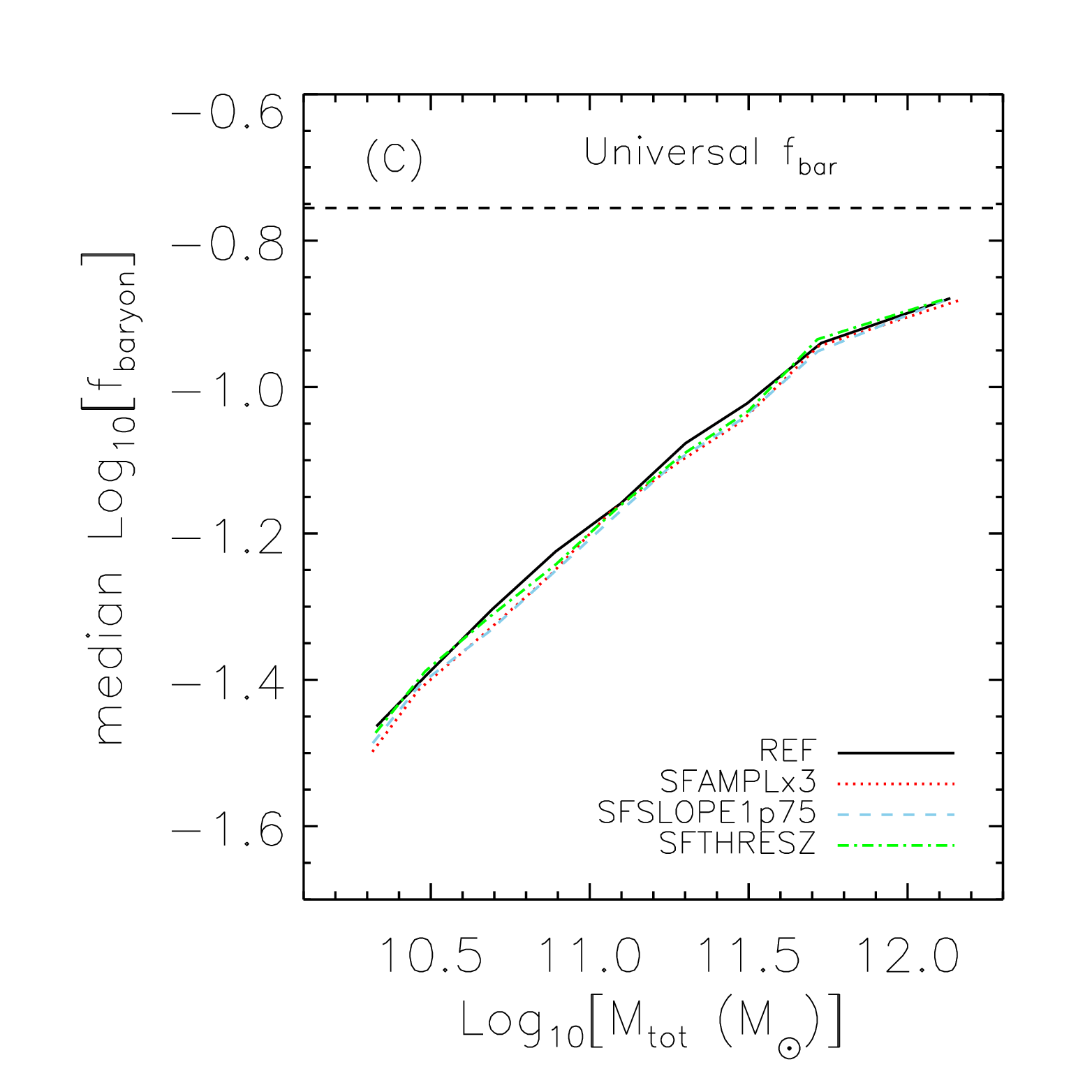} \\
\includegraphics[width=0.33\linewidth]{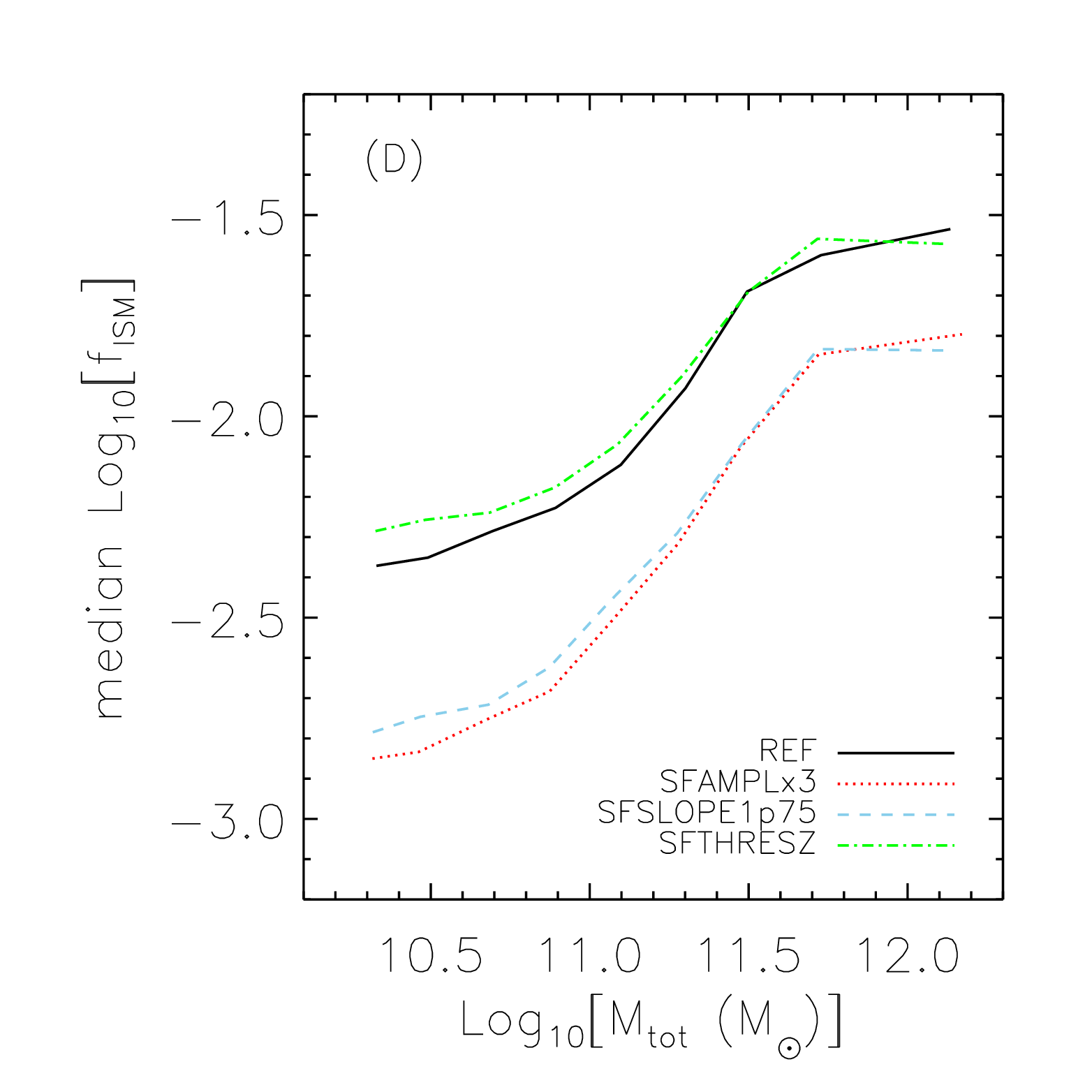}
\includegraphics[width=0.33\linewidth]{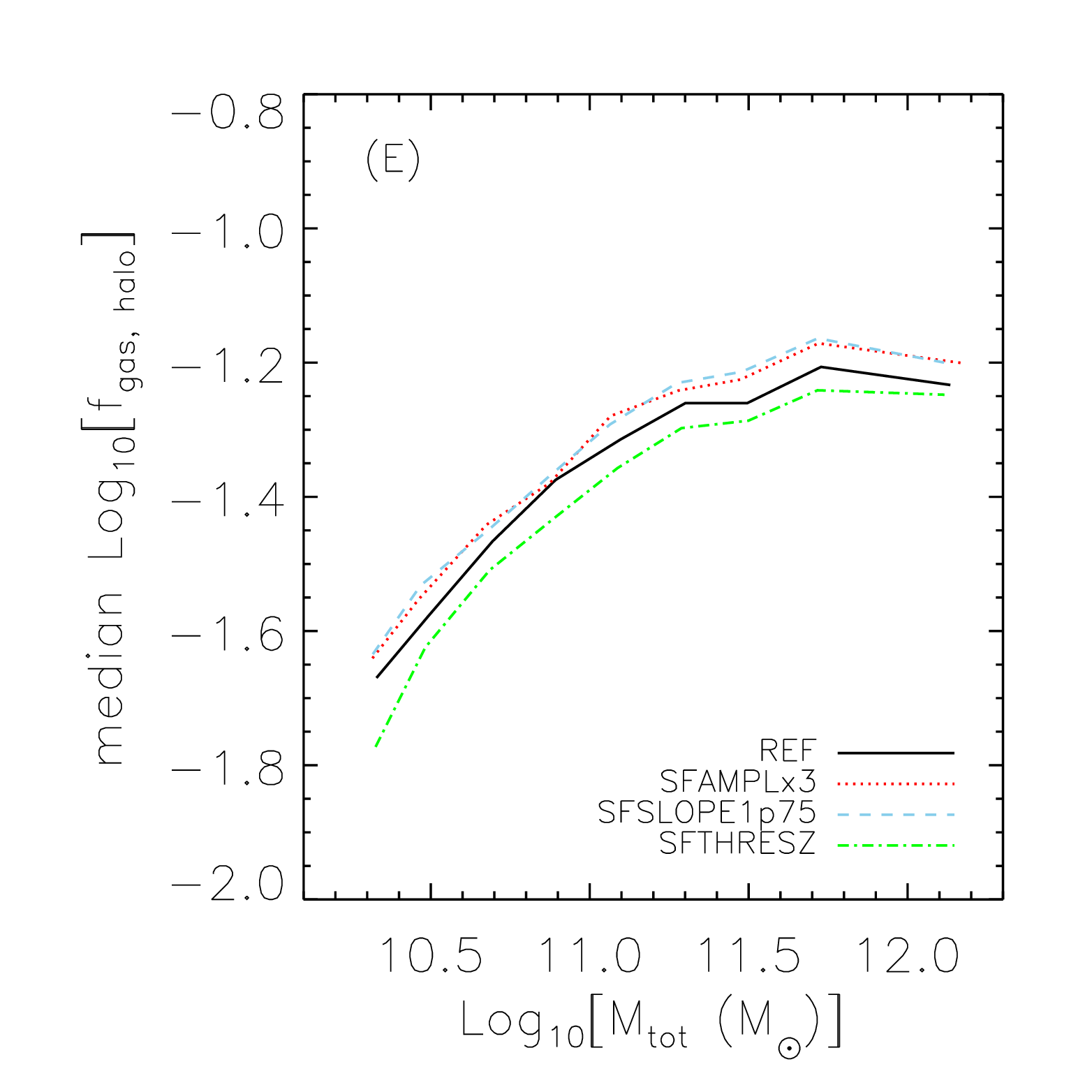}
\includegraphics[width=0.33\linewidth]{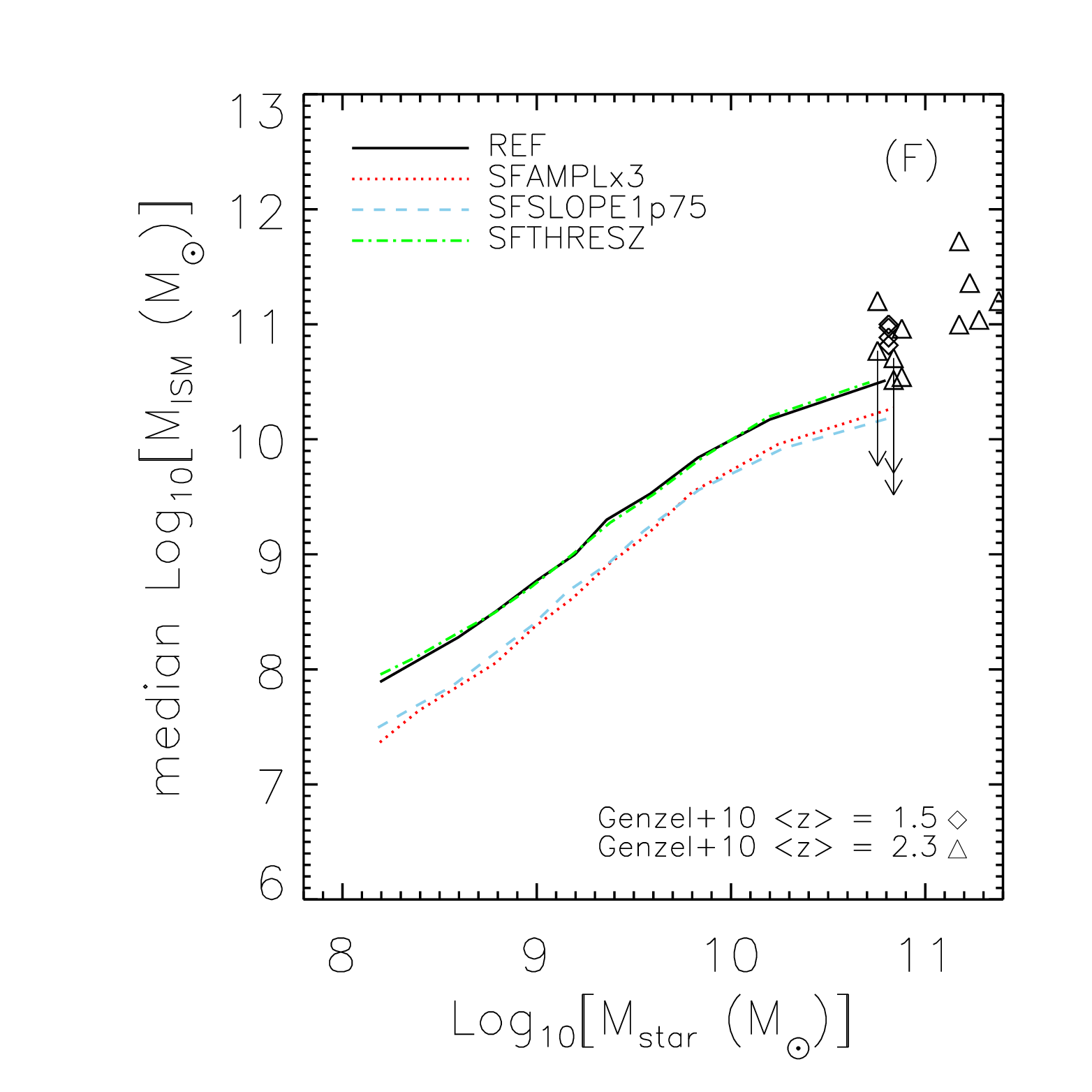} \\
\includegraphics[width=0.33\linewidth]{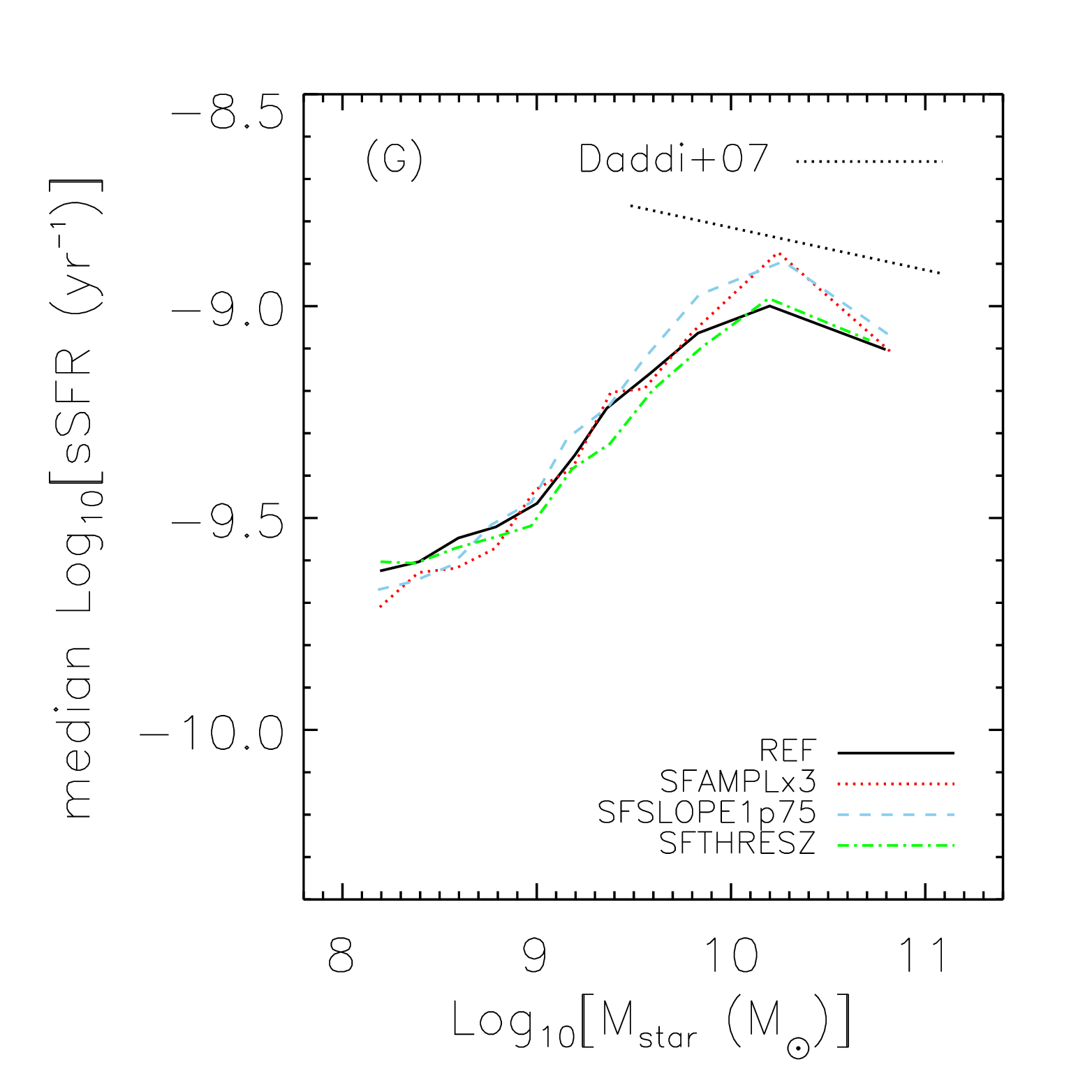}
\includegraphics[width=0.33\linewidth]{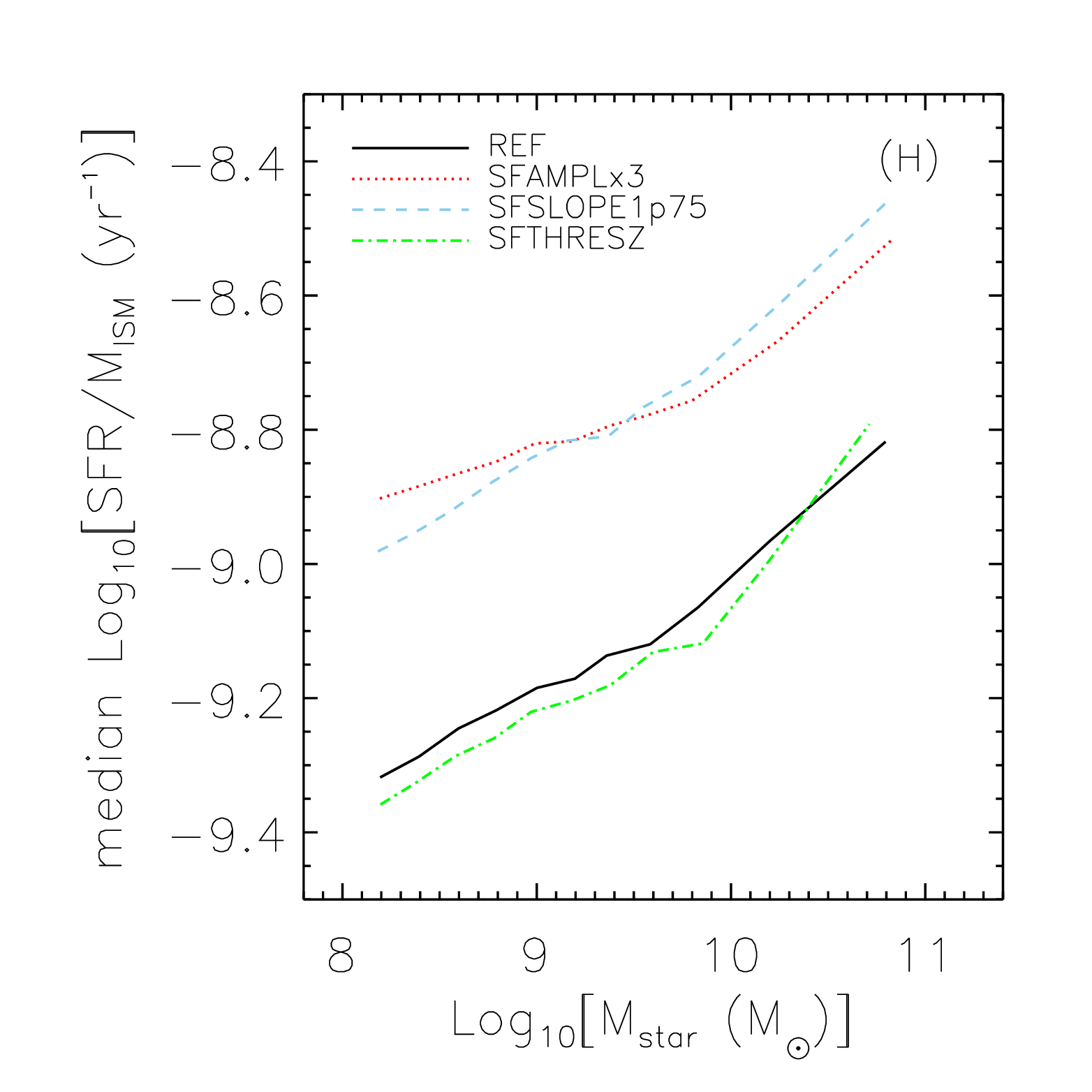}
\includegraphics[width=0.33\linewidth]{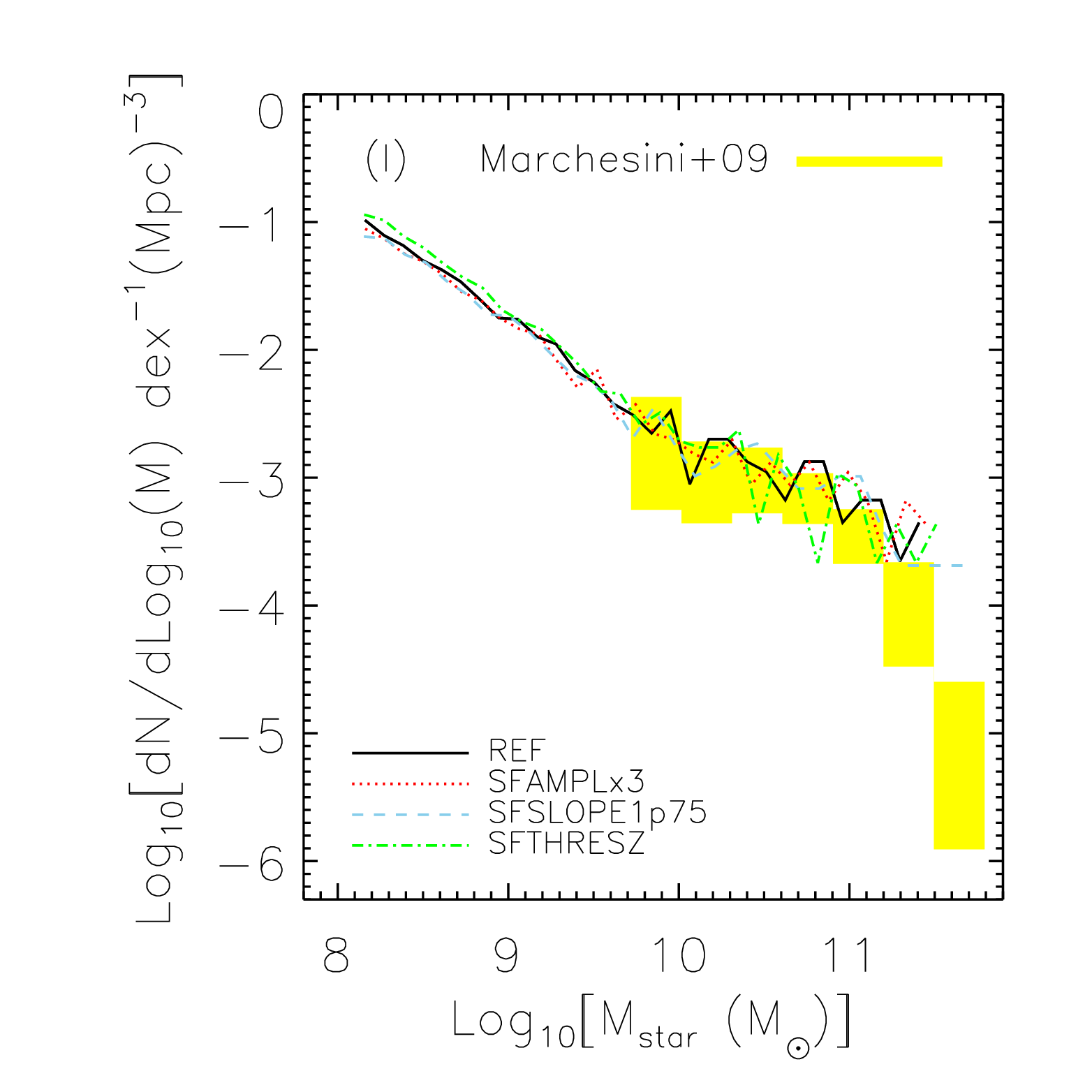} \\
\caption{As Fig.~\ref{fig:sims_cosmology}, but showing only the subset of simulations in which the parameters of the KS star formation law are changed. The `\textit{REF}' simulation (black, solid curve) uses a SF law that reproduces the observed KS law (Kennicutt 1998). The red, dotted curve shows a simulation where the amplitude of the KS law has been multiplied by a factor 3 (`\textit{SFAMPLx3}'), and the blue, dashed line shows the effect of changing the slope of the KS law from 1.4 to 1.75 (`\textit{SFSLOPE1p75}'). These two models both have more efficient SF than `\textit{REF}'. The green, dot-dashed curve shows a simulation in which the SF threshold is a function of the gas metallicity (`\textit{SFTHRESHZ}'), as predicted by \citet{schaye04}. The assumed star formation law affects the gas consumption time scale and the amount of star forming gas (ISM). It does not affect stellar masses or star formation rates, which can be understood if star formation is self-regulated (see text).} 
\label{fig:sims_sf} 
\end{figure*}

\noindent Because large-volume cosmological simulations have neither the resolution nor the physics to model SF, the standard approach is to implement a simple, stochastic volume density SF law and to calibrate it to match the observed Kennicutt-Schmidt surface density law, 
\begin{equation} \label{eq:KS}
\dot{\Sigma}_{*} = A (\Sigma_{\rm g} / 1 \textrm{\msun\ pc}^{-2})^n,
\end{equation}
where $n=1.4$ and $A=1.151\times 10^{-4}$ \msun yr$^{-1}$ kpc$^{-2}$ and the law steepens below a gas surface density $\Sigma_{\rm g} \sim 10$\,\msun pc$^{-2}$ \citep{kennicutt98}. In our simulations we do something similar, except that we implement SF as a pressure law \citep[see eqns 12, 14 and 18 in][]{schayedallavecchia08}, which has the advantage that no tuning is required, since we can convert the observed surface density law in a pressure law under the assumption of vertical hydrostatic equilibrium \citep{schayedallavecchia08}. As we have seen in the previous section, another advantage of this approach is that the predicted SFRs become independent of the assumed equation of state of the star forming gas \citep[see also][]{schayedallavecchia08,owls}. Following \citet{schayedallavecchia08}, the SFR of a star-forming gas particle of mass $m_g$ as a function of the pressure ($P$) is given by
\begin{equation}
\dot{m}_* = m_g A(1{\rm M_\odot\, pc}^{-2})^{-n} \Big( \frac{\gamma}{G} P \Big)^{(n-1)/2}
\end{equation}
where $\gamma=5/3$ is the ratio of the specific heats (not to be confused with the effective equation of state described in the previous section), and the parameters $A$ and $n$ are those of the KS-law in Eq.~\ref{eq:KS}.

In order to isolate the effect of changing the SF law, we compare the `\textit{REF}' simulation to three simulations in which the parameters of the Kennicutt-Schmidt are varied. Simulation `\textit{SFAMPLx3}' increases the normalization of the Kennicutt-Schmidt law, $A$, by a factor of three, which implies that for a gas particle at a given pressure, the SFR is a factor 3 higher in \textit{SFAMPLx3}' than in `\textit{REF}'. In simulation `\textit{SFSLOPE1p75}', the power-law slope of the KS-law is increased from $n=1.4$ to $1.75$. The normalisation of the Kennicutt-Schmidt law is chosen such that the SFR surface density is the same as `\textit{REF}' at $\Sigma_{\rm g} = 1$\, \msun $/$pc$^2$. As this is below the SF threshold, this KS-law forms stars more efficiently than the `\textit{REF}' simulation at all densities. The third variation on the `\textit{REF}' model, termed `\textit{SFTHRESHZ}' is a simulation in which the threshold density for SF, $n_{\textrm{\scriptsize H}}^*$, depends on the gas metallicity, $n_{\textrm{\scriptsize H}}^*\propto Z^{-0.64}$, such that the threshold density is equal to the reference simulation's SF threshold at a metallicity of $0.1 Z_\odot$ ($n_\textrm{H} = 0.1$ cm$^{-3}$) and there is a maximum allowed critical density of $n_{\textrm{\scriptsize H}}^*=10$\,cm$^{-3}$. This change mimics the metallicity dependence of the critical surface density for the formation of a cold, molecular phase predicted by \citet{schaye04}.

Panel (H) of Fig.~\ref{fig:sims_sf} shows that, as expected, the gas consumption time scale is significantly (about a factor of 3) shorter for `\textit{SFAMPLx3}' and `\textit{SFSLOPE1p75}' than for `\textit{REF}'. Model `\textit{SFTHRESHZ}' predicts nearly the same gas consumption time scale as `\textit{REF}', although SF is slightly less efficient in `\textit{SFTHRESHZ}' for all but the highest stellar masses. 

Intriguingly, the much shorter gas consumption time scale in models `\textit{SFAMPLx3}' and `\textit{SFSLOPE1p75}' has virtually no effect on the galaxy stellar mass (panels A and I) or its SFR (panels B and G). This agrees with \citet{owls}, who found that all these simulations predict nearly the same cosmic SF histories. 
This, initially surprising, result can be understood by considering the distribution of gas inside haloes.  Panels (C) and (E) demonstrate that the total baryon fraction inside haloes does not depend on the assumed SF law, but that the amount of ISM gas, i.e.\ the amount of star forming gas, is very sensitive to the same parameters (panels D and F).  Indeed, the response of the galaxies to a change in the gas consumption time scale, is to decrease the amount of star forming gas by a similar factor (compare panels D and H). This supports the explanation put forward by \citet{owls} that \emph{star formation is self-regulated by the interplay between the available fuel supply and feedback processes}: The feedback accompanying SF regulates the amount of gas condensing onto the ISM, and hence SF itself, so that outflow roughly balances inflow \cite[see also][]{bouche10,dave12}. As the outflow rate depends on the SFR rather than on the gas consumption time scale, the SFR, and hence the stellar mass, remain the same if the SF law is changed. To produce the same SFR with a different SF law, the galaxy will have to adjust the amount of star forming gas. 

Naively one would expect that the metallicity-dependent SF threshold should have a large effect on the amount of stars formed in a simulation (zero metallicity gas has a SF threshold 100 times higher than in the `\textit{REF}' simulation), but Fig.~\ref{fig:sims_sf} shows that this is not the case: the `\textit{SFTHRESHZ}' (green, dashed) curves closely track the `\textit{REF}' (black, solid) curved in all panels.  This is another manifestation of the self-regulation described above. If the gas in a galaxy is insufficiently dense to form stars, then there are no processes to prevent gas from condensing into the galaxy and increasing the gas density until SF becomes possible.  At this point the galaxy continues to form stars until the rate at which they inject energy into the galaxy is sufficient for the resulting galactic wind to balance the rate of gas inflow. Seemingly contrary to our results, e.g. \citet{gnedin11} found that a metallicity dependent cut-off is important for the SF properties of galaxies. The dependency on metallicity in those studies is, however, stronger than the one assumed here and they focus on low-mass galaxies. 

\citet{hopkins11} performed high-resolution galaxy simulations in which they varied the SF laws in a similar manner to that investigated in this section. Because of their much higher resolution ($M_{\rm star}\sim 10^2 - 10^3$\,\msun), which enables them to use a much higher SF threshold ($n_{\rm H} = 100~{\rm cm}^{-3}$), their feedback operates on a much smaller spatial scale. Because star particles form in the high-density peaks, their feedback is acting on much higher-density gas than it is in our simulations. \citet{hopkins11} find, in common with the results presented here, that the galaxy SFR is independent of the parameters of their SF law.  However, in apparent contrast to our results, they find little dependence of the amount of gas in the ISM on these same parameters. This apparent discrepancy can be understood by noting that in the simulations of \citet{hopkins11} there is a strong dependence of the high-density tail of the gas density distribution function on the SF law parameters.  They find that there is more gas at very high densities when the SF efficiency is lowered.  The difference, therefore, results from the scale at which the feedback is imposed on the gas. For their simulations this occurs at much higher densities, and the feedback therefore only regulates the high-density end of the gas distribution -- the gas where stars are actually forming. Our lower resolution simulations impose the feedback on larger scales and at densities comparable to our SF threshold, which coincides with our definition of the ISM (except for `\textit{SFTHRESHZ}'). Therefore, in both cases feedback from SF acts to regulate the amount of \emph{star-forming} gas. As we do not a priori know on what scales feedback regulates SF in real galaxies, it is not clear which gas densities are sensitive to the SF law.

\subsection{The stellar initial mass function} \label{sec:imf}

\begin{figure*}
\centering
\includegraphics[width=0.33\linewidth]{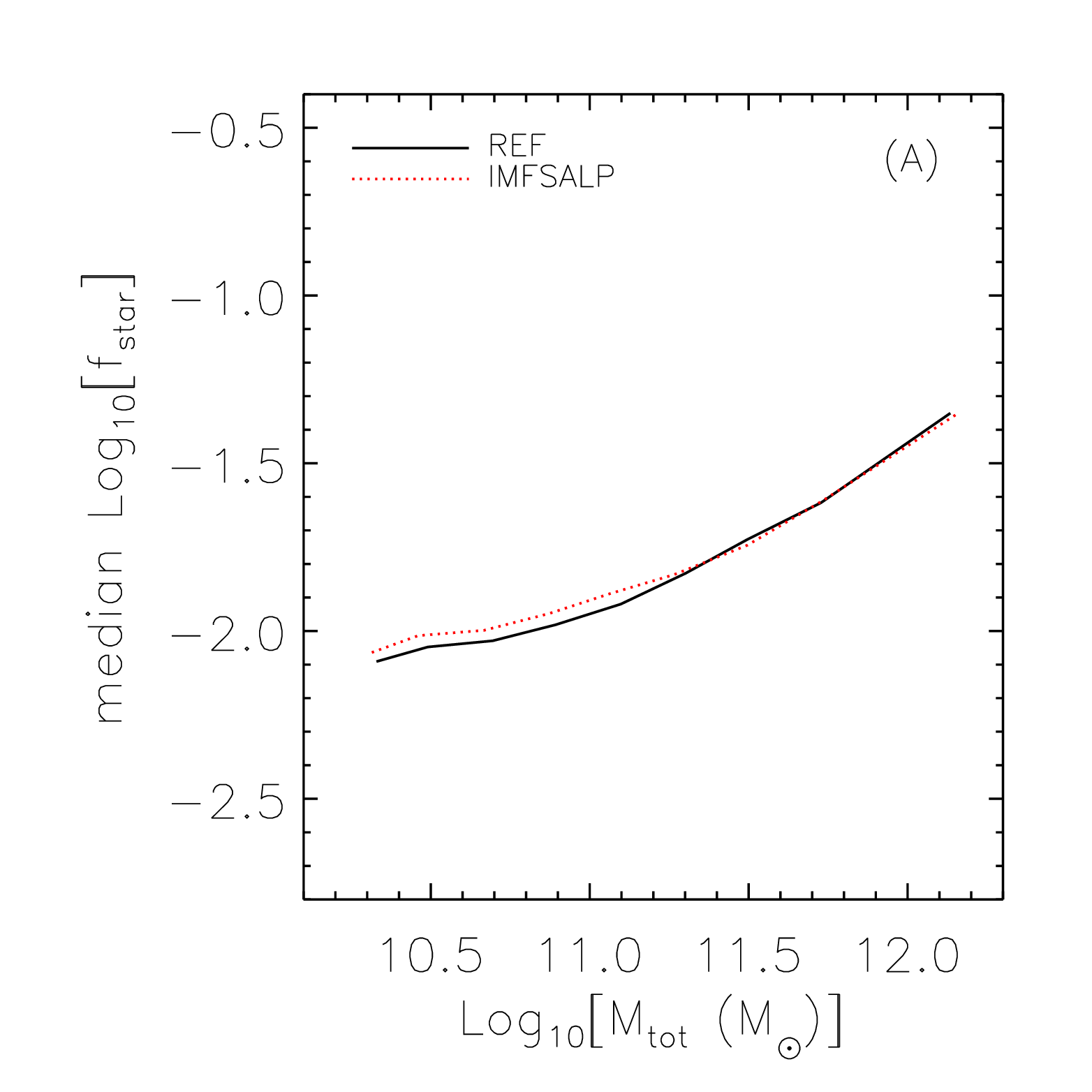}
\includegraphics[width=0.33\linewidth]{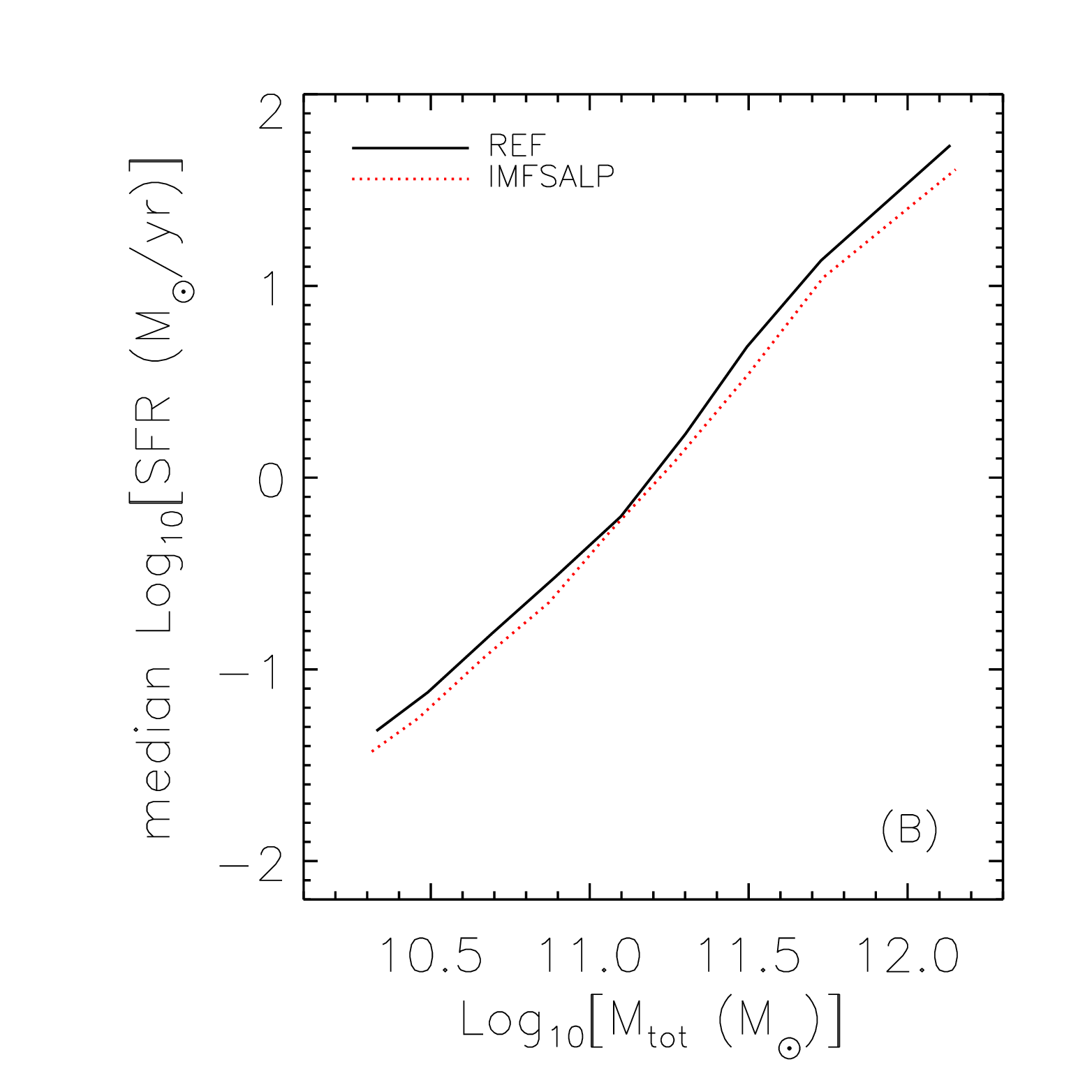}
\includegraphics[width=0.33\linewidth]{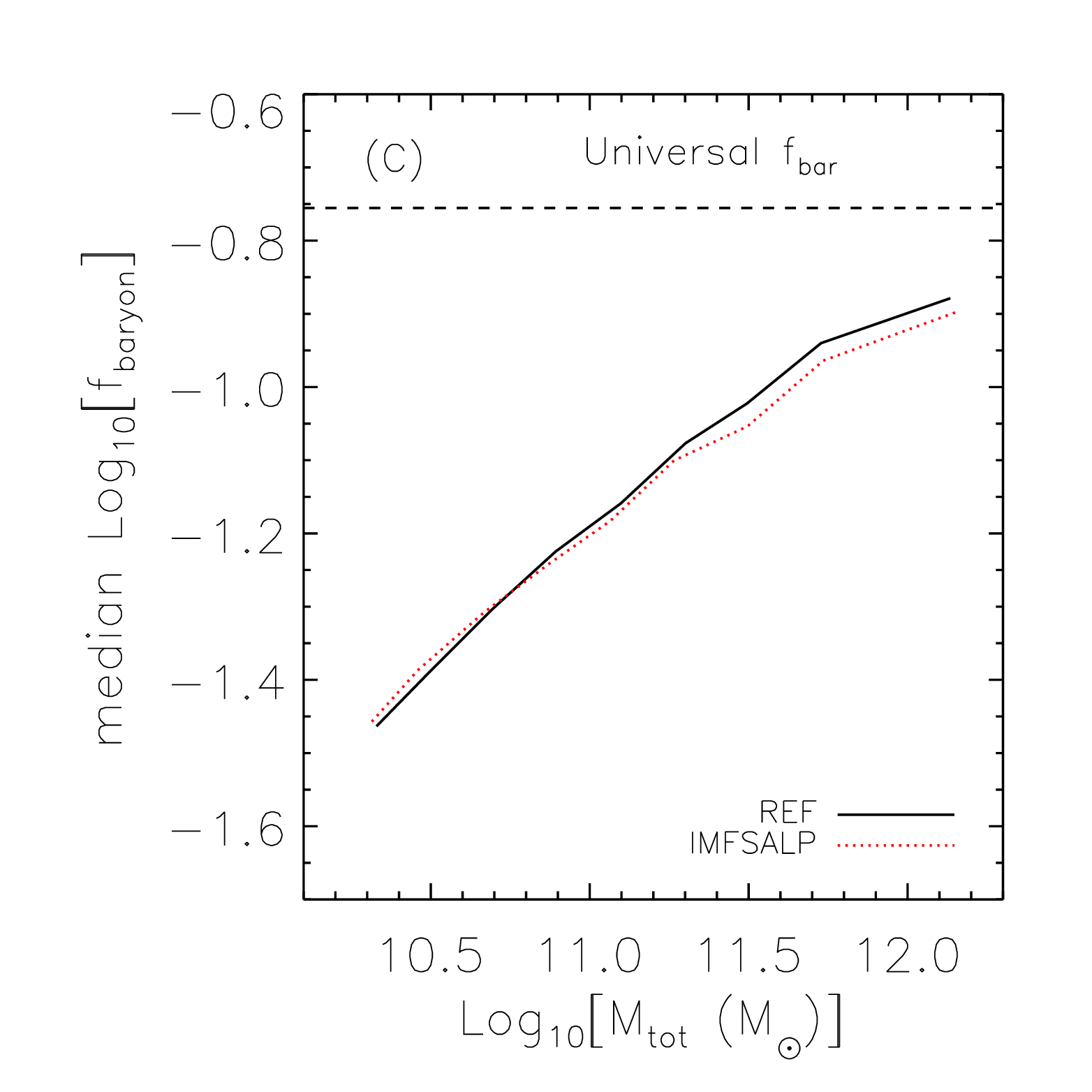} \\
\includegraphics[width=0.33\linewidth]{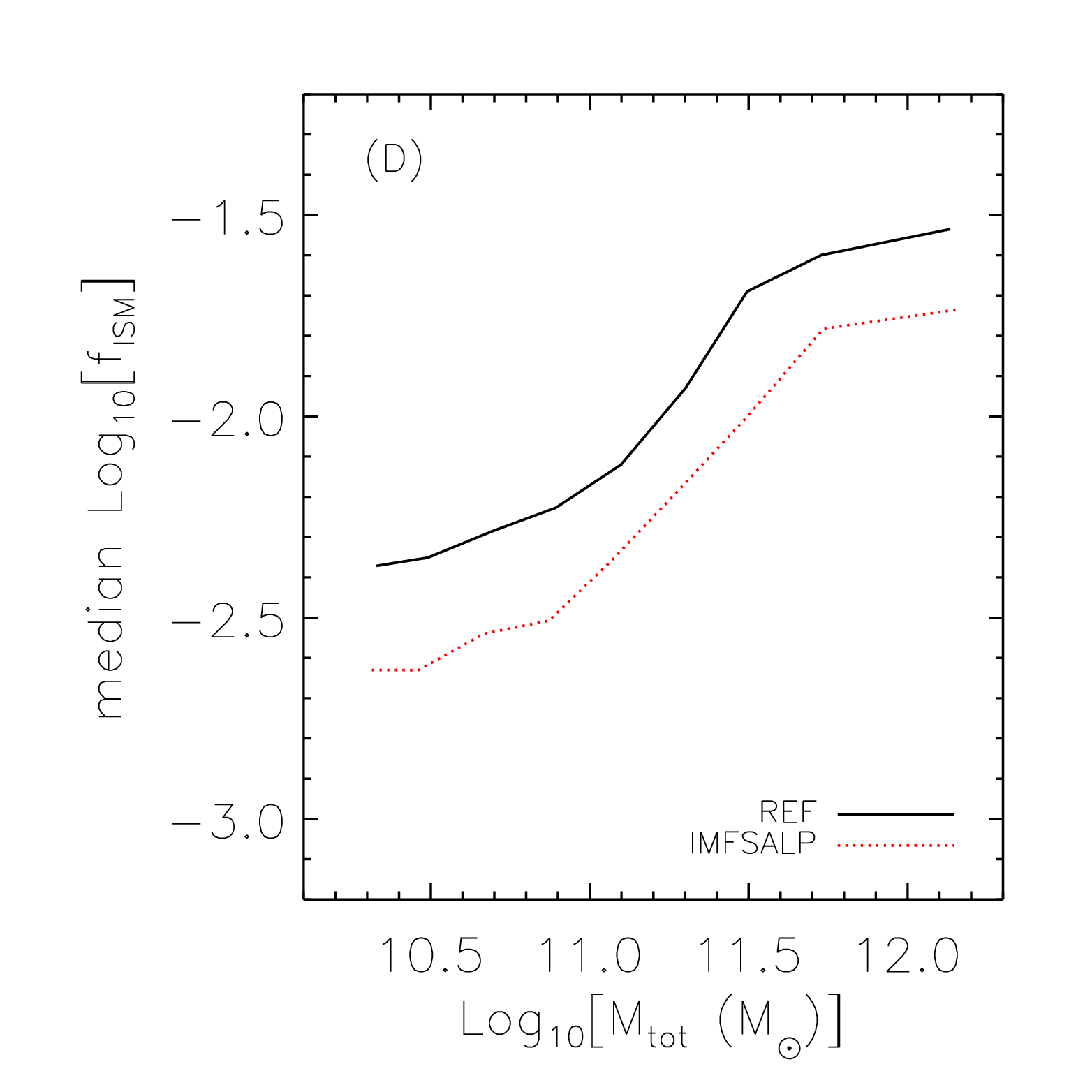}
\includegraphics[width=0.33\linewidth]{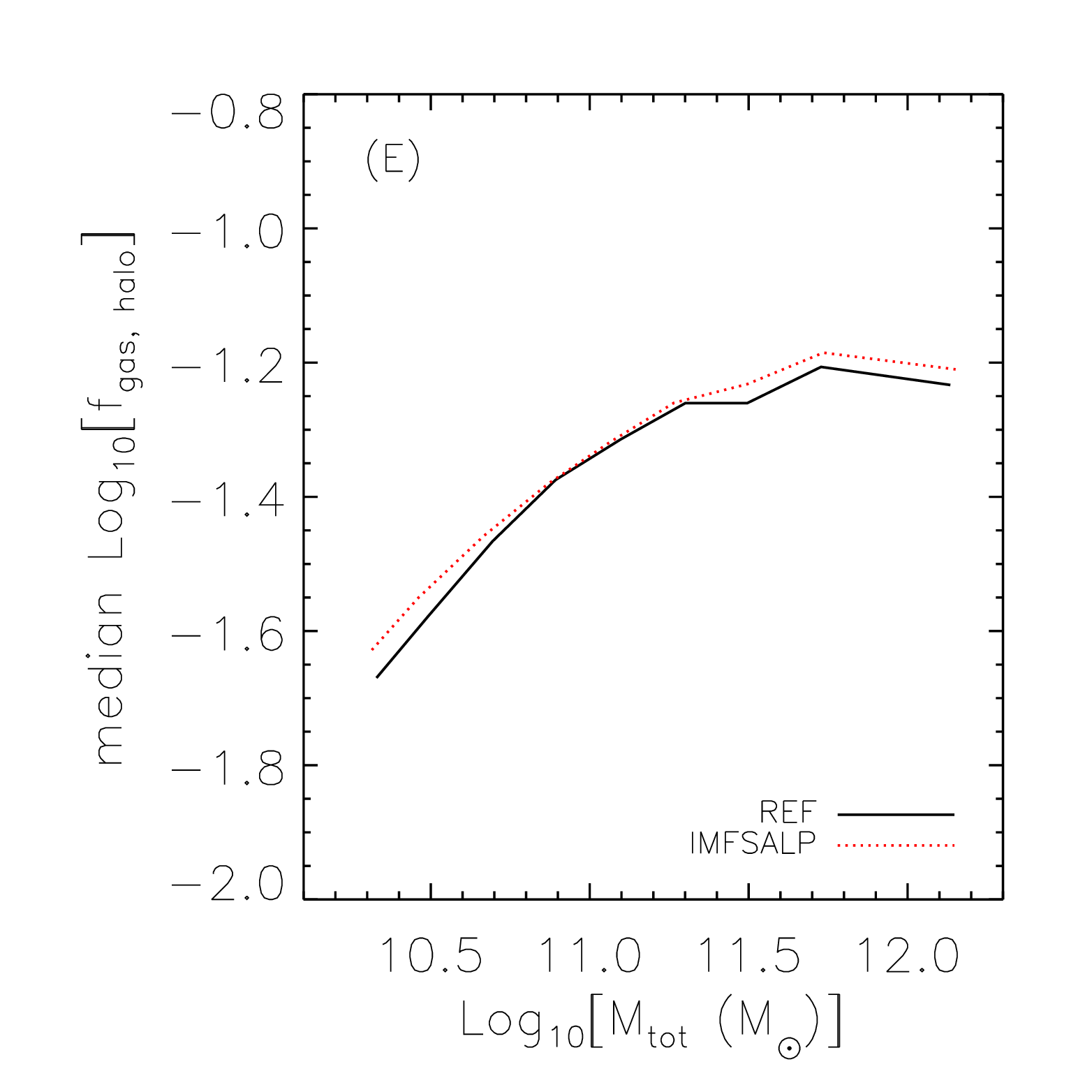}
\includegraphics[width=0.33\linewidth]{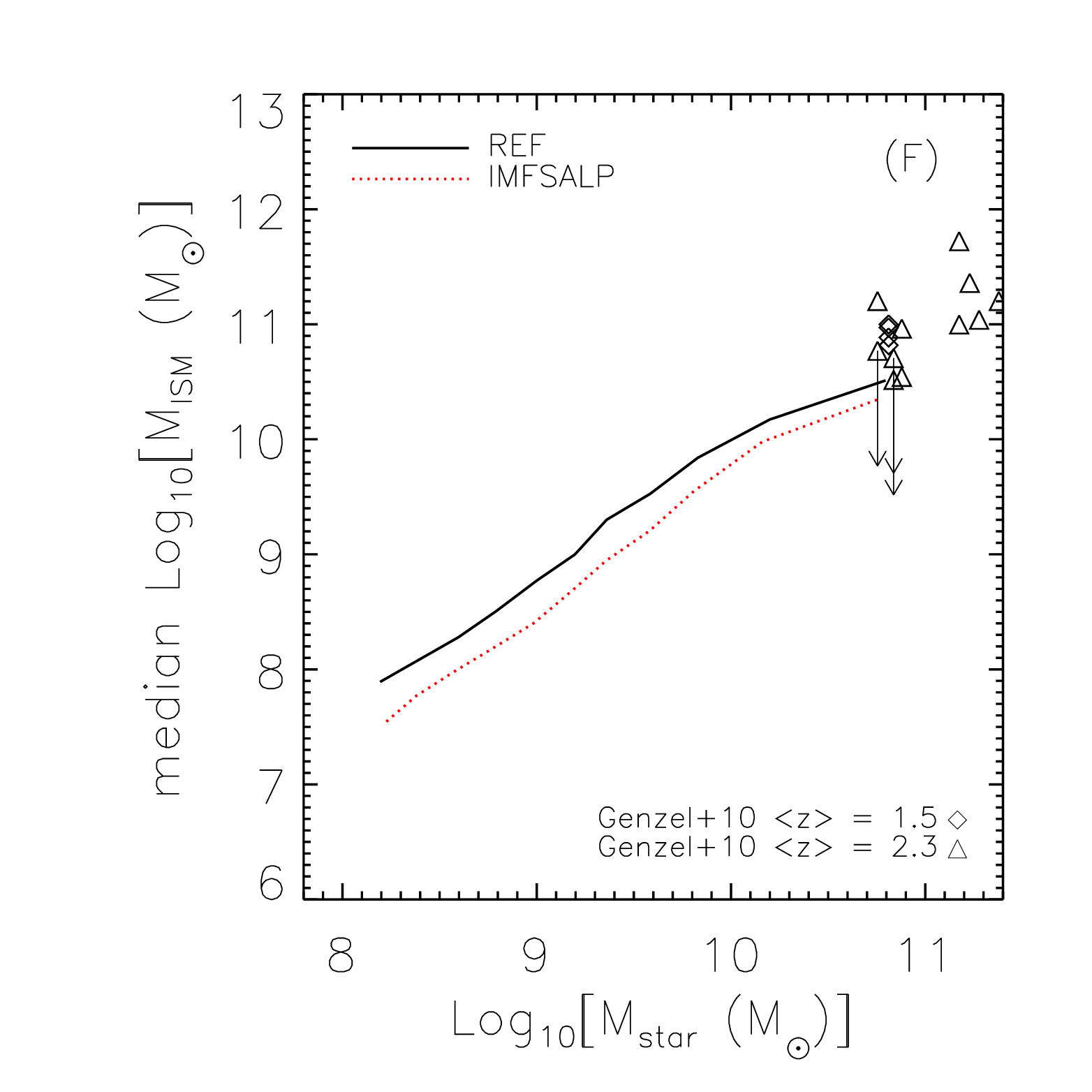} \\
\includegraphics[width=0.33\linewidth]{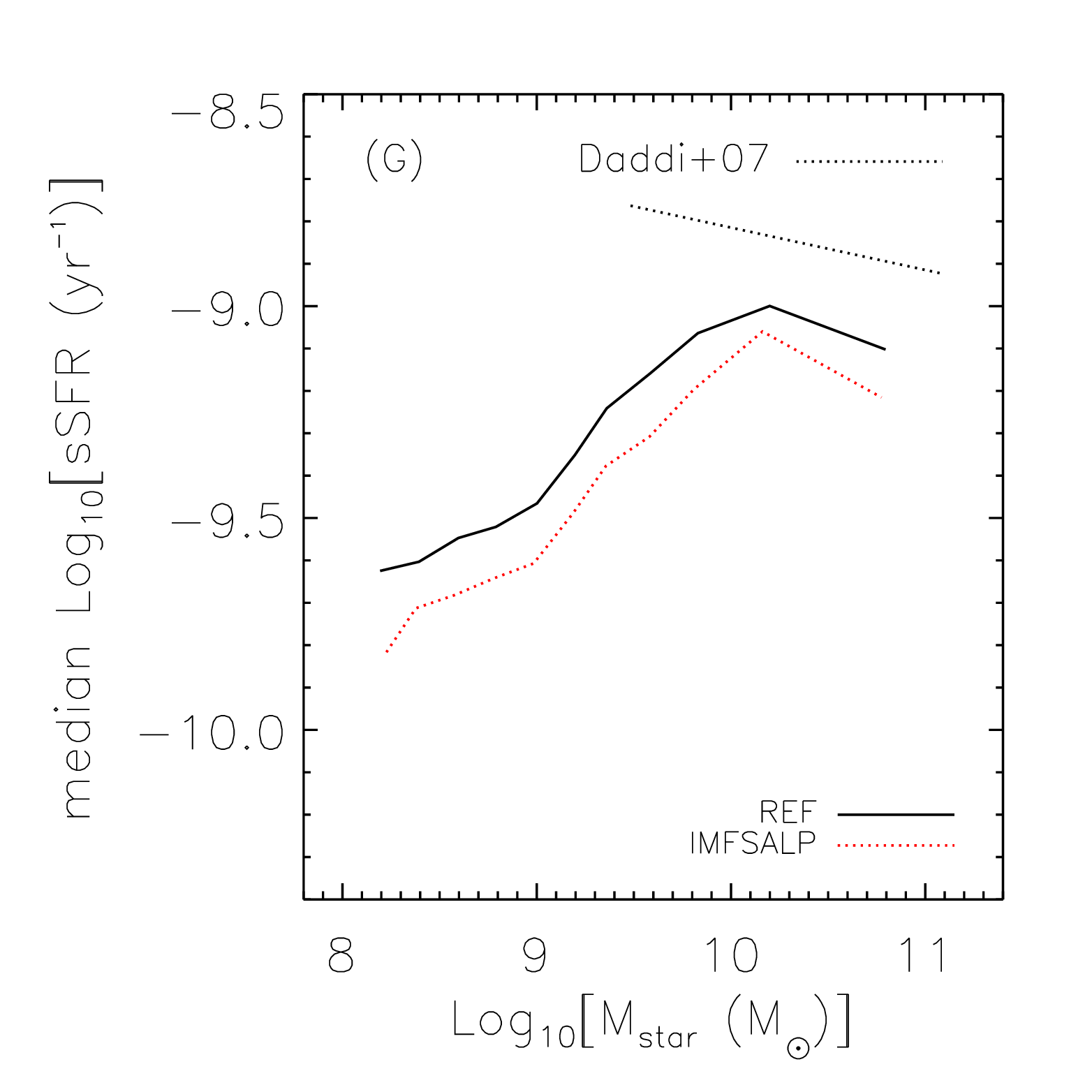}
\includegraphics[width=0.33\linewidth]{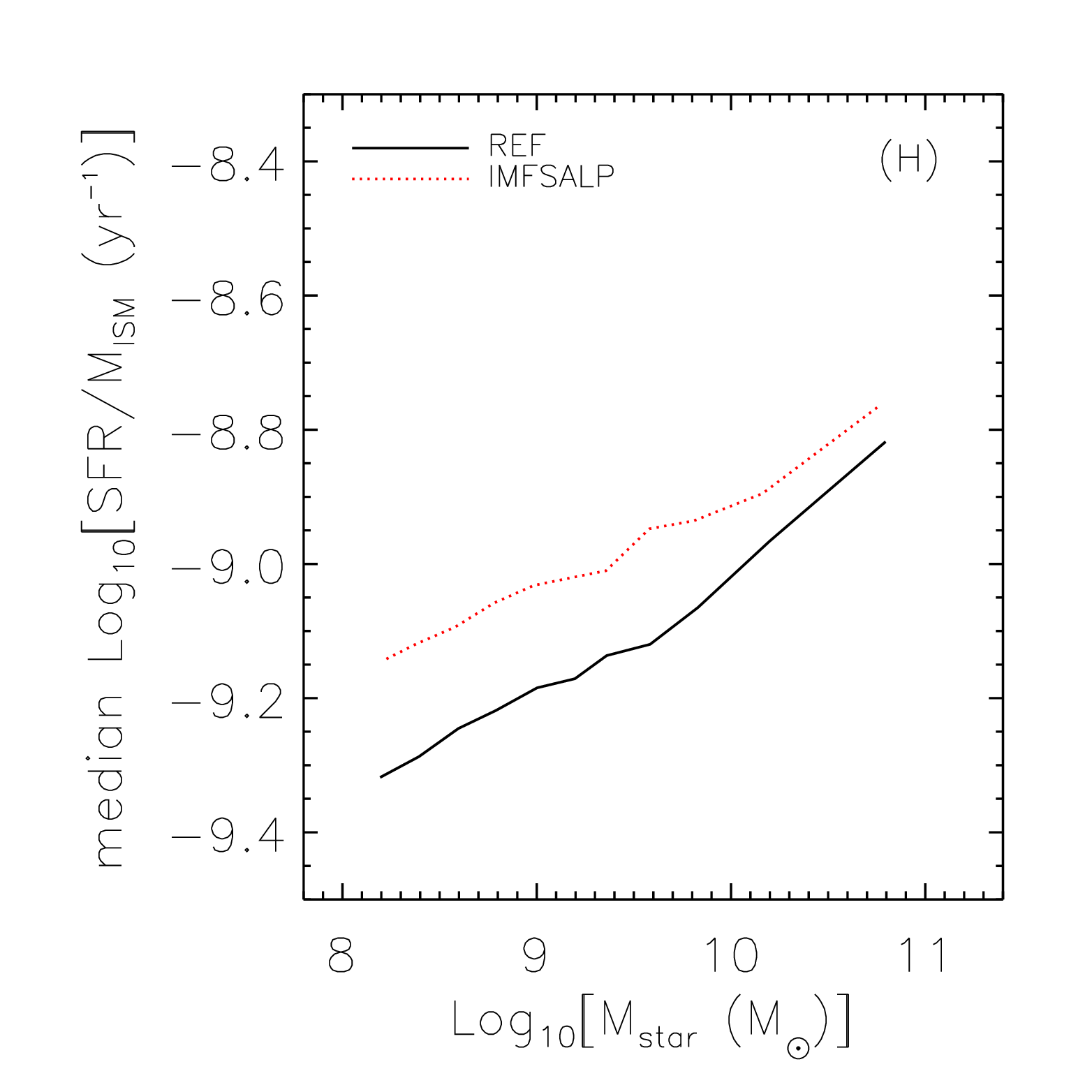}
\includegraphics[width=0.33\linewidth]{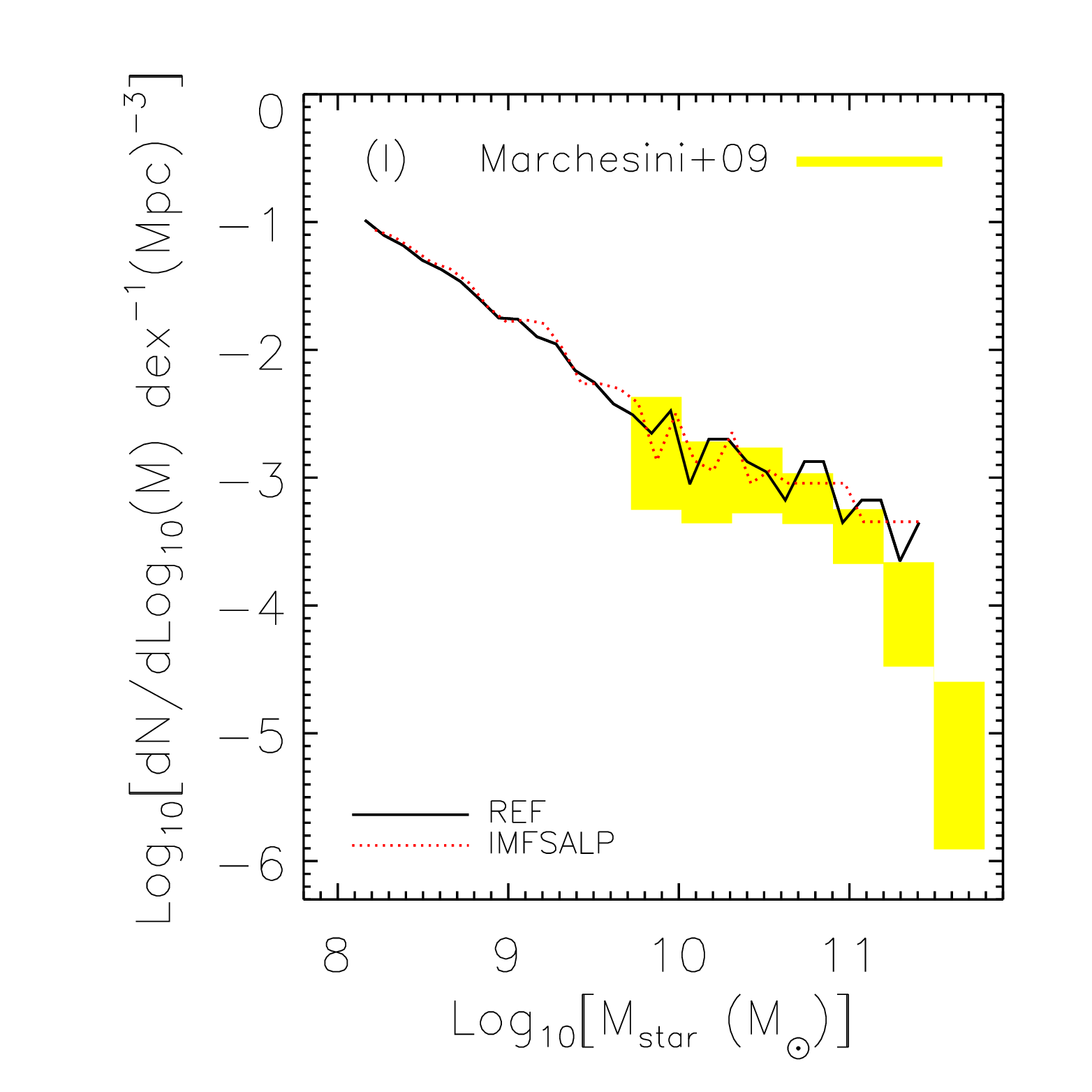} \\
\caption{As Fig.~\ref{fig:sims_cosmology}, but showing only the subset of simulations in which the stellar IMF is changed. The `\textit{REF}' simulation (black, solid curve) uses a \citet{chabrier03} IMF, whereas `\textit{IMFSALP}' (red, dotted curve) assumes a \citet{salpeter55} IMF. The star formation law has been rescaled for `\textit{IMFSALP}' to maintain consistency with the observations. This changes the gas consumption time scale (panel H) and, because of self-regulation, the ISM mass fraction (panels D and F). The other changes are small and can be attributed to the lower yields in `\textit{IMFSALP}'.} 
\label{fig:sims_imf} 
\end{figure*}

\noindent The assumed stellar IMF can potentially have a large effect on the galaxy population.  The IMF determines the ratio of low- to high-mass stars.  This, in turn, changes the integrated colours of stellar populations, their chemical yields and the number of SNe per unit stellar mass.

In this section we compare `\textit{REF}', which assumes a \citet{chabrier03} IMF to a simulation that assumes a \citet{salpeter55} IMF (`\textit{IMFSALP}').  Above 1 M$_\odot$ the two IMFs have similar shapes, but whereas the Salpeter IMF is a pure power law, the Chabrier IMF includes a log-normal decrease at the low-mass end, resulting in much lower mass-to-light ratios.  Both IMFs are integrated over the same mass range ($0.1-100$\,M$_\odot$).  We increase the normalization of the SF law by a factor of 1.65 in the `\textit{IMFSALP}' simulation to take into account the difference in the number of ionizing photons produced per unit stellar mass (the observed SF law is based on the flux from massive stars combined with an assumed IMF).  We do not, however, change the feedback parameters (wind velocity and mass loading are the same as in the reference model; for the effect of changing those parameters, see Paper I). Thus, because in a Salpeter IMF there are fewer high mass stars per unit stellar mass formed but we use the same feedback parameters, the SNe in `\textit{IMFSALP}' inject $40\% \cdot 1.65\approx66\%$ of the total available SN energy into their surroundings (i.e. $\sim 34\%$ is assumed to be lost radiatively).

Fig.~\ref{fig:sims_imf} demonstrates that the total stellar masses of the galaxies (panel A), as well as the total amount of baryons in the halo (panels C and E), are not strongly affected by the choice of IMF.  Inspection of Fig.~\ref{fig:prettypics} shows that the extent and morphology of the gaseous disk is also not strongly affected by the IMF.

However, at a given halo or stellar mass, SFRs (panel B) are somewhat lower in the `\textit{IMFSALP}' simulation.  This is because the Salpeter IMF produces less metals and returns less mass to the ISM than the Chabrier IMF.  This decreases the available fuel supply both by keeping more mass locked up in stars and by decreasing the effect of metal-line cooling (see also Paper I).

At a given mass, the amount of star-forming gas is significantly lower in `\textit{IMFSALP}' than in `\textit{REF}' (panels D and F).  This occurs because the higher SF efficiency implied by the change in the SF law means that for a Salpeter IMF more stars are formed at a given gas density (see panel H).  As discussed in Sec.~\ref{sec:sf} this means that the stellar population is injecting more energy into its surroundings at a given gas density (recall that the wind parameters are the same in the two simulations, only the SF efficiency has changed), and this allows the galaxy to self-regulate at lower gas densities.

%%%%%%%%%%%%%%%%%%%%%%%%%%%%%%%%%%%%%%%%%%%%%%%%%%%%%%%%%%%%%%%%%%%%%%%%%%%%%%%%%%%%%%%%%%%%%%%%

\section{Conclusions} \label{sec:conclusions}
We have analysed a large set of high-resolution cosmological
simulations from the \owls\ project \citep{owls}. We focused on the
baryonic properties of (friends-of-friends) haloes at redshift 2,
while varying parameters in the sub-grid models for  reionization, the
pressure of the unresolved multiphase ISM, star formation (SF), the stellar initial mass function, as well as the cosmology. 

This paper complements Paper I, which focuses on variations in the energetic feedback from star formation and the effects of metal-line cooling and feedback from AGN.

Our main conclusions can be summarised as follows:
\begin{itemize}
\item The stellar mass, SFRs, and gas fractions of galaxies are
  insensitive to the stiffness of the equation of state that we impose
  on the unresolved, multiphase ISM. This is a consequence of our use
  of the \citet{schayedallavecchia08} recipe for SF, which ensures
  that the observed surface density law for SF is satisfied
  independent of the assumed equation of state.
\item The gas fraction of galaxies, i.e. the fraction of the baryonic mass in star-forming gas, is sensitive to the assumed SF
  law. If SF is assumed to be more efficient, i.e.\ if the gas consumption time
  scale at a fixed gas surface density is shorter, then the gas fraction
  will become lower. This is a result of self-regulation: the gas fraction increases until the formation rate of massive stars is sufficient to drive galactic winds that can balance the rate at which gas accretes onto the galaxies. As a consequence, the SFRs and stellar masses are insensitive to the assumed SF law.
\item In a cosmology with a higher $\sigma_{8}$ structure formation
  happens earlier. Hence, galaxies residing in a halo of
  a fixed mass at a fixed time have somewhat higher stellar masses. The characteristic densities are also higher, which reflects the higher density of the Universe at the time the halo formed. These higher densities, in turn, cause feedback from SF to become inefficient at slightly lower masses if $\sigma_{8}$ is higher. The differences in halo properties between different cosmologies are, however, much smaller than the differences between the cosmic SF histories we found in \citet[][see also \citealt{springelhernquist03}]{owls}. This is because the halo mass function is sensitive to cosmology, which is more important for the SF history than the relatively small change in the internal properties of the galaxies at a fixed time and halo mass.
\item For low-mass haloes ($M\lesssim 10^{11}\,$M$_\odot$, $M_\ast \lesssim 10^9\,$M$_\odot$) the reheating associated with reionization is important, although by $z=2$ the results are insensitive to the redshift at which reionization happened, at least as long as it happened no later than $z=6$, as required by observations \citep{fan06}. Without reionization, these haloes would host higher-mass galaxies with higher gas fractions.
\item The IMF mainly affects the physical properties of galaxies
  because it determines the amount of energy and momentum that massive
  stars can inject into the ISM per unit stellar mass formed. If this
  feedback efficiency is kept fixed when the IMF is changed, then the
  assumed SF law must change if it is to remain consistent with
  observations. This will in turn modify the gas fractions, as
  explained above. The IMF also changes the metal yields and the rate at
  which mass is returned from stars to the ISM (recycling), but these
  effects play a smaller albeit still significant role.
\end{itemize}

There are inconsistencies between observations and our simulations. On the one hand, the simulated sSFRs at fixed stellar mass are lower than observed, and the relation between the two is steeper than observed in the observed range, which is in the mass range where SN feedback in the simulations in this paper is ineffective. In the regime where SN feedback is efficient, the slope of the relation is inverted (rising sSFR with stellar mass). The stellar mass function falls within the observational errors for almost all observed masses, except possibly the very lowest masses, where the simulations likely overproduce the number of galaxies.

We conclude, also on the basis of the results presented in Paper I, that the integrated physical properties of galaxies are mainly determined by the efficiency of the feedback from star formation and AGN, while the efficiency of star formation (i.e. the star formation law) determines the gas fractions. To make further progress, it is therefore most important to improve our understanding of galactic winds. 

%%%%%%%%%%%%%%%%%%%%%%%%%%%%%%%%%%%%%%%%%%%%%%%%%%%%%%%%%%%%%%%%%%%%%%%%%%%%%%%%%%%%%%%%%%%%%%%%

\section*{Acknowledgements}
The authors thank the anonymous referee for a helpful report. We are grateful to the members of the \owls\ collaboration for useful discussions about, and comments on, this work. The simulations presented here were run on Stella, the LOFAR BlueGene/L system in Groningen, on the Cosmology Machine at the Institute for Computational Cosmology in Durham (which is part of the DiRAC Facility jointly funded by STFC, the Large Facilities Capital Fund of BIS, and Durham University) as part of the Virgo Consortium research programme, and on Darwin in Cambridge. This work was sponsored by the National Computing Facilities Foundation (NCF) for the use
of supercomputer facilities, with financial support from the
Netherlands Organization for Scientific Research (NWO),
also through a VIDI grant. The research leading to these results has received funding from the European Research
Council under the European Union's Seventh Framework Programme (FP7/2007-2013) / ERC Grant agreement 278594-GasAroundGalaxies
and from the Marie Curie Training Network CosmoComp (PITN-GA-2009-238356). VS acknowledges support
through SFB 881, ‘The Milky Way System’, of the DFG.

\bibliographystyle{mn2e}
\bibliography{nonfb}

%%%%%%%%%%%%%%%%%%%%%%%%%%%%%%%%%%%%%%%%%%%%%%%%%%%%%%%%%%%%%%%%%%%%%%%%%%%%%%%%%%%%%%%%%%%%%%%%

\label{lastpage}

\end{document}